\documentclass[manuscript,screen, nonacm]{acmart}
\AtBeginDocument{%
  }

\usepackage{mathtools}
\usepackage{tikz}
\usepackage{hyperref}
\usepackage{todonotes}
\usepackage{calc}
\usetikzlibrary{fit,calc,positioning}
\usepackage{tikz-cd}
\usepackage[inline]{enumitem}
\newcommand{\TA}{{\mathsf{TA}}}

\newcommand{\FTA}{{\mathsf{FTA}}}

\newcommand{\Sig}{\mathsf{Sig}}
\newcommand{\Mod}{\mathsf{Mod}}
\newcommand{\Sen}{\mathsf{Sen}}

\newcommand{\Set}{\mathbb{S}\mathsf{et}}

\newcommand{\A}{\mathfrak{A}}
\newcommand{\B}{\mathfrak{B}}

\newcommand{\act}{\mathfrak{a}}
\newcommand{\bact}{\mathfrak{b}}
\newcommand{\cact}{\mathfrak{c}}
\newcommand{\Space}{~~~~~}

\newcommand{\et}{\mathsf{e}}
\newcommand{\pos}[1]{{\langle}#1{\rangle}}

\newcommand{\Forall}[1]{\forall #1\,{\cdot}\,}
\newcommand{\Exists}[1]{\exists #1\,{\cdot}\,}
\newcommand{\Frac}[2]{\displaystyle\frac{#1}{#2}}
\newcommand{\red}{\mathord{\upharpoonright}}

\usepackage[scr=boondoxo]{mathalpha}
\newcommand{\err}{\mathscr{r}}
\newcommand{\ltrue}{\top}

\newcommand{\bigland}{\bigwedge}

\newcommand{\lif}{\mathrel{\textsf{if\/}}}
\newcommand{\var}{\operatorname{var}}
\newcommand{\mgu}{\operatorname{mgu}}
\newcommand{\nf}[1]{\operatorname{nf}(#1)}
\usepackage{scalerel}
\newcommand{\bbsemicolon}{%
  \scalerel*{%
    \hbox{\usefont{U}{bbold}{m}{n} ;}%
  }{;}%
}
\newcommand{\comp}{\mathbin{\bbsemicolon}}
\newcommand{\gap}{\circ}
\newcommand{\case}[2][0pt]{%
  \setbox0=\hbox{\ensuremath{#2}}%
  {\smash{\fbox{\makebox[\maxof{\wd0}{#1}]{\strut\ensuremath{#2}}}}}
}
\newcommand{\tacase}[1]{%
  {\fbox{\makebox[1.2em][c]{\strut\ensuremath{#1}}}}%
}
\newcommand{\eq}{\textsc{eq}}
\newcommand{\tr}{\textsc{tr}}

\usetikzlibrary{decorations.markings}
\usetikzlibrary{decorations.pathmorphing}
\tikzset{
  negate/.style={
    decoration={
      markings,
      mark= at position 0.5 with {
        \node[transform shape] (tempnode) {\(/\)};
      },
    },
    postaction={decorate},
  },
  squiggly/.style={
    line join=round,
    decorate, decoration={
      zigzag,
      segment length=4,
      amplitude=.6,post=lineto,
      post length=2pt
    },
  },
}
\newcommand{\rewrite}[2][]{%
  \setbox0=\hbox{\(\scriptstyle #2\mkern3mu\)}%
  \mathrel{\tikz[baseline]
    \draw[->, >={Straight Barb[length=2pt, width=4pt]}, line width=.55pt, #1]
    (0pt,.5ex) -- (\wd0+.75em,.5ex)
    node [inner ysep=0pt, outer ysep=0pt, midway, above=1.5pt] {\(\scriptstyle #2\mkern3mu\)};}%
}
\newcommand{\revrewrite}[2][]{%
  \setbox0=\hbox{\(\scriptstyle \mkern3mu#2\)}%
  \mathrel{\tikz[baseline]
    \draw[<-, >={Straight Barb[length=2pt, width=4pt]}, line width=.55pt, #1]
    (0pt,.5ex) -- (\wd0+.75em,.5ex)
    node [inner ysep=0pt, outer ysep=0pt, midway, above=1.5pt] {\(\scriptstyle \mkern3mu#2\)};}%
}

\newcommand{\glrew}[2][]{%
  \rewrite[squiggly, #1]{#2}%
}

\newcommand{\rtmrew}{\rewrite}
\newcommand{\ltmrew}{\revrewrite}

\newcommand{\conv}{\downarrow}
\newcommand{\narrow}[2][]{%
  \setbox0=\hbox{\(\scriptstyle #2\mkern8mu\)}%
  \mathrel{\tikz[baseline]
    \draw[->>, >={Straight Barb[length=2pt, width=4pt]}, line width=.55pt, #1]
    (0pt,.5ex) -- (\wd0+.75em,.5ex)
    node [inner ysep=0pt, outer ysep=0pt, midway, above=1.5pt] {\(\scriptstyle #2\mkern8mu\)};}%
}
\theoremstyle{definition}
\newtheorem{definition}{Definition}
\theoremstyle{plain}
\newtheorem{remark}{Remark}
\newtheorem{proposition}{Proposition}
\newtheorem{theorem}{Theorem}
\newtheorem{corollary}{Corollary}
\newtheorem{lemma}{Lemma}
\newtheorem{example}{Example}

\begin{document}

\title{Executable Specifications in Transition Algebra}

\author{Daniel G\u{a}in\u{a}}
\email{daniel@imi.kyushu-u.ac.jp}
\orcid{0000-0002-0978-2200}
\affiliation{%
  \institution{Kyushu University}
  \city{Fukuoka}
  \country{Japan}
}

\author{Adri{\'{a}}n Riesco}
\email{ariesco@fdi.ucm.es}
\orcid{0000-0002-9716-4612}
\affiliation{%
  \institution{Universidad Complutense de Madrid}
  \city{Madrid}
  \country{Spain}
}

\author{Ionu\c{t} \c{T}u\c{t}u}%
\email{ittutu@gmail.com}
\orcid{0009-0004-4280-3036}
\affiliation{%
  \institution{Simion Stoilow Institute of Mathematics of the Romanian Academy}
  \city{Bucharest}
  \country{Romania}
}

\renewcommand{\shortauthors}{G\u{a}in\u{a}, Riesco and \c{T}u\c{t}u}

\begin{abstract}
  Transition Algebra can be viewed as an extension of many-sorted first-order logic by borrowing essential features from dynamic logic.
  This leads us to new, more expressive sentences that include transitions labelled with actions, which are built from terms using union, composition, and iteration operators.
  We define a suitable notion of Horn clause, develop a system of proof rules for reasoning about clauses, and subsequently show that the proof system is sound and complete.
  We then introduce a rewriting relation based on Horn clauses, capturing transitions along established paths expressed as actions.
  This rewriting relation is naturally extended to a narrowing procedure for solving queries, thereby laying foundation for logic programming in Transition Algebra.
  As usual, soundness for both rewriting and narrowing is straightforward, whereas completeness requires additional conditions.
  Finally, we provide a prototype tool that demonstrates the practical aspects and shows the effectiveness of the proposed method.
\end{abstract}

\ccsdesc[500]{Theory of computation~Logic and verification}
\ccsdesc[500]{Theory of computation~Equational logic and rewriting}
\ccsdesc[300]{Theory of computation~Constraint and logic programming}
\ccsdesc[300]{Theory of computation~Finite Model Theory}

\keywords{Transition algebra, Horn clause, rewriting, narrowing, calculi, tool support}


\maketitle

\section{Introduction}\label{section:intro}

\paragraph{Motivation}
As software systems become more complex and safety-critical, ensuring their correctness becomes not only desirable but essential. Subtle defects in such systems can lead to significant economic losses, security vulnerabilities, or even threats to human life, making rigorous verification a fundamental concern~\cite{bugs}. Traditional testing techniques, while valuable, are often insufficient to guarantee correctness across all possible scenarios~\cite{Dijkstra1970StructuredProgramming}. Consequently, there is a growing need for specification languages that are sufficiently expressive to capture intricate system behaviours, yet executable so that models can be validated, simulated, and analyzed efficiently. Such languages bridge the gap between formal reasoning and practical validation, enabling early detection of inconsistencies and providing scalable mechanisms to assess system correctness.
In this regard, a novel denotational semantics called Transition Algebra ($\TA$)~\cite{go-icalp24} was introduced for algebraic specification languages that are executable through rewriting. This semantics is defined by extending first-order logic with constructs inspired by dynamic logic~\cite{Harel00}, thereby enabling a more expressive and structured framework for describing state transitions.
%
Within the new framework, $\TA$ supports actions, which are composed from binary relations using standard relational operators: union~($\cup$), composition~($\comp$), and iteration~($*$).
Moreover, the algebraic structure of $\TA$ can be easily extended further to include additional action operators (e.g., tests, converse, or parallel composition).
To simplify the presentation and highlight the core ideas of the proposed method, this initial formulation focuses only on the original three relational operators.
Another notable feature of $\TA$ is that it offers a rigorous foundation, in logical terms, for a core fragment of the Maude strategy language~\cite{EkerMMRV23}. 
Maude~\cite{conf/maude/2007} is a specification language that implements Rewriting Logic~\cite{Meseguer90}, and also a state-of-the-art and highly efficient tool for equational reasoning and term rewriting.
The strategy language significantly enhances its expressive power by enabling fine-grained control over the application of transition rules—that is, the rules that govern how terms, representing system states, change through rewriting. By modelling strategies as algebraic constructs, $\TA$ offers a natural and uniform framework for reasoning about strategic control in rewriting systems.
It is worth noting that iteration $(*)$ significantly increases the expressive power of $\TA$ at the expense of other good properties such as compactness. 
For example, \cite{DBLP:conf/mfcs/GoG25} establishes a categoricity property for $\TA$, roughly showing that theories formulated in $\TA$ can control the cardinality of any of their models, including those of singular cardinality. This result holds under the Generalized Continuum Hypothesis and assuming the nonexistence of inaccessible cardinals.

\paragraph{Institutional approach}
The formulation of $\TA$ adheres to the principles of institution theory~\cite{gog-ins}, a general and abstract model-theoretic framework that formalizes the notion of a logical system through four key components: signatures, sentences, models, and a satisfaction relation between models and sentences.
A cornerstone of this framework is the satisfaction condition, a fundamental requirement for supporting modularity and composability in logical systems. The satisfaction condition ensures that the truth of sentences is preserved under translations induced by signature morphisms---structure-preserving mappings between signatures that coherently translate syntax and reduce semantics.
Much of the theoretical foundation of modern specification languages has been developed within the institutional framework~\cite{Diaconescu2025,SannellaT2012,Diaconescu12}, grounded in the principle that formal specification should be based on a well-defined underlying institution. 
Two prominent examples of specification languages that were explicitly designed following the institutional approach are CafeOBJ~\cite{dia-caf}, developed in Japan, and CASL~\cite{AstesianoBKKMST02}, developed in Europe. Both languages are built upon logical systems that have been rigorously formulated as institutions, ensuring strong support for modularity, compositionality, and interoperability between different logics.
A notable exception to this institution-based design philosophy is the Maude system~\cite{conf/maude/2007}. As mentioned above, Maude is based on Rewriting Logic~\cite{Meseguer90}, a framework which does not have, in full, an institutional formalization. As a result, the development of Maude has taken, in some respects, a different path from that of mainstream institution-theoretic approaches---particularly in areas such as modularization and logical heterogeneity~\cite{CodescuMRM10}.

\paragraph{Quantification}
In \cite{go-icalp24}, each signature is equipped with a separate set of plain, unstructured \emph{transition labels}, which are interpreted in models as \emph{atomic transition relations}.
In contrast, the present work adopts a more minimalistic approach: the signatures are ordinary algebraic signatures, and we use terms both to denote elements in models and to form atomic transition labels.
Concretely, given a term $t$ and a model $\A$ defined over a signature $\Sigma$, the interpretation $t^\A$ is an element of the carrier of $\A$, as in standard algebraic semantics.
In addition to this, for every sort $s$ in $\Sigma$, we associate $t^\A_s$ with a binary transition relation on the carrier set $\A_s$; that is, $t^\A_s \subseteq \A_s \times \A_s$.
This dual semantic role of terms---as both elements and binary relations---enables a flexible treatment of transitions within the model-theoretic framework.
This approach has a powerful consequence: it allows us to quantify over atomic transitions using ordinary first-order variables. This feature proves particularly useful in applications such as the formal specification of process calculi like Milner’s Calculus of Communicating Systems (CCS)~\cite{Milner80,Milner89}, where the ability to abstract over transitions is essential.
Effectively, this mechanism serves as a method of encoding second-order quantification (over transition relations) using first-order quantifiers. By avoiding the complexity of a second-order logic, we retain a simpler logical foundation while still enabling expressive reasoning about behaviours and interactions in concurrent systems.

\paragraph{Entailments}
In the present work, we focus on defining a fragment of $\TA$ with good computational properties, making specifications written in this logical framework executable via rewriting and narrowing.
To that end, we introduce a notion of Horn clause tailored to $\TA$, which appears in two distinct forms: 
\begin{enumerate*}[label=(\alph*)] 
\item equational, $\Forall{X}\ell = \err \lif H$, and
\item transitional, $\Forall{X}\tau(\ell, \err) \lif H$,
\end{enumerate*}
where $\ell$ and $\err$ are terms with variables from $X$, $\tau$ is also a term with variables from $X$ representing a transition label, and $H$ is a set of conditions consisting of equations and transitions along actions. Importantly, for transitional clauses, the head $\tau(\ell, \err)$ needs to be atomic, meaning that $\tau$ is not constructed from any other actions using operators such as union, composition, or iteration.
This differs from the approach taken by systems such as Maude and CafeOBJ.
The latter does not take into account labelled transitions, whereas the former admits them, but does not support explicit references to labels in conditions.
In contrast, in $\TA$, the use of labelled atomic transitions is unconstrained, as they can occur anywhere in clauses.
Moreover, the framework allows the conditions in $H$ to include actions that are arbitrarily complex and not necessarily atomic.
That is a key feature of our approach, allowing more expressive and flexible specifications.
We define a system of proof rules for reasoning about clauses in $\TA$, and we establish that the resulting entailment system is both sound and complete. Notably, our completeness proof does not rely on the use of a \emph{cut} rule. This means that properties of systems specified by clauses in $\TA$ can be proved without requiring the discovery of intermediate results---a process often referred to as lemma discovery in the formal-methods community.
However, permitting arbitrary actions in the conditions of clauses introduces a distinctive challenge: to establish the validity of a clause, it may be necessary to reason beyond the space of clauses themselves. This requirement compels us to go beyond traditional clause-based reasoning techniques, leading to both theoretical insights and practical consequences for the design of executable specification languages.
As a result, the completeness theorem established in this work departs significantly from those found in earlier approaches (e.g.,~\cite{journals/sigplan/GoguenM82,gai-cbl,CodescuG08}) where clause conditions are restricted to simpler forms, typically limited to equations or plain relations.

\paragraph{Rewriting}
The rewrite rules we employ in $\TA$ are derived from general Horn clauses by imposing restrictions on the occurrences of variables in order to ensure that the rules are executable---that is, to ensure that rules can be applied in rewriting and narrowing steps. We distinguish between two kinds of rewrite rules:
\begin{enumerate*}[label=(\alph*)]
\item equational rewrite rules of the form $\Forall{X}\ell = \err \lif H$, where the set of variables in the right-hand side is restricted by the condition $\var(\err) \subseteq \var(\ell) \cup \var(H)$, and
\item transitional rewrite rules of the form $\Forall{X}\tau(\ell, \err) \lif H$, with a similar restriction: $\var(\err) \subseteq \var(\ell) \cup \var(\tau) \cup \var(H)$.
\end{enumerate*}
In both cases, the hypothesis $H$ is a finite set of equations and transitions, which act as side conditions that must be satisfied for the rule to be applied. These conditions, while not themselves rewrite rules, may include complex compositions of actions and thus require their own operational semantics.
To manage this, we proceed in two stages.

First, we define a rewriting relation on sets of conditions.
To that end, it is more convenient to regard those sets of conditions as goals $G$ that are meant to be proved by rewriting, where proving a goal amounts to rewriting it to the empty set (meaning that all the conditions get discharged).
This preliminary definition is intentionally broad: to prove $G$ we may need to apply other rewrite rules, which may be conditional; unlike term rewriting, for goal rewriting we do not check the conditions of rules straight away; instead, we add them to $G$ so that they are checked at some point during rewriting.
Moreover, we do not commit to a specific mechanism for computing matching substitutions, nor do we assume that the matches are unique.
The purpose, at this stage, is to develop a framework for rewriting that is as general as possible, independent of any particular computational strategy.

Second, we define term rewriting using equational rewrite rules and, separately, term transitions using transitional rewrite rules. The latter captures the idea of rewriting along a path described by an action, thereby giving an operational interpretation to transitions in the system. This separation clarifies how terms transition in response to rules and how side conditions are validated through structured rewriting.
Within this framework, we establish a completeness result for rewriting, thus ensuring that a goal $G$ follows (semantically) from a set $\Gamma$ of clauses only if $G$ can be operationally reduced to the empty set through a sequence of rewriting steps. In symbols, $\Gamma\models G$ implies $G\glrew{*}\emptyset$. 
This result bridges the gap between the model-theoretic semantics of $\TA$ and its operational semantics defined via rewriting.
To achieve it, the set of rules needs to satisfy two key assumptions:
\begin{itemize}
\item \emph{Confluence} of equational rewriting rules. This guarantees the consistency and determinism of computations based on equational reasoning.
Confluence together with \emph{termination} are crucial for ensuring that the rewriting of terms with respect to equational rules yields unique normal forms, irrespective of the rewriting path taken.

\item \emph{Coherence} of transitional rewrite rules. This property ensures that the interaction between equational and transitional reasoning behaves uniformly, i.e., the application of transitional rules respects the structure induced by equational rewriting. Coherence avoids discrepancies between the logical entailment of transitions and their realizability via rewriting sequences.
\end{itemize}

\paragraph{Narrowing}
The definition of narrowing can be obtained directly from rewriting by replacing matching with unification.
We use the same rewrite rules as those used for rewriting, without restricting ourselves to unconditional transitional rewrite rules. 
We consider queries defined as existentially quantified goals, revisit the classical notion of solution to a query, and propose a suitable narrowing relation.
In this setting, we show that a query $\Exists{X} Q$ is satisfiable with respect to a set $\Gamma$ of clauses, meaning that there exists a substitution $\theta$ such that $Q\theta$ follows from $\Gamma$, if and only if $\Exists{X} Q$ can be reduced to a trivial query of the form $\Exists{X'} \emptyset$ through a sequence of narrowing steps that `compute' $\theta$ (or another suitable substitution that is more general than $\theta$).
In symbols, $\Gamma \models Q\theta$ if and only if $\Exists{X} Q\narrow{*}_{\psi} \Exists{X'} \emptyset$ for some substitutions $\psi$ and $\delta$ such that $\theta = \psi \comp \delta$.
The completeness of narrowing holds under similar conditions as those required for the completeness of rewriting: we need confluence, coherence, and also an additional termination requirement for goal rewriting.

To the best of our knowledge, the standard narrowing algorithm implemented in Maude 3.5 operates over rewrite theories that use only unconditional rules. This provides a solid foundation for equational reasoning and symbolic execution, but it limits the expressiveness of the rewrite theories that the system can handle.
To address this limitation, the framework has been extended in recent work to support rewrite theories with conditional rules, allowing more complex computational behaviours to be modeled directly. Notably, Meseguer~\cite{Meseguer20} and López-Rueda et al.~\cite{LopezRuedaES23} extended the standard narrowing algorithm with conditional rules that consist of SMT constraints, thereby broadening the scope of symbolic reasoning in Maude.
Roughly, the conditional rules include quantifier-free first-order formul\ae. 
This extension significantly enhances the expressiveness of the framework, enabling the formal specification and analysis of systems with rich logical constraints within the narrowing-based symbolic reasoning paradigm~\cite{MeseguerT05,EscobarMT05,ThatiM06,EscobarSM12}.
The narrowing relation proposed in this paper also supports unrestricted transitional rewrite rules, allowing for greater flexibility in the kinds of rewrite theories it can handle.
However, we achieve that flexibility entirely within the scope of Transition Algebra, meaning that, instead of SMT constraints we use plain $\TA$ conditions that get solved by rewriting or narrowing, thus avoiding calls to external SMT solvers.

\paragraph{Theorem proving}
In \cite{DBLP:conf/mfcs/GoG25}, we showed that $\TA$ is sufficiently expressive to control the cardinality of its own models. In particular, $\TA$ can naturally model both reachable and finite systems, two classes of structures that frequently arise in computer science. In this paper, we propose a verification method for proving properties of such models. The method is based on $\TA$ clauses and employs rewriting and narrowing techniques to discharge proof obligations. We illustrate the approach through two case studies: the natural numbers and the integers modulo a natural number. Although simple, these examples are particularly instructive, as they highlight both the strengths and the limitations of our framework. On the one hand, the method provides a high degree of automation in the verification process. On the other hand, the expressiveness and non-compactness of the underlying logical system entail certain inherent challenges, such as the necessity of employing the cut rule when reasoning about properties formalized by employing the transitive closures of relations despite the fact that completeness does not require the cut rule. 
This apparent contradiction arises from the fact that we employ infinitary proof rules, that is, proof rules with infinitely many premises. Such rules cannot, in general, be applied by a purely mechanical proof search procedure and must instead be justified by inductive arguments. Consequently, while our framework admits a highly automated treatment of many verification tasks through rewriting and narrowing, inductive reasoning remains indispensable for establishing the soundness of certain proof steps and for handling properties involving transitive closure. In this sense, the need for the cut rule should not be viewed as a deficiency of the method, but rather as a reflection of the intrinsic complexity of reasoning in a highly expressive setting, where finite proof objects are required to capture computational behaviours, possibly infinite.

\paragraph{Tool support} We have implemented a tool prototype for writing Transition Algebra specifications, reducing terms, checking transitions, and solving queries.
This was made possible by SpeX~\cite{Tutu22}, which is both a heterogeneous logical environment and a collection of Maude libraries that facilitate the development of new specification languages.
We also took advantage of the fact that Rewriting Logic is reflective~\cite{DBLP:journals/tcs/ClavelMP07}, and that Maude supports meta-level computations where modules and terms can be used as standard data.
Thus it has been possible to define a dedicated syntax for $\TA$ specifications, to parse specifications, transform them into internal Maude representations, and then use those representations to reduce terms and solve queries.
It is worth noting, however, that both the sentences and the execution mechanisms of Rewriting Logic and Transition Algebra are different, hence no direct translation is possible. For that reason, the tool that we have developed builds on the features provided by Maude to obtain specially designed resolution procedures for Transition Algebra, in line with the theoretical foundations described in the following sections.

Finally, we have worked in detail a medium-sized case study on process algebra in order to assess the effectiveness of both the logical framework and of the tool.
We have conducted the case study at two levels of abstraction.
First, at a more general level, we have developed a thorough $\TA$ specification of CCS; this showcases the expressiveness of $\TA$, as it allows us to capture the operational semantics of CCS with minimal conceptual and notational overhead.
Second, at a more particular level, building on the $\TA$ specification of CCS, we have given a formalization of the MapReduce programming model~\cite{DeanG08}, and then we have checked that an instance of it is correct using narrowing; this demonstrates the potential of the $\TA$ framework for formal reasoning.

\paragraph{Organization of the paper}
The rest of the paper is structured as follows: Section~\ref{section:transition-algebra} presents Transition Algebra.
Some preliminary concepts required to understand the rest of the paper are described in Section~\ref{section:preliminaries}.
Section~\ref{section:query} discusses conditions to ensure completeness of the proof rules for reasoning about queries. Sections~\ref{section:op-semantics} and \ref{sec:strict} 
are devoted to the operational semantics, while Sections~\ref{sec:op-complete} and~\ref{sec:birkhoff} discuss completeness. 
Section~\ref{sec:rifm} demonstrates the expressiveness of TA by presenting methods for modeling reachability and finiteness, as well as verification techniques for ensuring the correctness of reachable and finite systems.
Section~\ref{section:imp} presents the tool support developed for Transition Algebra and an important case study from the domain of concurrent and distributed systems, namely MapReduce, a widely used framework for processing large datasets across distributed computing infrastructures.
In Section~\ref{section:conc}, we present our conclusions and outline several directions for future work.


\section{Transition algebra} \label{section:transition-algebra}

In this section, we define \emph{the logic of many-sorted transition algebras}, or \emph{transition algebra}~($\TA$), for short.
We present, in order: signatures, models, sentences, and the \(\TA\) satisfaction relation.
\paragraph{Signatures}
We work with many-sorted algebraic signatures of the form $(S,F)$, where $S$ is a set of \emph{sorts} and $F=\{F_{w\to s}\}_{(w,s)\in S^*\times S}$ is an $S^*\times S$-sorted set of \emph{function symbols},
where $S^*$ denotes the set of strings over $S$.
We often write \(\sigma \colon w \to s \in F\) to indicate that \(\sigma \in F_{w, s}\), and we refer to \(w \in S^{*}\) and \(s \in S\) as the \emph{arity} and \emph{sort}, respectively, of the symbol \(\sigma\).
Under this notation, \(F\) can also be regarded as an ordinary set consisting of \emph{function declarations} of the form \(\sigma \colon w \to s\).
When \(w\) is the empty arity, we may speak of \(\sigma \colon \to s\) as a \emph{constant} (symbol) of sort \(s\).
Throughout the paper, we let 
$\Sigma$, $\Sigma'$, and $\Sigma_i$ range over arbitrary signatures of the form 
$(S, F)$, $(S', F')$, and $(S_i, F_i)$, respectively.

As usual in institution theory~\cite{Diaconescu2025,SannellaT2012}, important constructions such as signature extensions with constants as well as open formul\ae\ and quantifiers are realized in a multi-signature setting, so moving between signatures is common.
A $\TA$ \emph{signature morphism} \(\chi \colon \Sigma \to \Sigma'\) is just an ordinary algebraic signature morphism \(\chi \colon (S, F) \to (S', F')\).

\begin{remark}
  Signature morphisms compose componentwise.
  Their composition has identities and is associative, thus leading to a category $\Sig$ of signatures.
\end{remark}

\paragraph{Models}
Given a signature \(\Sigma\), a \emph{\(\Sigma\)-model} \(\mathfrak{A}\) is a $\Sigma$-algebra \(\mathfrak{A}\) equipped,
for each sort $s_1\in S$ and every element $\et \in \A_{s_1}$, with a family of \emph{$S$-sorted transition relations}
$\{\et_s  ~\subseteq~ \A_s\times \A_s\}_{s \in S}$.
That is, every element of the algebra is mapped to a family of transition relations indexed by the sorts.
A \emph{$\Sigma$-homomorphism} $h:\A\to \B$ is an ordinary algebraic \((S, F)\)-homomorphism that preserves transitions: 
$h(\et_s) ~\subseteq~ h(\et)_s$ for all  sorts $s,s_1\in S$ and all elements \(\et \in \A_{s_1}\).
$\Sigma$-homomorphisms form a subcategory, which we denote by $\Mod(\Sigma)$, of the category of algebraic homomorphisms.
\begin{remark}
Each signature morphism \(\chi \colon \Sigma \to \Sigma'\) determines a \emph{model-reduct functor} $\_\red_\chi\colon\Mod(\Sigma')\to\Mod(\Sigma)$ where:
\begin{enumerate}

\item For every \(\Sigma'\)-model \(\A'\), the $\Sigma$-model $\A=\A'\red_\chi$ is defined as follows:
\begin{itemize}
\item \(\A_s=\A'_{\chi(s)}\) for all sorts $s\in S$;

\item \(\sigma^\A = \chi(\sigma)^{\A'}\) for each function symbol \(\sigma \in F\); and

\item $\et_s \coloneqq \et_{\chi(s)}$ 
for all sorts $s,s_1\in S$ and all elements $\et\in \A_{s_1}=\A'_{\chi(s_1)}$.
\end{itemize}
\item For every \(\Sigma'\)-homomorphism \(h' \colon \A' \to \B'\), \((h' \red_{\chi})_{s} = h'_{\chi(s)}\) for any sort \(s \in S\).    
\end{enumerate}
 Moreover, the mapping \(\chi \mapsto \_\red_\chi\) is functorial.
\end{remark}

For any signature morphism $\chi:\Sigma\to\Sigma'$, any $\Sigma$-model $\A$, and any $\Sigma'$-model $\A'$, if $\A=\A'\red_\chi$, we say that $\A$ is the \emph{$\chi$-reduct} of $\A'$, and that $\A'$ is a \emph{$\chi$-expansion} of  $\A$.
For example, for any many-sorted set \(X\) (say, of variables) that is disjoint from the set of constant-function symbols in \(\Sigma\), consider the inclusion morphism $\iota_X:\Sigma\hookrightarrow\Sigma[X]$, where \(\Sigma[X] = (S, F[X])\) is the signature obtained from \(\Sigma\) by adding the elements of \(X\) to \(F\) as new constants of appropriate sort.
Then an expansion of a $\Sigma$-model $\A$ along $\iota_X$ can be seen as a pair $\A'=\pos{\A,\nu:X\to \A}$, where $\nu$ is a valuation of $X$ into the elements of $\A$.

As in many-sorted algebra, there is a special, initial model in \(\Mod(\Sigma)\), which we denote by \(T_{\Sigma}\), whose elements are ground terms built from function symbols, and whose transitions are all empty.
The \(\Sigma\)-model \(T_{\Sigma}(X)\) of terms with variables from \(X\) is defined as the $\iota_X$-reduct of $T_{\Sigma[X]}$. That is, $T_\Sigma(X)=T_{\Sigma[X]}\red_{\iota_X}$.
The following property is an immediate consequence of the initiality of \(T_{\Sigma}\).

\begin{remark}
  Any signature morphism \(\chi \colon \Sigma \to \Sigma'\) determines uniquely a \(\Sigma\)-ho\-mo\-mor\-phism \(T_{\Sigma} \to T_{\Sigma'} \red_{\chi}\).
  In order to simplify notations later on, we denote that homomorphism by \(\chi \colon T_{\Sigma} \to T_{\Sigma'} \red_{\chi}\).
  Therefore, for any \(\Sigma\)-term \(\sigma(t_{1}, t_{2}, \dotsc, t_{n})\), we have \(\chi(\sigma(t_{1}, t_{2}, \dotsc, t_{n})) = \chi(\sigma)(\chi(t_{1}), \chi(t_{2}), \dotsc, \chi(t_{n}))\).
\end{remark}
\paragraph{Sentences}
Given a signature $\Sigma$,
the set $A$ of \emph{actions} is defined by the following grammar:
\begin{center}
$\act ~\Coloneqq~
\tau ~\mid~
\act \comp \act ~\mid~
\act \cup \act ~\mid~
\act^*$
\end{center}
where $\tau$ is any term.
Notice that actions are heterogeneous, meaning that we can compose, take unions, and iterate over terms of different sorts.
We extend the notational convention that we use for the components of signatures to their corresponding sets of actions; that is, we usually denote
by \(A'\), \(A_{i}\), and \(A[X]\) the sets of actions over signatures \(\Sigma'\), \(\Sigma_{i}\), and \(\Sigma[X]\), respectively.
Moreover, through a slight abuse of notation, we also denote by \(\chi \colon A \to A'\) the obvious canonical map determined by a signature morphism \(\chi \colon \Sigma \to \Sigma'\).

To define sentences, we assume a countably infinite set of \emph{variable names} $\{v_i\mid i \in \mathbb{N}\}$.
A \emph{variable} for a signature $\Sigma$ is a triple $\pos{v_i,s,\Sigma}$, where 
\begin{enumerate*}[label=(\alph*)]
\item $v_i$ is a variable name and 
\item $s$ is a sort in $\Sigma$.
\end{enumerate*}
The third component is used only to ensure that variables are distinct from the constant-operation symbols declared in \(\Sigma\), which is essential when dealing with quantifiers; but in most situations, unless there is some danger of confusion, we omit the third component and refer to variables either in a qualified form, as in $v_{i} \colon s$, or just by their name, $v_{i}$.
A \emph{block of variables} for $\Sigma$ is a set of variables $X$ for $\Sigma$ such that for all $\pos{v_i,s_1,\Sigma},\pos{v_j,s_2,\Sigma}\in X$ we have $s_1\neq s_2$ implies $i\neq j$.
The set $\Sen(\Sigma)$ of \emph{sentences} over $\Sigma$ is given by the following grammar:
\begin{center}
  $\phi \Coloneqq  t_{1} = t_{2} \mid \act(t_1,t_2)\mid \neg\phi \mid \lor \Phi \mid \Exists{X}\phi'$
\end{center}
where
\begin{enumerate*}[label=(\alph*)]
\item $t_1$ and $t_2$ are $\Sigma$-terms of the same sort;
\item $\act$ is any action;
\item $\Phi$ is a finite set of $\Sigma$-sentences; and 
\item $X$ is a finite block of variables for $\Sigma$ and $\phi'$ is a $\Sigma[X]$-sentence.
\end{enumerate*}
A sentence of the form $\act(t_1,t_2)$ is called a \emph{transition}, which can also be written in infix notation $t_1 \mathrel{\act} t_2$.
An \emph{atomic sentence} is a ground equation  $t=t'$ or a ground transition $t_1\mathrel{\tau}t_2$ such that $\tau$ is an atomic action (i.e., a term).

When \(\Phi = \{\phi_{1}, \phi_{2}, \dotsc, \phi_{n}\}\), we may write \(\phi_{1} \vee \phi_{2} \vee \dotsb \vee \phi_{n}\) instead of \(\lor \Phi\).
Besides the above core connectives, we also make use of the following convenient (and standard) abbreviations:
$\land\Phi\coloneqq\neg\lor_{\phi\in\Phi}\neg\phi$ for finite conjunctions;
$\bot\coloneqq\lor\emptyset$ for falsity;
$\top\coloneqq\land\emptyset=\neg\bot$ for truth;
$\phi_1 \lif \phi_2 \coloneqq \phi_1 \lor \neg \phi_2$ for implications; and
\(\Forall{X} \phi' \coloneqq \neg \Exists{X} \neg \phi'\) for universally quantified sentences.

\begin{remark}
Any signature morphism $\chi\colon\Sigma\to\Sigma'$ can be canonically extended to a \emph{sentence-translation function} $\chi\colon\Sen(\Sigma)\to\Sen(\Sigma')$ given by:
\begin{itemize}
\item $\chi(t_{1} = t_{2}) \coloneqq (\chi(t_{1})=\chi(t_{2}))$; 
\item $\chi(t_1 \mathrel{\act} t_2) \coloneqq \chi(t_1) \mathrel{\chi(\act)} \chi(t_2)$;
\item $\chi(\neg\phi) \coloneqq \neg\chi(\phi)$; 
\item $\chi(\lor\Phi) \coloneqq \lor\chi(\Phi)$; and
\item $\chi(\Exists{X}\phi') \coloneqq \Exists{X'}\chi'(\phi')$, where \(X' = \{ \pos{v_i,\chi(s),\Sigma'} \mid \pos{v_i,s,\Sigma} \in X \}\) and $\chi':\Sigma[X]\to\Sigma'[X']$ is the extension of $\chi$ mapping each variable $\pos{v_i,s,\Sigma}\in X$ to $\pos{v_i,\chi(s),\Sigma'}\in X'$.

\end{itemize}
Moreover, this sentence-translation mapping is functorial in \(\chi\).
\end{remark}

Provided that there is no danger of confusion, if we identify variables only by their name and sort,
then each inclusion morphism $\iota\colon\Sigma\hookrightarrow\Sigma'$ determines an inclusion function $\iota\colon\Sen(\Sigma)\hookrightarrow\Sen(\Sigma')$.
This corresponds to the classical approach in model theory and simplifies the presentation greatly. 
A situation where we cannot apply this convention arises when translating a $\Sigma$-sentence $\Exists{X}\phi$ along the inclusion $\iota_X:\Sigma\hookrightarrow\Sigma[X]$.

\paragraph{Satisfaction relation}
Every $\Sigma$-model $\A$ interprets actions as binary relations as follows:
\begin{itemize}
\item for each atomic action $\tau$, the interpretation $\tau^\A_s\subseteq \A_s\times \A_s$ is already part of the model;
\item $(\act_1\comp\act_2)^\A=\act_1^\A \comp \act_2^\A$ (i.e., diagrammatic composition of $S$-sorted binary relations);
\item $(\act_1\cup\act_2)^\A=\act_1^\A \cup \act_2^\A$ (the union of $S$-sorted binary relations); and
\item $(\act^*)^\A= (\act^\A)^*$ (the reflexive and transitive closure of $S$-sorted binary relations).
\end{itemize}

We define the \emph{satisfaction relation} between models and sentences as follows:
\begin{itemize}
\item $\A\models t_{1}=t_{2}$ iff $t_{1}^\A=t_{2}^\A$;
\item $\A\models t_1 \mathrel{\act} t_2$ iff $(t_1^\A, t_2^\A) \in \act^\A$;
\item $\A\models \neg\phi$ iff $\A\not\models\phi$;
\item $\A\models\lor\Phi$ iff $\A\models\phi$ for some sentence $\phi\in \Phi$; and
\item $\A\models\Exists{X}\phi'$ iff $\A'\models\phi'$ for some expansion $\A'$ of $\A$ to the signature $\Sigma[X]$.
\end{itemize}

Notice that, since every term can be regarded as an action, we can quantify over transitions.
For example, given a signature $\Sigma$, a variable $x$ for $\Sigma$ and two $\Sigma[x]$-terms $t_1$ and $t_2$ of the same sort, $\Forall{x} x(t_1, t_2)$ is a well-defined sentence.
However, even though it may seem so at a glance, this does not mean that we can quantify over arbitrary relations.
Instead, Transition Algebra allows us to quantify only over those relations that are determined by the elements of models (which form a small fraction of the set all possible binary relations).
Hence this form of quantification is still of first order.
We present concrete examples of $\TA$ sentences that involve quantification over transitions in Section~\ref{section:imp}.

Let $\Phi,\Phi'$ be sets of $\Sigma$-sentences and $\A$ a $\Sigma$-model.
We also use the following notations:
\begin{itemize}
\item $\A\models\Phi$  ~iff~ $\A\models\phi$ for all sentences $\phi\in\Phi$;
\item $\Gamma\models\Phi$ ~iff~ $\A\models\Gamma$ implies $\A\models\Phi$ for all $\Sigma$-models $\A$.
\end{itemize}
In particular, we write $\Gamma\models\phi$ instead of $\Gamma\models\{\phi\}$ for any set of sentences $\Gamma$ and any single sentence $\phi$.
\begin{proposition} [Satisfaction condition] \label{proposition:trans_models}
For all signature morphisms $\chi:\Sigma\to\Sigma'$,
all $\Sigma'$-models $\A$ and
all $\Sigma$-sentences $\phi$,
we have:
$\A\red_{\chi}\models\phi
\quad\text{iff}\quad
\A\models\chi(\phi)$.
\end{proposition}
\paragraph{Finite models}
Many applications in computer science involve properties that hold only for finite models, or for models in which certain sorts are interpreted as finite sets.
To account for this, we introduce Finite Transition Algebra $(\FTA)$, a variation of $\TA$ with signatures of the form $(S \supseteq S^f, F)$, consisting of a $\TA$ signature together with a distinguished subset $S^f$ of sorts.
The corresponding models are transition algebras in which every sort in $S^f$ is interpreted as a finite set.
We write $\Gamma\models^f\Phi$ iff 
for all models $\A\in |\Mod(S\supseteq S^f,F)|$, 
we have $\A\models\Gamma$ implies $\A\models\Phi$.

\section{Technical preliminaries}\label{section:preliminaries}

In this section, we introduce a series of previously developed concepts upon which the rest of the paper is built and we review some existing results to help contextualize our contributions.
\paragraph{Substitutions}
Consider a signature $\Sigma$ and two sets of new constants $C_1$ and $C_2$.
A \emph{substitution} $\theta : C_1 \to C_2$ over $\Sigma$ is a sort-preserving mapping from $C_1$ to $T_\Sigma(C_2)$.
As in the case of signature morphisms, a substitution $\theta : C_1 \to C_2$ determines:
\begin{itemize}
\item a sentence functor $\Sen(\theta):\Sen(\Sigma[C_1])\to\Sen(\Sigma[C_2])$, which preserves $\Sigma$ and maps each constant $c\in C_1$ to a term $c\theta$, and 
\item a reduct functor $\_\red_\theta:\Mod(\Sigma[C_2])\to\Mod(\Sigma[C_1])$, which preserves the interpretation of $\Sigma$ and interprets each constant $c\in C_1$ the same as the term $\theta(c)\in T_\Sigma(C_2)$;
  that is, for all $\A\in|\Mod(\Sigma[C_2])|$ we have 
\begin{enumerate*}[label=(\alph*)]
\item $(\A\red_\theta)_s=\A_s$ for all sorts $s$ in $\Sigma$,
\item $\sigma^{(\A\red_\theta)}=\sigma^\A$ for all function symbols $\sigma$ in $\Sigma$, and
\item $c^{(\A\red_\theta)}=\theta(c)^\A$ for all constants $c\in C_1$.
\end{enumerate*} 
\end{itemize}
By a slight abuse of notion, we let $\theta$ denote both the substitution $\theta:C_1\to C_2$ and the sentence functor $\Sen(\theta):\Sen(\Sigma[C_1])\to\Sen(\Sigma[C_2])$.
The following result is a straightforward generalization of \cite[Proposition 5.8]{Diaconescu2025}.
\begin{proposition} [Satisfaction condition for substitutions]  \label{prop:sat-cond-subst} 
For all substitutions $\theta:C_1\to C_2$,
all $\Sigma[C_2]$-models $\A$ and
all $\Sigma[C_1]$-sentences $\phi$
we have:
$\A\red_{\theta}\models\phi
\quad\text{iff}\quad
\A\models\phi\theta$.
\end{proposition}
For any substitution $\theta:C_1\to C_2$ and any subset $C\subseteq C_1$,
we let $\theta_C$ denote the restriction of $\theta$ to $C$.
This notation is useful to define the operational semantics for transition algebra.
\paragraph{Reachable models}
A $\Sigma$-model $\A$ is \emph{reachable}~\cite{Petria07} when the unique homomorphism $h:T_\Sigma\to \A$ is surjective.
The following result gives an abstract characterization of the notion of reachable model based on substitutions.
\begin{proposition}\label{prop:reach}
A model $\A$ defined over a signature $\Delta$ is \emph{reachable} 
iff 
for each set of new constants $C$ for $\Sigma$
and any expansion $\B$ of $\A$ to $\Sigma[C]$, 
there exists a substitution  $\theta : C \to \emptyset$ over $\Sigma$ such that 
$\A\red_\theta=\B$.
\end{proposition}
The proof of Proposition~\ref{prop:reach} can be found, for example, in \cite[Proposition 8]{gai-cbl}.
\begin{example}[Natural numbers] \label{ex:nat}
Consider a signature $\Sigma$ with two sorts, $Nat$ and $Label$, and the following function symbols:
$0:\to \mathit{Nat}, s\_:\mathit{Nat}\to \mathit{Nat}, \_+\_:\mathit{Nat}~\mathit{Nat}\to \mathit{Nat}, \mathit{succ}:\to \mathit{Label}$.
Let $\Gamma$ be a set of sentences consisting of:
\begin{enumerate}
\item $\Forall{x}succ(x,s~x)$ and $\Forall{x,y} y = s~x \lif succ(x,y)$,
which state that $succ$ denotes the successor relation; and
\item $\{\Forall{x} x + 0 = x \lif succ^n(0,x)\mid n\in \mathbb{N}\}$ and $\{\Forall{x,y} x + s~y = s (x+y) \lif succ^n(0,x) \land succ^m(0,y) \mid n,m\in\mathbb{N}\}$,
which define addition by induction on the first argument for all elements reachable from $0$ via the successor relation.
\end{enumerate}
This presentation of natural numbers can also be written using the $\TA$ specification language, as we describe next.
The syntax that we employ follows the tradition of the OBJ family of languages~\cite{Goguen00} and is similar to that of Maude~\cite{conf/maude/2007}.
We use keywords such as \texttt{spec} to introduce specifications, \texttt{protecting} to import/reuse existing specifications\footnote{Here, \texttt{protecting} indicates that the newly defined specification should not introduce new elements on `old' sorts, originating from the imported specification, nor should it identify `old' elements of those sorts. That is, we adhere to the `no junk' and `no confusion' principles of OBJ~\cite{Goguen2000}.}, \texttt{sort} or \texttt{sorts} to declare sorts, \texttt{op} or \texttt{ops} for operations, \texttt{vars} for universally quantified variables,
\texttt{ax} for axioms (Horn clauses), and so on.
In the listing below, \texttt{NAT} is just the signature of natural numbers, \texttt{SUCC} introduces the successor relation, and \texttt{NAT-ADD-REACH} defines addition for elements that are reachable along the successor relation.
The reason we split the overall specification into these three parts is that it enables us to reuse some of them later on in the paper.
{\small \begin{verbatim}
spec NAT is
  sort Nat .
  op 0 : () -> Nat .
  op s_ : Nat -> Nat .
endspec

spec SUCC is
  protecting NAT .
  sort Label .
  op succ : () -> Label .
  vars x, y : Nat .
  ax x =(succ)> s x .
  ax y = s x if x =(succ)> y .
endspec

spec NAT-ADD-REACH is
  protecting NAT .
  protecting SUCC .
  op _+_ : Nat Nat -> Nat .
  vars x, y : Nat .
  ax x + 0 = x if 0 =((succ)*)> x .
  ax x + s y = s (x + y) if 0 =((succ)*)> x and 0 =((succ)*)> y .
endspec
\end{verbatim}}
\end{example}
Note that the second axiom from the \texttt{SUCC} module cannot be specified in Maude, as it would use a rewrite condition in an equation.
Also notice that $\mathit{succ}$ is just the relational counterpart of the successor operation $\mathit{s\_}$.
Hence, any model over the signature in Example~\ref{ex:nat} which satisfies the sentence $\Forall{x}succ^*(0,x)$ is reachable, having elements obtained from $0$ by repeated applications of the successor operation. 
As shown in finite model theory, reachability/connectedness is not first‑order definable; equivalently, first‑order logic cannot define the transitive closure of a binary relation. 
For standard references and an in‑depth discussion, consult Libkin~\cite{DBLP:books/sp/Libkin04}.
Also, $\Gamma\models \Forall{x,y}x + y = y + x \lif succ^*(0,x) \land succ^*(0,y)$, which means that all reachable models of $\Gamma$ satisfies the commutativity of addition. 
However, there is no finite subset of $\Gamma$ which satisfies  $ \Forall{x,y}x + y = y + x \lif succ^*(0,x) \land succ^*(0,y)$.
Therefore, $\TA$ is not compact.
\paragraph{Basic sentences}
A set of sentences $B$  defined over a signature $\Sigma$ is called \emph{basic} \cite{dia-ult} if there exists a $\Sigma$-model $\A^B$ such that, for all $\Sigma$-models $\A$,
\begin{center} 
$\A\models B$ iff there exists a morphism $\A^B\rightarrow \A$. 
\end{center}
We say that $\A^B$ is a \emph{basic model} of $B$. If in addition the morphisms $\A^B\rightarrow \A$ is unique then the set $B$ is called \emph{epi basic}. It is known that any set of  atoms in first-order logic is epi basic and the basic model is reachable
(see, for example, \cite[Fact 5.23]{Diaconescu2025}).
Similarly, any set of atomic sentences from $\TA$ is epi basic and the basic model is reachable.
We denote by $\Sen_b:\Sig\to \Set$ the sub-functor of $\Sen:\Sig\to \Set$
that maps each signature $\Sigma$ in $\TA$ to the set $\Sen_b(\Sigma)$ of atomic sentences from $\Sen(\Sigma)$.
\paragraph{Entailment systems}
The syntactic counterpart of the satisfaction relation for reasoning about the logical consequences of sentences is defined below.
\begin{figure}[h]\centering
\begin{tabular}{|l l|}
\hline
&  \\
$(\mathit{Monotonicity})~\Frac{}{\Gamma\cup\{\gamma\}\vdash\gamma}$  & 
$(\mathit{Cut})~\Frac{\Gamma\vdash\gamma_2 \Space \Gamma\cup \{ \gamma_2\} \vdash \gamma_1}{\Gamma\vdash\gamma_1}$ \\
& \\
$(\mathit{Translation})~\Frac{\Gamma \vdash \gamma}{\chi(\Gamma)\vdash\chi(\gamma)}$ & where $\chi:\Sigma\to\Sigma'$ \\
&\\
\hline
\end{tabular}
\caption{Entailment systems}
\label{table:entail}
\end{figure}

\begin{definition}[Entailment system]
An \emph{entailment system}  is a family of relations ${\vdash}=\{{\vdash_\Sigma}\}_{\Sigma\in|\Sig|}$ between sets of sentences and sentences (that is, ${\vdash_\Sigma}\subseteq \mathcal{P}(\Sen(\Sigma))\times \Sen(\Sigma)$ for all signatures $\Sigma\in|\Sig|$) satisfying $(\mathit{Monotonicity})$ and $(\mathit{Translation})$ defined in Fig.~\ref{table:entail}.
We drop the subscript $\Sigma$ from the notion $\vdash_\Sigma$ when there is no danger of confusion.
\end{definition}

\begin{remark}
  Entailment system are generally required to also satisfy $(\mathit{Cut})$, also called \emph{transitivity}~\cite{Meseguer:General-logics-1989}, as defined in Fig.~\ref{table:entail}.
  We omit it from the definition of entailment systems to emphasize the fact that lemma discovery is not needed for the proof calculus developed in this paper.
\end{remark}

\begin{definition}[Entailment properties]
Let $\vdash$ be an entailment system.
\begin{enumerate}
\item $\vdash$ is \emph{sound} if for all signatures $\Sigma$ we have $\vdash_\Sigma \ \subseteq\ \models_\Sigma$; and \emph{complete} when the opposite inclusion holds: $\models_\Sigma \ \subseteq\ \vdash_\Sigma$.
\item $\vdash$ is \emph{compact} if for all signatures $\Sigma$, all sets of sentences $\Gamma$ and all sentences $\gamma$, we have:
$\Gamma\vdash_\Sigma\gamma$ implies $\Gamma_f\vdash_\Sigma\gamma$ for some finite $\Gamma_f\subseteq\Gamma$.
\end{enumerate}
\end{definition}
\paragraph{Contexts}
We recall the notions of context and position from the term rewriting literature~\cite{BaaderN98}, which are fundamental to the development of an operational semantics.
\begin{definition}[Context]
Let $\gap$ be a distinguished variable name used to indicate a gap within a term.
A \emph{context} is a term that contains exactly one occurrence of a variable named $\gap$.
We typically denote contexts by $c$, or by $c[\gap \colon s]$ when we want to make explicit the sort $s$ of the gap.
For any context $c[\gap \colon s]$ and any sort $t$ of sort $s$, we write $c[t]$ for the term $c(\gap \mapsto t)$, where $(\gap \mapsto t) \colon \{\gap \colon s\} \to \emptyset$ is the substitution that maps $\gap$ to $t$.

We say that an equation $t_1=t_2$ is an \emph{extended context} if either $t_1$ or $t_2$ is an ordinary term context.
Similarly, a transition $\act(t_{1}, t_{2})$ is an \emph{extended context} if either $t_{1}$ is a context, $t_{2}$ is a context, or $\act$ contains exactly one occurrence of an atomic sub-action (i.e., term) that is a context.
A set $G$ of equations and transitions is also called an \emph{extended context} if exactly one of the formul\ae\ $\varphi \in G$ is an extended context.
We denote extended contexts by $C$, and we write $C[t]$ for the (set of) formul\ae\ obtained by substituting a term $t$ for $\gap$ within an extended context $C$.
\end{definition}
\begin{definition}[Position]
Assume a signature $\Sigma$.
The set of positions $\mathit{Pos}(t)$ of a term $t\in T_\Sigma$ is a set of strings with elements from the natural numbers, which is inductively defined as follows:
\begin{center}
\begin{enumerate*}[label=(\alph*)]
\item
$\mathit{Pos}(c)=\{\varepsilon\}$ ~~~and~~~
\item 
$\mathit{Pos}(\sigma(t_1,\dots,t_n))=\bigcup_{i=1}^n\{i p\mid p\in \mathit{Pos}(t_i)\}$.
\end{enumerate*}
\end{center}
For $p\in \mathit{Pos}(t)$, the subterm of $t$ at position $p$, denoted by $t|_p$, is defined by induction on the length of $p$:
\begin{center}
\begin{enumerate*}[label=(\alph*)]
\item $t|_\varepsilon=t$ ~~~and~~~
\item $\sigma(t_1,\dots,t_n)|_{ip}=t_i|_p$.
\end{enumerate*}
\end{center}
Let $p\in \mathit{Pos}(t)$ and $\tau\in T_\Sigma$ such that the sort of $\tau$ coincides with the sort of $t|_p$.
We denote by $t[\tau]_p$ the term obtained from $t$ by replacing the subterm at position $p$ by $\tau$, that is:
\begin{center}
\begin{enumerate*}[label=(\alph*)]
\item $t[\tau]_\varepsilon=\tau$ ~~~and~~~
\item $\sigma(t_1,\dots,t_n)[\tau]_{ip}=\sigma(t_1, \dots, t_i[\tau]_{p},\dots,t_n)$.
\end{enumerate*}
\end{center}
\end{definition}

\section{Query completeness}\label{section:query}

In this section, we propose a notion of Horn clause that is more expressive than the classical one defined for first-order logic.
A Horn clause is a sentence of the form $\Forall{X}\varphi \lif H $ where 
\begin{enumerate*}[label=(\alph*)]
\item $X$ is a finite block of variables;
\item $H$ is a finite set of equations and transitions;\footnote{Since $H$ and $\land H$ are semantically equivalent, for the sake of simplicity, we let  $\Forall{X}\varphi \lif H$ denote $\Forall{X}\varphi \lif \land H$.} and
\item $\varphi$ is an atomic sentence.
\end{enumerate*} 
A query is an existentially quantified sentence $\Exists{X}Q$, where $Q$ is a finite set of equations and transitions.
We define a system of proof rules for deducing queries from a set of Horn clauses. 
We develop a two-layered approach to the completeness of the proof rules proposed:
\begin{enumerate}
\item We begin with a basic layer composed of atomic sentences. At this initial stage, we consider the deduction of an atomic sentence from a set of atomic sentences---that is, $\Gamma \vdash \gamma$, where both $\Gamma$ and $\gamma$ consist solely of atomic sentences.
\item In the second stage, we increase the complexity by allowing arbitrary clauses on the left-hand side of the turnstile and queries on the right-hand side of the turnstile. 
That is, $\Gamma$ may be any set of clauses, while $\gamma$ is a query.
Another important result here is the fact that any set of Horn clauses $\Gamma$ has an initial models and the satisfaction of a query $\Exists{X}Q$ by a program modeled as a set of Horn clauses $\Gamma$, can be reduced to the computation of a substitution $\theta: X\to Y$ such that $\Gamma\models \Forall{Y}Q\theta$ which is furthermore equivalent with the satisfaction in the initial model, that is, $\A^\Gamma\models \Exists{X}Q$.
\end{enumerate}
\subsection{Basic clauses}
In this section, we prove an important result about atomic sentences. 
\begin{table}
\centering
\begin{tabular}{|l l|}
\hline
& \\
$(R)~\Frac{}{\Gamma\vdash t = t}$ \Space
$(S)~\Frac{\Gamma\vdash t_1 = t_2}{\Gamma\vdash t_2 = t_1}$ & 
$(T)~\Frac{\Gamma\vdash t_1 = t_2 \Space \Gamma\vdash t_2 = t_3}{\Gamma\vdash t_1 = t_3}$ \\
& \\
$(F)~\Frac{\Gamma\vdash t_1 = t_1' ~\dots~ \Gamma\vdash t_n = t_n' }{\Gamma\vdash \sigma(t_1,\dots,t_n) = \sigma(t_1',\dots, t_n')} $ & 
$(L)~\Frac{\Gamma\vdash t_i=t_i' \text{ for both } i\in\{1,2\} \Space \Gamma\vdash \tau = \tau' \Space \Gamma\vdash \tau(t_1,t_2)}{\Gamma\vdash \tau'(t_1',t_2')}$ \\
& \\
\hline
\end{tabular}
\caption{Basic proof rules}
\label{tab:basic}
\end{table}
The proof rules for reasoning about atomic sentences are defined in Table~\ref{tab:basic}.
Notice that $\Gamma$ from Table~\ref{tab:basic} stands for any set of sentences not only just for atomic sentences.
\begin{lemma}[Basic soundness]
The proof rules defined in Table~\ref{tab:basic} are sound. That is, the least entailment system closed under the proof rules defined in Table~\ref{tab:basic} is sound.
\end{lemma}
We show that each set of sentences and any sound entailment relation closed under the proof rules in Table~\ref{tab:basic} gives rise to a model that preserves formal deductions.
\begin{theorem}\label{th:basic}
Let $\vdash$ be any sound entailment system closed under the proof rules defined in Table~\ref{tab:basic}.
Any set of sentences $\Gamma$ over a signature $\Sigma$ defines a model $\A^\Gamma$ as follows:
\begin{enumerate}
\item Let $\equiv^\Gamma~\coloneqq~ \{t_1=t_2\mid \Gamma\vdash t_1 = t_2\}$ be the congruence on $T_\Sigma$ determined by $\Gamma$.

\item Given a $\Sigma$-term $t$, we let $[t]$ denote the equivalence class of $t$ under the congruence $\equiv^\Gamma$. 
Then $\A^\Gamma$ is obtained from $T_\Sigma/_{\equiv^\Gamma}$ by interpreting each binary relation $[\tau]_s$, where $s\in S$ and $\tau\in T_{\Sigma,s'}$ for some $s'\in S$,
as a binary relation $[\tau]_s~\coloneqq~ \{ ([t_1],[t_2]) \mid t_1,t_2\in T_{\Sigma,s}\text{ and }\Gamma \vdash \tau(t_1, t_2) \}$.
\end{enumerate}
Then $\Gamma\vdash\varphi$ iff $\A^\Gamma\models\varphi$ for all atomic sentences $\varphi$ defined over $\Sigma$.
\end{theorem}
\begin{proof}
It is straightforward to prove that $\equiv^\Gamma$ is a congruence relation on $T_\Sigma$.
By $(L)$, the definition of $[\tau]_s$  does not depend on the representatives $\tau$, $t_1$, and $t_2$ of the classes $[\tau]$, $[t_1]$, and $[t_2]$, respectively.
In other words, $\Gamma\vdash \tau(t_1, t_2)$ iff $([t_1],[t_2]) \in [\tau]_s$.
Therefore, $[\tau]_s$ is well-defined.
There are two cases for proving the statement of the theorem:
\begin{description}[topsep=1em, itemsep=1em]
\item [{\case[3em]{t_1=t_2}}]
 $\Gamma\vdash t_1 = t_2$ iff 
 $t_1\equiv^\Gamma t_2$ iff 
 $\A^\Gamma\models t_1 = t_2$.
 \item [{\case[3em]{t_1~\tau~t_2}}]
 $\Gamma\vdash \tau(t_1,t_2)$ iff 
 $([t_1] ,[t_2]) \in [\tau]_s$ 
 iff
 $\A^\Gamma\models t_1~\tau~t_2$.
 \qedhere
\end{description}
\end{proof}
Completeness for atomic sentences follows from Theorem~\ref{th:basic}, showing that to reason about the consequences of a set of atomic sentences, it suffices to consider only its initial model.
\begin{corollary}[Basic completeness]\label{cor:basic}
Let $\vdash$ be any sound entailment system closed under the proof rules defined in Table~\ref{tab:basic}.
For any set of atomic sentences $\Phi$ and any atomic sentence $\varphi$ defined over the same signature $\Sigma$,
the following are equivalent:
\vspace{-\baselineskip}
\begin{center}
\begin{enumerate*}[label=(\alph*)]
\item\label{cor:basic-a}~$\Phi\vdash \varphi$, \quad
\item\label{cor:basic-b}~$\Phi\models \varphi$, ~~and~~
\item\label{cor:basic-c}~$\A^\Phi\models \varphi$,
\end{enumerate*}
\end{center}
where $\A^\Phi$ is the model described in Theorem~\ref{th:basic}.
\end{corollary}
\begin{proof}
We proceed as follows:
\begin{description}
\item [$\ref{cor:basic-a}\implies\ref{cor:basic-b}$] 
By soundness.
\item [$\ref{cor:basic-b}\implies\ref{cor:basic-c}$]
For any $\phi\in\Phi$, since $\Phi\vdash\phi$, by Theorem~\ref{th:basic}, $\A^\Phi\models\phi$.
Therefore, $\A^\Phi\models\Phi$. 
Since $\Phi\models \varphi$, 
we get $\A^\Phi\models\varphi$.
\item [$\ref{cor:basic-c}\implies\ref{cor:basic-a}$]
  By Theorem~\ref{th:basic}.
  \qedhere
\end{description}
\end{proof}
Notice that, by Corollary~\ref{cor:basic}, $\A^\Phi$ is a model of $\Phi$, hence the initial model of $\Phi$.
\subsection{Queries}
Table~\ref{tab:init} defines the proof rules used to deduce queries.
\begin{table}
\centering
\begin{tabular}{|l l|}
\hline
& \\
$(\mathit{Comp})~\Frac{\Gamma\vdash \act_1 (t_1,t_2) \Space \Gamma\vdash \act_2(t_2, t_3)}{\Gamma\vdash (\act_1\comp\act_2)(t_1, t_3)}$ & 
$(\mathit{Union})~\Frac{\Gamma\vdash \act_i(t_1, t_2) }{\Gamma\vdash (\act_1\cup\act_2)(t_1, t_2)}$ \\
& \\
$(\mathit{Eq})~\Frac{\Gamma\vdash t_1=t_2 }{\Gamma\vdash \act^0(t_1 , t_2)}$  &
$(\mathit{Star})~\Frac{\Gamma\vdash \act^n(t_1, t_2)}{\Gamma\vdash \act^*(t_1 ,t_2)}$ \\
&\\
$(\mathit{MP})~\Frac{\Gamma\cup \{ \Forall{X}\varphi\lif H \} \vdash \theta(\phi) \text{ for all } \phi\in H }{\Gamma \cup \{ \Forall{X}\varphi\lif H \} \vdash \theta(\varphi)}$  
 &
$(\mathit{Subst})~\Frac{\Gamma\vdash \phi\theta \text{ for all } \phi\in Q}{\Gamma\vdash \Exists{X}Q}$ \\
where $\theta:X\to \emptyset$ &  where $\theta:X\to \emptyset$ \\
& \\
\hline
\end{tabular}
\caption{Actions, modus ponens and substitutivity}
\label{tab:init}
\end{table}
\begin{lemma}
  The proof rules defined in Tables~\ref{tab:basic} and~\ref{tab:init} are sound.
  That is, the least entailment system closed under the proof rules defined in Tables~\ref{tab:basic} and~\ref{tab:init} is sound.
\end{lemma}
We show that any set of clauses has an initial reachable model, which preserves formal deductions of ground equations and transition rules.
\begin{theorem}[Initiality \& query completeness] \label{th:init} 
Let $\vdash$ be any sound entailment system closed under the proof rules defined in Tables~\ref{tab:basic} and~\ref{tab:init}.
Let \(\Gamma\) be a set of clauses defined over a signature \(\Sigma\).
\begin{enumerate}
\item \label{th:init-2} $\A^\Gamma$ is the initial model of $\Gamma$, where $\A^\Gamma$ is the model defined in Theorem~\ref{th:basic}.
\item \label{th:init-3} 
For all queries $\Exists{X}Q$, we have:
\begin{enumerate*}[label=(\alph*)]
\item \label{th:init-a} $\Gamma\vdash \Exists{X}Q$ ~iff~
\item \label{th:init-b} $\Gamma\models \Exists{X}Q$ ~iff~
\item \label{th:init-c} $\A^\Gamma\models\Exists{X}Q$.
\end{enumerate*}
\end{enumerate}
\end{theorem}
\begin{proof}
First, we prove the following statement:

\begin{equation} \tag{$\dagger$}\label{init-a} 
\Gamma\vdash \act(t_1,t_2) \text{~~if~~}  \A^\Gamma \models \act(t_1, t_2)
\end{equation}
For \emph{statement~(\ref{init-a})}, 
we proceed by induction on the structure of actions:
\begin{description}[leftmargin=4.5em, topsep=1em, itemsep=1em]
\item[{\case[3em]{\tau\in T_\Sigma}}] By Theorem~\ref{th:basic}.
\item[{\case[3em]{\act_1\cup\act_2}}] 
We have 
$\A^\Gamma\models (\act_1\cup\act_2)(t_1, t_2)$ 
iff
$(t_1^{(\A^\Gamma)},t_2^{(\A^\Gamma)})\in \act_1^{(\A^\Gamma)} \cup \act_2^{(\A^\Gamma)}$
iff
$(t_1^{(\A^\Gamma)},t_2^{(\A^\Gamma)})\in \act_1^{(\A^\Gamma)}$ or 
$(t_1^{(\A^\Gamma)},t_2^{(\A^\Gamma)})\in \act_2^{(\A^\Gamma)}$
iff
$\A^\Gamma\models \act_1(t_1, t_2)$ or
$\A^\Gamma\models \act_2(t_1, t_2)$. 
By the induction hypothesis,
$\Gamma\vdash \act_1(t_1, t_2)$ or
$\Gamma\vdash \act_2(t_1, t_2)$.
By $(\mathit{Union})$, $\Gamma\vdash (\act_1\cup\act_2)(t_1, t_2)$.
\item[{\case[3em]{\act_1\comp\act_2}}] 
We have $\A^\Gamma\models (\act_1\comp\act_2)(t_1, t_2)$.
Since $\A^\Gamma$ is reachable, we have 
$\A^\Gamma\models \act_1(t_1, t)$ and $\A^\Gamma\models \act_2(t, t_2)$ for some term $t\in T_\Sigma$.
By induction hypothesis,
$\Gamma\vdash \act_1(t_1, t)$ and
$\Gamma\vdash \act_2(t, t_2)$.
Hence, by $(\mathit{Comp})$, $\Gamma\vdash (\act_1\comp\act_2)(t_1, t_2)$.
\item [{\case[3em]{\act^*}}]
We have $\A^\Gamma\models \act^*(t_1, t_2)$.
By semantics, $\A^\Gamma\models \act^n(t_1, t_2)$ for some natural number $n\in\mathbb{N}$.
By induction hypothesis, $\Gamma\vdash \act^n(t_1, t_2)$.
By $(\mathit{Star})$, $\Gamma\vdash \act^*(t_1, t_2)$.
\end{description}

\smallskip\noindent
\emph{For the first statement of the theorem}, 
by Corollary~\ref{cor:basic},
it suffices to prove that $\A^\Gamma\models \Gamma$.
Let $\Forall{X}\varphi\lif H$ be a clause in $\Gamma$.
Let $\B$ be an expansion of $\A^\Gamma$ to $\Sigma[X]$ such that $\B\models_{\Sigma[X]} H$.
Since $\A^\Gamma$ is reachable,
there exists a substitution $\theta:X\to\emptyset$ such that $\A^\Gamma\red_\theta=\B$.
By the satisfaction condition for substitutions,
since $\A^\Gamma\red_\theta \models_{\Sigma[X]} H$,
we have $\A^\Gamma\models \theta(H)$.
By statement~(\ref{init-a}), $\Gamma\vdash\theta(\phi)$ for all $\phi\in H$.
By $\mathit{(MP)}$, $\Gamma\vdash \theta(\varphi)$.
By Corollary~\ref{cor:basic}, $\A^\Gamma\models \theta(\varphi)$.
By the satisfaction condition for substitutions, $\B\models_{\Sigma[X]}\varphi$.

\smallskip\noindent
\emph{The second statement} is proved as follows:
\begin{description}
\item [$\ref{th:init-a}\implies\ref{th:init-b}$] 
By soundness.
\item [$\ref{th:init-b}\implies\ref{th:init-c}$] 
Since $\A^\Gamma\models\Gamma$.
\item [$\ref{th:init-c}\implies\ref{th:init-a}$] 
Since $\A^\Gamma\models_\Sigma\Exists{X}Q$, we have $\B\models_{\Sigma[X]} Q$ for some expansion $\B$ of $\A^\Gamma$ to $\Sigma[X]$. 
Since $\A^\Gamma$ is reachable, 
there exists a substitution $\theta:X\to \emptyset$ such that $\A^\Gamma\red_\theta=\B$.
Since $\B\models_{\Sigma[X]}Q$, 
by the satisfaction condition for substitutions, 
$\A^\Gamma\models_\Sigma Q\theta$. 
By the first part of the proof, 
$\Gamma\vdash \phi\theta$ for all $\phi\in Q$.
By $\mathit{(Subst)}$,
$\Gamma\vdash \Exists{X}Q$.
\qedhere
\end{description}
\end{proof}
In the next three sections, we focus on providing a framework for proving queries automatically.

\section{Operational semantics}\label{section:op-semantics}

In this section, we present the operational semantics of Transition Algebra, which will be used to formally describe system behaviour from the computational point of view.
For this purpose, we consider two kinds of rewrite rules to ‘execute’ specifications.
An \emph{equational rewrite rule} is a Horn clause of the form $\Forall{X} (\ell=\err) \lif H $ such that
\begin{enumerate*}[label=(\alph*)]
\item $\ell$ is not a variable and
\item $X=\var(\ell) \cup \var(H)$.
\end{enumerate*}
Similarly, a \emph{transitional rewrite rule} is a Horn clause of the form $\Forall{X} \tau(\ell,\err) \lif H $ such that
\begin{enumerate*}[label=(\alph*)]
\item $\ell$ is not a variable and
\item $X=\var(\tau) \cup \var(\ell) \cup \var(H)$.
\end{enumerate*}
Hereafter, whenever we refer to a \emph{term rewriting system} in $\TA$ we mean a pair $(\Sigma,\Gamma)$ consisting of a signature $\Sigma$ and a set $\Gamma\subseteq\Sen(\Sigma)$ of equation and transition rewrite rules.

\subsection{Rewriting}
In this section, we define a rewriting relation on Transition Algebra \emph{goals}, which are sets of (ground) equations and transitions, and then prove its soundness with respect to the denotational semantics presented in Section~\ref{section:transition-algebra}.
\begin{definition}[Goal rewriting step] 
Let $(\Sigma,\Gamma)$ be a term rewriting system.
The \emph{one-step-rewriting} relation on ground equations and transitions -- referred to as \emph{goals} in certain contexts -- is defined as follows (we use different labels for the relation to indicate different cases):
\begin{description}[topsep=.5em, itemsep=.25em, font=\normalfont]
\item[\tacase{\eq}] $C[\ell\theta] \glrew{\eq} C[\err\theta] \cup H\theta$
for all equational rules $\Forall{Y} (\ell=\err) \lif H \in \Gamma$ and all substitutions $\theta:Y\to \emptyset$;
\item[\tacase{=}] $\{t=t\} \cup G \glrew{=} G$;
\item[\tacase{\tr}] $\{\kappa\theta(\ell\theta,t)\} \cup G \glrew{\tr} \{\err\theta=t\} \cup H\theta \cup G$
for all transition rules $\Forall{Y}\kappa(\ell,\err) \lif H \in \Gamma$ and all substitutions $\theta:Y\to\emptyset$;
\item[\tacase{\comp}] $\{(\act_1\comp\act_2)(t_1,t_2)\} \cup G \glrew{\comp} \{\act_1(t_1,t), \act_2(t,t_2)\} \cup G$,
for any $\Sigma$-term $t$;
\item[\tacase{\cup}]
$\{(\act_1\cup\act_2)(t_1,t_2)\} \cup G \glrew{\cup} \{\act_i(t_1,t_2)\} \cup G$,
for any $i\in\{1,2\}$;
\item[\tacase{\circlearrowright}]
$\{\act^*(t_1,t_2)\} \cup G  \glrew{\circlearrowright} \{\act^n(t_1,t_2)\} \cup G$,
for any $n\in\mathbb{N}$.
\end{description}
\end{definition}
This form of rewriting, defined directly on formul\ae, is similar to the treatment of pararewriting in~\cite{CazanescuD09}.
We use it instead of the more conventional approach (see, e.g.,~\cite{Terese03} and also~\cite{Dershowitz90,Klop92,Plaisted93,BaaderN98}) because it provides a more straightforward way to proving completeness and relating it to narrowing in the latter parts of the section.
The connection with the conventional approach to term rewriting is established in Lemma~\ref{lemma:tr+tt} and Proposition~\ref{prop:coherence}.

\begin{definition}[Goal rewriting]
  The general \emph{rewriting} relation $\glrew{*}$ on Transition Algebra goals is the reflexive and transitive closure of the one-step-rewriting relation.
  We define it by ${\glrew{*}} = \bigcup \{{\glrew{n}} \mid n \in \mathbb{N}\}$, where:
\begin{enumerate*}[label=(\alph*)]
\item $G \glrew{0} G$, and
\item if $G_1\glrew{\textit{lb}} G_2$ and $G_2\glrew{n} G_3$,
then 
$G_1 \glrew{n+1} G_3$, for each label $\textit{lb}\in \{\eq,=,\tr,\comp,\cup,\circlearrowright\}$.
\end{enumerate*}
When the label is irrelevant, we may omit it and write simply $G_{1} \glrew{} G_{2}$ for a single rewriting step.
\end{definition}
\begin{remark}
So far, we have defined rewriting on sets of ground equations and transitions.
To extend it to equations and transitions with variables from some given block $X$, we treat the variables in $X$ as new constants that we add to the underlying signature.
Concretely, for any term rewriting system $(\Sigma,\Gamma)$ and block $X$ of variables, rewriting equations and transitions with variables from $X$ amounts to applying rewriting over $(\Sigma[X],\Gamma)$.
\end{remark}

This notion of goal rewriting is sound in the following sense: given the rewriting rules (Horn clauses) provided by $\Gamma$, the outcome of a rewrite is always stronger, with respect to the satisfaction relation, than the original goal.

\begin{theorem}[Soundness of rewriting]%
  \label{theorem:soundness-of-rewriting}
  Let $(\Sigma,\Gamma)$ be a term rewriting system.
  For any goals $G, G' \subseteq \Sen(\Sigma)$,
  $$
  G \glrew{*} G'
  \quad\text{implies}\quad
  \Gamma \cup G' \models G.$$
\end{theorem}
\begin{proof}
  We proceed by induction on the number of rewriting steps.
  For the base case, it suffices to notice that $G\glrew{0}G'$ if and only if $G=G'$, which leads to the obvious entailment $\Gamma \cup G \models G$.
  
  For the induction step, there are six cases to consider.
  Three of them, for $=$, $\cup$, and $\circlearrowright$, are straightforward.
  Therefore, we focus only on the remaining ones:
  \begin{description}[topsep=.5em, itemsep=.25em, font=\normalfont]
  \item[\tacase{\eq}]
    Suppose $C[t] \glrew{\eq} C[\err\theta] \cup H\theta \glrew{n} G'$,
    where $t=\ell\theta$ for some rule $\Forall{Y} (\ell=\err) \lif H \in \Gamma$ and some substitution $\theta:Y\to\emptyset$.
    By the induction hypothesis,
    $\Gamma \cup G' \models C[\err\theta] \cup H\theta$.
    Since $\Gamma \cup G' \models (\ell\theta=\err\theta) \lif H\theta$ and $\Gamma \cup G' \models H\theta$,
    it follows that $\Gamma \cup G'\models \ell\theta=\err\theta$.
    Moreover, since  $\Gamma \cup G' \models C[r\theta]$ and $\Gamma \cup G' \models \ell\theta=r\theta$, 
    we have $\Gamma \cup G' \models C[\ell\theta]$.
    Hence, $\Gamma \cup G' \models C[t]$.
  \item[\tacase{\tr}] Assume that 
    $\{\tau(t_1,t_2)\} \cup G \glrew{\tr} \{\err\theta=t_2\} \cup H\theta \cup G \glrew{n} G'$,
    where $\tau=\kappa\theta$ and $t_1=\ell\theta$,
    for some rewrite rule $\Forall{Y}\kappa(\ell,r) \lif H \in \Gamma$ and some substitution $\theta:Y\to \emptyset$.
    By the induction hypothesis, $\Gamma \cup G' \models \{\err\theta=t_2 \} \cup H\theta \cup G$.
    Since $\Gamma \cup G' \models (\kappa\theta)(\ell\theta,\err\theta) \lif H\theta$ and $\Gamma \cup G' \models H\theta$,
    it follows that $\Gamma \cup G'\models (\kappa\theta)(\ell\theta,\err\theta)$.
    And because $\tau=\kappa\theta$, $t_1=\ell\theta$, and $\Gamma \cup G' \models \err\theta=t_2$, 
    we infer $\Gamma \cup G' \models \tau(t_1,t_2)$. 
    Hence, $\Gamma \cup G' \models \{\tau(t_1,t_2)\} \cup G$. 
  \item[\tacase{\comp}]
    Suppose $\{(\act_1\comp\act_2)(t_1,t_2)\} \cup G \glrew{\comp} \{\act_1(t_1,t) ,\act_2(t,t_2)\} \cup G \glrew{n} G'$ for some $\Sigma$-term $t$.
    By the induction hypothesis, we have $\Gamma \cup G' \models \{\act_1(t_1,t) ,\act_2(t,t_2)\} \cup G$.
    Since $\Gamma \cup G' \models \{\act_1(t_1,t) ,\act_2(t,t_2)\}$, by the semantics of Transition Algebra, $\Gamma \cup G' \models(\act_1\comp\act_2)(t_1,t_2)$.
    Therefore, $\Gamma \cup G' \models \{(\act_1\comp\act_2)(t_1,t_2)\} \cup G$.
    \qedhere    
  \end{description}
\end{proof}

\begin{corollary}
  For any term rewriting system $(\Sigma,\Gamma)$ and goal $G \subseteq \Sen(\Sigma)$,
  $G \glrew{*} \emptyset$ implies $\Gamma \models G$.
\end{corollary}
\subsection{Narrowing}
Next, we look for solutions to \emph{queries}, which are defined as existentially quantified goals and are usually denoted $\Exists{X} Q$, where $X$ is a block of variables and $Q$ is a goal with variables from $X$.\footnote{As in the case of clauses, $\Exists{X} Q$ abbreviates $\Exists{X} \bigland Q$.}
As in equational logic programming~\cite{GoguenM87}, a \emph{solution} to $\Exists{X} Q$ is a substitution $\theta \colon X \to X'$ such that $\Gamma \models \Forall{X'} Q\theta$.
We solve queries by narrowing, which is similar to rewriting, but employs most general unifiers (mgu) instead of matching substitutions in order to compute `values' for the variables in $X$ such that the equalities and transitions in $Q$ hold true.
\begin{definition}[Narrowing step] 
  \label{definition:narrowing-step}
Let $(\Sigma,\Gamma)$ be a term rewriting system.
The \emph{one-step-narrowing} relation on queries, indexed by so-called \emph{computed substitutions}, is defined as follows:
\begin{description}[topsep=.5em, itemsep=.25em, font=\normalfont]
\item[\tacase{\eq}] 
$\Exists{X} C[t] \narrow{\eq}_{\psi_{X}} \Exists{X'} (C[\err] \cup H)\psi$ when 
$t$ is not a variable and 
$\psi = \mgu\{t \doteq \ell\} \colon X \uplus Y \to X'$ for some rule $\Forall{Y} (\ell=\err) \lif H \in \Gamma$;
\item[\tacase{=}] $\Exists{X} \{t_{1}=t_{2}\} \cup Q \narrow{=}_{\psi} \Exists{X'} Q\psi$
  when $\psi = \mgu\{t_{1} \doteq t_{2}\} \colon X \to X'$;
\item[\tacase{\tr}] $\Exists{X} \{\tau(t_1,t_2)\} \cup Q \narrow{\tr}_{\psi_{X}} \Exists{X'} (\{\err=t_2\} \cup H \cup Q)\psi$
when $\psi = \mgu\{\tau \doteq \kappa, t_{1} \doteq \ell\} \colon X \uplus Y \to X'$ for some rule $\Forall{Y} \kappa(\ell, \err) \lif H \in \Gamma$;
\item[\tacase{\comp}] $\Exists{X} \{(\act_1\comp\act_2)(t_1,t_2)\} \cup Q \narrow{\comp}_{\iota_{x}} \Exists{X \cup \{x\}} \{\act_1(t_1,x) ,\act_2(x,t_2)\} \cup Q$,
for any new variable $x$, distinct from those in $X$, and for the obvious inclusion $\iota_{x} \colon X \hookrightarrow X \cup \{x\}$;
\item[\tacase{\cup}]
$\Exists{X} \{(\act_1\cup\act_2)(t_1,t_2)\} \cup Q \narrow{\cup}_{1_{X}} \Exists{X} \{\act_i(t_1,t_2)\} \cup Q$,
for any $i\in\{1,2\}$; \footnote{$1_{X} \colon X \to X$ is the identity.}
\item[\tacase{\circlearrowright}]
$\Exists{X} \{\act^*(t_1,t_2)\} \cup Q \narrow{\circlearrowright}_{1_{X}} \Exists{X} \{\act^n(t_1,t_2)\} \cup Q$,
for any $n\in\mathbb{N}$.
\end{description}
\end{definition}
The narrowing relation corresponds to the reflexive and transitive closure of the one-step-narrowing relation.
\begin{definition}[Narrowing]
The general \emph{narrowing} relation is defined by ${\narrow{*}_{\psi}} = \bigcup \{{\narrow{n}_{\psi}} \mid n \in \mathbb{N}\}$, where:
  \begin{enumerate}
  \item $\Exists{X} Q \narrow{0}_{1_{X}} \Exists{X} Q$, and

  \item if $\Exists{X_{1}} Q_1 \narrow{\textit{lb}}_{\psi_{1}} \Exists{X_{2}} Q_2$ and $\Exists{X_{2}} Q_2 \narrow{n}_{\psi_{2}} \Exists{X_{2}} Q_3$,
    then 
    $\Exists{X_{1}} Q_1 \narrow{n+1}_{\psi_{1} \comp \psi_{2}} \Exists{X_{3}} Q_3$, for each label $\textit{lb}\in \{\eq,=,\tr,\comp,\cup,\circlearrowright\}$.
  \end{enumerate}
  Similarly to rewriting, we may drop the label $\textit{lb}$ when we want to refer to an arbitrary narrowing step.
\end{definition}
The following result establishes soundness of narrowing based on the soundness of rewriting.
\begin{theorem}[Reduction lemma \& soundness of narrowing] \label{th:narrow-sound}
  Let $(\Sigma, \Gamma)$ be a term rewriting system.
\begin{enumerate}
\item If $\Exists{X} Q \narrow{n}_{\psi} \Exists{X'} Q'$, then $Q\psi \glrew{n} Q'$.
\item If $\Exists{X} Q \narrow{*}_{\psi} \Exists{X'} Q'$, then for any solution $\theta \colon X' \to X''$ to $\Exists{X'} Q'$, the substitution $\psi \comp \theta$ is a solution to $\Exists{X} Q$.
\end{enumerate}
\end{theorem}
\begin{proof}
We prove the first statement by induction on the number of steps.
 The base case is trivial, hence we deal only with the induction step. To shorten the proof, we present just the case of $\eq$, since all the other ones are similar.
\begin{description}[topsep=.5em, itemsep=.25em, font=\normalfont]
\item[\tacase{\eq}]
Suppose $\Exists{X} C[t] \narrow{\eq}_{\psi_{X}} \Exists{X_{2}} (C[\err] \cup H)\psi \narrow{n}_{\psi_{2}} \Exists{X'} Q'$, where $\psi = \mgu\{t \doteq \ell\} \colon X \uplus Y \to X_{2}$ for some rule $\Forall{Y} (\ell=\err) \lif H \in \Gamma$ and $\psi_{X} \colon X \to X_{2}$ is the restriction of $\psi$ to $X$.
  Let $\psi_{Y} \colon Y \to X_{2}$ be the restriction of $\psi$ to $Y$, and $\psi_{X,2}^{\gap} \colon X \uplus \{\gap\} \to X' \uplus \{\gap\}$ be the unique extension of the composed substitution $\psi_{X} \comp \psi_{2}$ that maps $\gap$ to $\gap$.
  It is easy to see that $C[t]\psi_{X}\psi_{2} = C\psi_{X,2}^{\gap}[t\psi_{X}\psi_{2}]$ and $C\psi_{X,2}^{\gap}[\err\psi_{Y}\psi_{2}] \cup H\psi_{Y}\psi_{2} = (C[\err] \cup H)\psi\psi_{2}$.
  Moreover, since $\psi$ unifies $t$ and $\ell$, it follows that $t\psi_{X}\psi_{2} = \ell\psi_{Y}\psi_{2}$.
  So, by the definition of rewriting, we have $C\psi_{X,2}^{\gap}[t\psi_{X}\psi_{2}] \glrew{\eq} C\psi_{X,2}^{\gap}[\err\psi_{Y}\psi_{2}] \cup H\psi_{Y}\psi_{2}$, from which we get $C[t]\psi_{X}\psi_{2} \glrew{\eq} (C[\err] \cup H)\psi\psi_{2}$.
  Chained with $(C[\err] \cup H)\psi\psi_{2} \glrew{n} Q'$, which holds by the induction hypothesis, this $\eq$ rewriting step entails $C[t]\psi_{X}\psi_{2} \glrew{n+1} Q'$. 
\end{description}
  For the second statement, notice that by the soundness of rewriting (Theorem~\ref{theorem:soundness-of-rewriting}), $\Gamma \cup Q' \models_{\Sigma[X']} Q\psi$.
  By the general translation property of the semantic entailment system, and because $\theta$ leaves the clauses in $\Gamma$ unchanged, it follows that $\Gamma \cup Q'\theta \models_{\Sigma[X'']} Q\psi\theta$.
  But $Q'\theta$ is already a consequence of $\Gamma$ because the substitution $\theta$ is a solution to $\Exists{X'} Q'$.
  Therefore, $\Gamma \models_{\Sigma[X'']} Q\psi\theta$.
  By applying once more the general properties of the semantic entailment system, it follows that $\Gamma \models_{\Sigma} \Forall{X''} Q\psi\theta$, which shows that \(\psi \comp \theta\) is a solution to $\Exists{X} Q$.
\end{proof}

\begin{corollary}
  If\/ $\Exists{X} Q \narrow{*}_{\psi} \Exists{X'} \emptyset$, then $\psi$ is a solution to $\Exists{X} Q$.
\end{corollary}
The converse of the Reduction Lemma also holds. This kind of result is generally known in the literature as a Lifting Lemma, and it plays a crucial role in proving the completeness of narrowing (see Section~\ref{sec:op-complete}).
\subsection{Term rewriting}
In this section, we revisit the rewriting of terms and show how it relates to goal rewriting.
\begin{definition}[Term rewriting]
Let $(\Sigma,\Gamma)$ be a term rewriting system.
We overload the notation used for rewriting goals and define a \emph{one-step-rewriting} relation on $\Sigma$-terms as follows:
for any context $c$, 
\begin{description}[font=\normalfont]
\item[\tacase{\eq}]$
c[\ell\theta]\rtmrew{\eq} c[\err\theta]$ if there exist an equational rule $\Forall{Y} (\ell=\err) \lif H \in \Gamma$ and a substitution $\theta \colon Y \to \emptyset$ such that $H\theta \glrew{*} \emptyset$.
\end{description}
The family of \emph{$n$-step rewriting} relations $\{{\rtmrew{n}} \subseteq T_{\Sigma} \times T_{\Sigma} \mid n \in \mathbb{N}\}$ is defined inductively as follows:
\begin{enumerate*}[label=(\alph*)]
\item $t\rtmrew{0} t$, and
\item if $t_{1}\rtmrew{\eq} t_{2}$ and $t_{2}\rtmrew{n} t_{3}$, then $t_{1} \rtmrew{n+1} t_{3}$.
\end{enumerate*}
We write $t\rtmrew{*}t'$ whenever $t\rtmrew{n}t'$ for some $n\in\mathbb{N}$.
\end{definition}
The same idea of decoupling condition evaluation from actual term rewriting appears in \cite{Meseguer20}, where background theories are used to evaluate more complex conditions such as quantifier-free formulas.
\begin{remark} \label{remark:eq-vars}
Let $(\Sigma,\Gamma)$ be a term rewriting system.
\begin{enumerate}
\item Typically, the conditional rules $\Forall{Y} (\ell=\err) \lif H \in \Gamma$ used in term rewriting are required to satisfy $\var(H) \cup \var(\err) \subseteq \var(\ell) = X$ (see~\cite{Goguen06}).
  That is useful because it allows the block of variables to be omitted with no risk of confusion.
  Moreover, the matching substitution $\theta$ used during rewriting gets uniquely determined by the subterm $t$ that is being rewritten and the left-hand side $\ell$ of the (head of the) rule, thus eliminating a potential source of non-confluence---which we will revisit shortly.
\item The condition on variables can sometimes be relaxed by introducing some additional mechanism to deal with the variables that do not occur in $\ell$; see, for example, the \emph{matching conditions} of Maude~\cite{conf/maude/2007}.
  The condition we employ in this paper is $\var(\err) \subseteq \var(\ell) \cup \var(H) = X$.
  From an operational-semantics perspective, we can use narrowing to compute the action of $\theta$ on $\var(H) \setminus \var(\ell)$, provided that the rules in $\Gamma$ are written in such a way that, if a solution to $\Exists{X} \{t = \ell\theta\} \cup H\theta$ exists, then it is unique.
\item Term rewriting extends canonically from terms to actions:
\begin{enumerate*}[label=(\alph*)]
\item $\act_1\comp\act_2\rtmrew{*}\act_1'\comp\act_2'$ when
$\act_1\rtmrew{*}\act_1'$ and $\act_2\rtmrew{*}\act_2'$;
\item $\act_1\cup\act_2\rtmrew{*}\act_1'\cup\act_2'$ when
$\act_1\rtmrew{*}\act_1'$ and $\act_2\rtmrew{*}\act_2'$; and
\item $\act^*\rtmrew{*}\bact^*$ when
$\act\rtmrew{*}\bact$.
\end{enumerate*}
\end{enumerate}
\end{remark}
Term rewriting together with action rewriting allows us to define a transition relation on terms.
\begin{definition}[Term transition]
  Let $(\Sigma,\Gamma)$ be a term rewriting system.
  The relation $\rtmrew{\act}$ on $\Sigma$-terms is defined by induction on the structure of $\act$:
  \begin{description}[topsep=.5em, itemsep=.25em, font=\normalfont]
  \item [\tacase{\tr}]
    $t_1\rtmrew{\tau} t_2$
    if there exist
    a rewrite rule $\Forall{Y}\kappa(\ell,\err) \lif H \in \Gamma$ and a substitution $\theta:Y\to \emptyset$ 
    such that
    \begin{enumerate*}[label=(\alph*)]
    \item  $t_1\rtmrew{*} \ell\theta$,
    \item $\tau\rtmrew{*}\kappa\theta$,
    \item $\err\theta\rtmrew{*}t_2$, and
    \item $H\theta\glrew{*}\emptyset$;
    \end{enumerate*}
  \item[\tacase{\comp}]
    $t_1\rtmrew{\act_1\comp\act_2}t_2$ 
    if $t_1\rtmrew{\act_1}t$ and $t\rtmrew{\act_2}t_2$ for some term $t\in T_\Sigma$;
  \item[\tacase{\cup}]
    $t_1\rtmrew{\act_1\cup\act_2}t_2$ 
    if $t_1\rtmrew{\act_i}t$ for some $i\in\{1,2\}$; and
  \item[\tacase{\circlearrowright}]
    $t_1\rtmrew{\act^*}t_2$
    if $t_1\rtmrew{\act^n}t_2$ for some $n\in\mathbb{N}$.
  \end{description}
\end{definition}
\begin{remark}
  A note similar to that of Remark~\ref{remark:eq-vars} applies to the transitional rewrite rules $\Forall{Y}\kappa(\ell,\err) \lif H \in \Gamma$ used for computing transitions between terms.
  Here, we require that $\var(\err) \subseteq \var(\kappa) \cup \var(\ell) \cup \var(H) = X$, and we use once more narrowing to determine how the substitution $\theta$ acts on variables in $\var(H) \setminus (\var(\kappa) \cup \var(\ell))$ --- i.e., on variables whose `values' cannot be computed purely by matching.
\end{remark}
The next result shows that the rewriting of goals can be reduced to rewriting terms and checking term transitions.
\begin{lemma}\label{lemma:tr+tt}
Let $(\Sigma,\Gamma)$ be a term rewriting system.
\begin{enumerate}
\item \label{lemma:tr+tt-1} If $\act\rtmrew{*}\bact$ and $t_1\rtmrew{\bact}t_2$, then $t_1\rtmrew{\act}t_2$.
\item \label{lemma:tr+tt-2}  $t_1 \rtmrew{*} u \ltmrew{*} t_2$ for some term $u$ ~iff~ 
$\{t_1=t_2\} \glrew{*} \emptyset$.
\item \label{lemma:tr+tt-3} If $t_1\rtmrew{\act} u \ltmrew{*} t_2$, then $\{\act(t_1,t_2)\}\glrew{*} \emptyset$.
\end{enumerate}
\end{lemma}
\begin{proof}
The proofs of the above statements are straightforward by induction.
\end{proof}
For transitions, the implication in the opposite direction to that of Lemma~\ref{lemma:tr+tt} (\ref{lemma:tr+tt-3})requires some additional constraints on $(\Sigma, \Gamma)$, namely confluence and coherence.
The following notions of confluence and termination are standard in the term rewriting literature (see, e.g.,~\cite{BaaderN98}).
The only difference is that here we apply them to a new setting where arbitrary transitions can occur among the conditions of rewrite rules.
\begin{definition}[Confluence, termination, and canonicity]
A term rewriting system $(\Sigma,\Gamma)$ is:
\begin{itemize}
\item \emph{ground confluent} when for all $\Sigma$-terms $t_1$, $t_2$, and $t$ such that $t_1\ltmrew{*} t\rtmrew{*}t_2$ there exists a term $t'$ such that $t_1\rtmrew{*}t'\ltmrew{*} t_2$; 

\item \emph{confluent} when $(\Sigma[X],\Gamma)$ is ground confluent for all finite blocks of variables $X$ for $\Sigma$;

\item  \emph{ground terminating} when there is no infinite chain of $\Sigma$-term rewrites $t_0\rtmrew{\eq}t_1 \rtmrew{\eq}t_2 \rtmrew{\eq} t_3 \rtmrew{\eq} \dotsb$.

\item \emph{terminating} if $(\Sigma[X],\Gamma)$ is ground terminating for all finite blocks of variables $X$ for $\Sigma$.

\item \emph{canonical} when it is confluent and terminating.
\end{itemize}
\end{definition}
The following notion of coherence is closely related to the strong coherence property introduced in \cite{VIRY2002487}.
Coherence property guarantees that every transition which is possible in the equational semantics can also be computed using the oriented equations~\cite{DBLP:conf/concur/Noll99}.

\begin{definition}[Coherence]
A term rewriting system $(\Sigma,\Gamma)$ is coherent if for 
\begin{enumerate}
\item all transition rules $\Forall{Y} \kappa(\ell,\err) \lif H \in \Gamma$ and 
\item all substitutions $\theta:Y\to \emptyset$ 
\end{enumerate}
such that  $H\theta\glrew{*}\emptyset$ and $\ell\theta\rtmrew{*}t$,
there exist a rule $\Forall{Y'} \kappa'(\ell',\err') \lif H' \in \Gamma$ and 
a substitution $\theta':Y'\to \emptyset$ such that
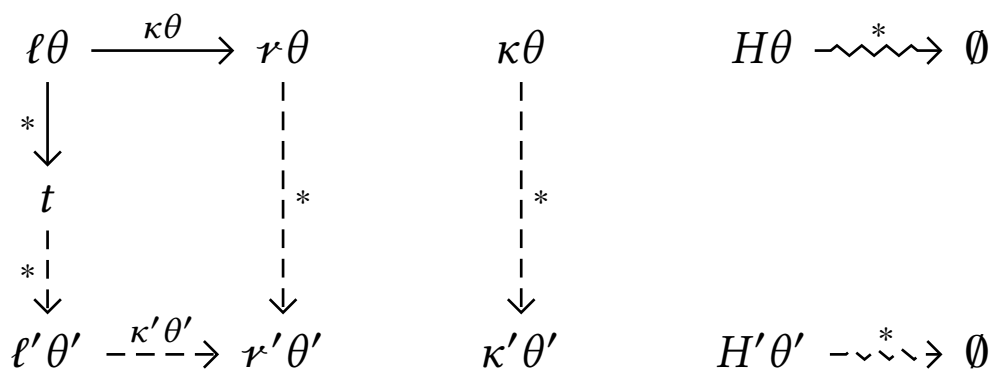
\begin{center}
\begin{enumerate*}[label=(\alph*)]
\item $H'\theta'\glrew{*}\emptyset$,
\item $t \rtmrew{*}\ell'\theta'$, 
\item $\kappa\theta \rtmrew{*} \kappa'\theta'$ and 
\item $\err\theta\rtmrew{*}\err'\theta'$. 
\end{enumerate*}
\end{center}
  \begin{center}
    \begin{tikzcd}[>=Straight Barb]
      \ell\theta \ar[d,->,"*",swap] \ar[r,->,"\kappa\theta"] & 
      \err\theta  \ar[dd,->,"*",dashed] & 
      \kappa\theta \ar[dd,->,dashed,"*"] & 
      H\theta \ar[r,->,"*",squiggly] & \emptyset\\
      t \ar[d,->,"*"{left},dashed] & & & \\
      \ell'\theta' \ar[r,->,"\kappa'\theta'",dashed] & \err'\theta' & 
      \kappa'\theta'  &
      H'\theta' \ar[r,->,"*",dashed,squiggly]& \emptyset
    \end{tikzcd}
  \end{center}
\end{definition}
We present an example of a coherent term rewriting system implemented using our syntax (recall Example~\ref{ex:nat}).
%
\begin{example}[Bubble sort] \label{ex:bubble}
Consider the following term rewriting systems:
{\small \begin{verbatim}
spec BOOL is
  sort Bool .
  ops true, false : () -> Bool .
endspec

spec NAT-LEQ is
  protecting NAT .
  protecting BOOL .
  op _<=_ : Nat Nat -> Bool .
  vars x, y : Nat .
  ax (0 <= y) = true .
  ax (s x <= 0) = false .
  ax (s x <= s y) = (x <= y) .
endspec
\end{verbatim}}
The signature of\/ $\mathtt{BOOL}$ consists of a single sort, $\mathtt{Bool}$, and two constants, $\mathtt{true}$ and $\mathtt{false}$, of sort $\mathtt{Bool}$. 
The term rewriting system $\mathtt{BOOL}$ has no rules.
$\texttt{NAT-LEQ}$ is built on top of\/ $\mathtt{BOOL}$ and $\mathtt{NAT}$ (from Example~\ref{ex:nat}); that is, the signatures of\/ $\mathtt{BOOL}$ and $\mathtt{NAT}$ are included in the signature of\/ $\texttt{NAT-LEQ}$.
In addition, $\texttt{NAT-LEQ}$ introduces a relation (as a Boolean-valued operation) \verb"_<=_: Nat Nat -> Bool", which is defined inductively with respect to both arguments.

Next, we define lists of natural numbers.
These are often presented using subsorting (by letting \texttt{Nat} be a subsort of a newly declared sort of lists of natural numbers) and a list-concatenation operator that is declared associative and with an identity element using special operator attributes such as \texttt{assoc} and \texttt{id:} (e.g.,~\cite{conf/maude/2007}).
Both of those features are compatible with $\TA$. However, in order to keep the conceptual framework as simple as possible, we avoid them for now and define lists of natural numbers using an explicit injection of numbers as singleton lists, denoted by an operation \texttt{i}, and we also make explicit, using equation, the associativity and the identity properties of the concatenation.
Last but not least, we label the axioms (using the keyword \texttt{label:}) to make it easier to reference them later in the text.
{\small\begin{verbatim}
spec NAT-LIST is
  protecting NAT .
  sort NatList .
  op empty : () -> NatList .
  op i : Nat -> NatList .
  op _;_ : NatList NatList -> NatList .
  vars z, z1, z2, z3 : NatList .
  ax empty ; z = z [label: id0] .
  ax z ; empty = z [label: id1] .
  ax (z1 ; z2) ; z3 = z1 ; (z2 ; z3) [label: dt] .
endspec
\end{verbatim}}
The specification $\texttt{NAT-LIST}$ includes both the signature and the rules of\/ $\mathtt{NAT}$.
In addition, $\texttt{NAT-LIST}$ introduces a new sort $\mathtt{NatList}$ together with three constructors (that is, function symbols that build the sort $\mathtt{NatList}$): $\mathtt{empty : \to NatList}$, $\mathtt{i: Nat \to NatList}$ and 
$\verb"_;_: NatList NatList -> NatList"$.
The first two equations state that $\mathtt{empty}$ is the identity for $\verb"_;_"$, while the third equation orients associativity so that terms are normalized to a right-associated form.

Finally, we write a specification of a comparison sort algorithm for lists of natural numbers inspired by bubble sort -- see~\cite{Goguen06} for an earlier algebraic-specification approach to bubble sort.
In the specification \texttt{BUBBLE} below, the added conditional transitions capture
the compare-and-swap steps of bubble sort.
Each of the three conditional transitions applies to a different possible occurrence of a misordered pair of adjacent numbers: in the middle, at the end, or at the beginning of a list.

{\small\begin{verbatim}
spec BUBBLE is
  protecting NAT-LIST .
  protecting NAT-LEQ .
  vars x, y : Nat .
  vars z1, z2 : NatList .
  ax z1 ; (i(x) ; (i(y) ; z2)) =(true)> z1 ; (i(y) ; (i(x) ; z2)) if (y <= x) = true [label: tr1] .
  ax z1 ; (i(x) ; i(y)) =(true)> z1 ; (i(y) ; i(x)) if (y <= x) = true [label: tr2] .
  ax i(x) ; i(y) =(true)> i(y) ; i(x) if (y <= x) = true [label: tr3] .
endspec
\end{verbatim}}
\end{example}
To show that $\mathtt{BUBBLE}$ is coherent, we define a notion of critical pairs formed from a transition rule $\Forall{Y_1}\kappa_1(\ell_1,\err_1) \lif H_1$ and an equation rule $\Forall{Y_2}\ell_2=\err_2 \lif H_2$.
\begin{description}[font=\itshape]
\item [Critical overlap.]
Assume a substitution $\theta:Y_1\uplus Y_2\to \emptyset$ and a position $p\in \mathit{Pos}(\ell_1)$ such that 
\begin{center}
\begin{enumerate*}[label=(\alph*)]
\item $\theta$ is a unifier for $\{\ell_1|_p\doteq\ell_2\}$,
\item $\ell_1|_p$ is not a variable,
\item  $H_1\theta\glrew{*}\emptyset$ and 
\item $H_2\theta\glrew{*}\emptyset$.
\end{enumerate*}
\end{center}
We have the following \emph{critical pair}:
\begin{equation}
(\ell_1\theta)[\err_2\theta]_{p} \ltmrew{\eq} \ell_1\theta \rtmrew{\kappa_1\theta} \err_1\theta.
\end{equation}
The critical pair $(\ell_1\theta)[\err_2\theta]_{p} \ltmrew{\eq} \ell_1\theta \rtmrew{\kappa_1\theta} \err_1\theta$ is \emph{joinable} if 
there exist a transition rule $\Forall{Y'}\kappa'(\ell',\err') \lif H'$ and a substitution $\theta':Y'\to \emptyset$ such that 
\begin{center}
\begin{enumerate*}[label=(\alph*)]
\item $(\ell_1\theta)[\err_2\theta]_p\rtmrew{*}\ell'\theta'$,
\item $\err_1\theta\rtmrew{*} \err'\theta'$,
\item $\kappa_1\theta\rtmrew{*}\kappa'\theta'$ and 
\item $H'\theta'\glrew{*}\emptyset$.
\end{enumerate*}
\end{center}
\end{description}
Another ingredient needed to show that a term rewriting system is coherent is a notion of condition that is preserved under rewriting.
\begin{definition}[Preservation under rewriting]
Consider a term rewriting system $(\Sigma,\Gamma)$ and a set of equations and/or transitions $H$ with variables from $Y$.
We say that $H$ is \emph{preserved under rewriting} if for any substitutions $\theta,\theta':Y\to \emptyset$ such that 
\begin{enumerate*}[label=(\alph*)]
\item $H\theta\glrew{*}\emptyset$ and
\item $\theta(y)\rtmrew{*}\theta'(y)$ for all $y\in Y$,
\end{enumerate*} 
we have $H\theta'\glrew{*}\emptyset$.
\end{definition}
If the underlying term rewriting system is ground confluent and $H$ consists solely of equations, then by Lemma~\ref{lemma:tr+tt}(\ref{lemma:tr+tt-2}), $H$ is preserved under rewriting.
\begin{lemma}[Coherence] \label{lemma:r-coherence}
Let $(\Sigma,\Gamma)$ be a term rewriting system that is ground confluent and ground terminating.
In addition, assume that
\begin{enumerate}
\item all critical pairs arising from the interaction between equation rules and transition rules are joinable, and
\item every transition rule  $\Forall{Y}\kappa(\ell,\err)\lif H\in \Gamma$ that forms a critical pair with an equation rule has a condition $H$ that is preserved under rewriting.
\end{enumerate}
Then $(\Sigma,\Gamma)$ is coherent. 
\end{lemma}
\begin{proof}
Assume that $t\ltmrew{*} \ell\theta \rtmrew{\kappa\theta} \err\theta$ for some transition rule $\Forall{Y}\kappa(\ell,\err)$ and some substitution $\theta:Y\to \emptyset$.
Since the term rewriting relation $(T_\Sigma,\rtmrew{+})$ is terminating, we proceed by well-founded induction on $\ell\theta\rtmrew{*}t$.
Since the base case is straightforward, we focus on the induction step.
There are two cases to consider.
\begin{description}[font=\itshape]
\item[Critical overlap.] 
In this case, we assume 
a ground term $t\in T_\Sigma$,
a transition rule $\Forall{Y_1}\kappa_1(\ell_1,\err_1)\lif H_1\in \Gamma$ and 
an equation rule $\Forall{Y_2}\ell_2=\err_2\lif H_2\in \Gamma$ such that 
$t \ltmrew{*} (\ell_1\theta)[\err_2\theta]_{p} \ltmrew{\eq} \ell_1\theta \rtmrew{\kappa_1\theta} \err_1\theta$, 
where 
\begin{enumerate*}[label=(\alph*)]
\item $p\in Pos(\ell_1)$ is a position such that $\ell_1|_p$ is not a variable, and
\item $\theta:Y_1\uplus Y_2\to \emptyset$ is a unifier for $\{\ell_1|_p\doteq\ell_2\}$ such that
$H_1\theta\glrew{*}\emptyset$ and $H_2\theta\glrew{*}\emptyset$.
\end{enumerate*}
Since all critical pairs are joinable,
there exist a transition rule $\Forall{Y'}\kappa'(\ell',\err') \lif H'$ and a substitution $\theta':Y'\to \emptyset$ such that 
\begin{enumerate*}[label=(\alph*)]
\item $(\ell_1\theta)[\err_2\theta]_p\rtmrew{*}\ell'\theta'$,
\item $\err_1\theta\rtmrew{*} \err'\theta'$,
\item $\kappa\theta\rtmrew{*}\kappa'\theta'$ and 
\item $H'\theta'\glrew{*}\emptyset$.
\end{enumerate*}
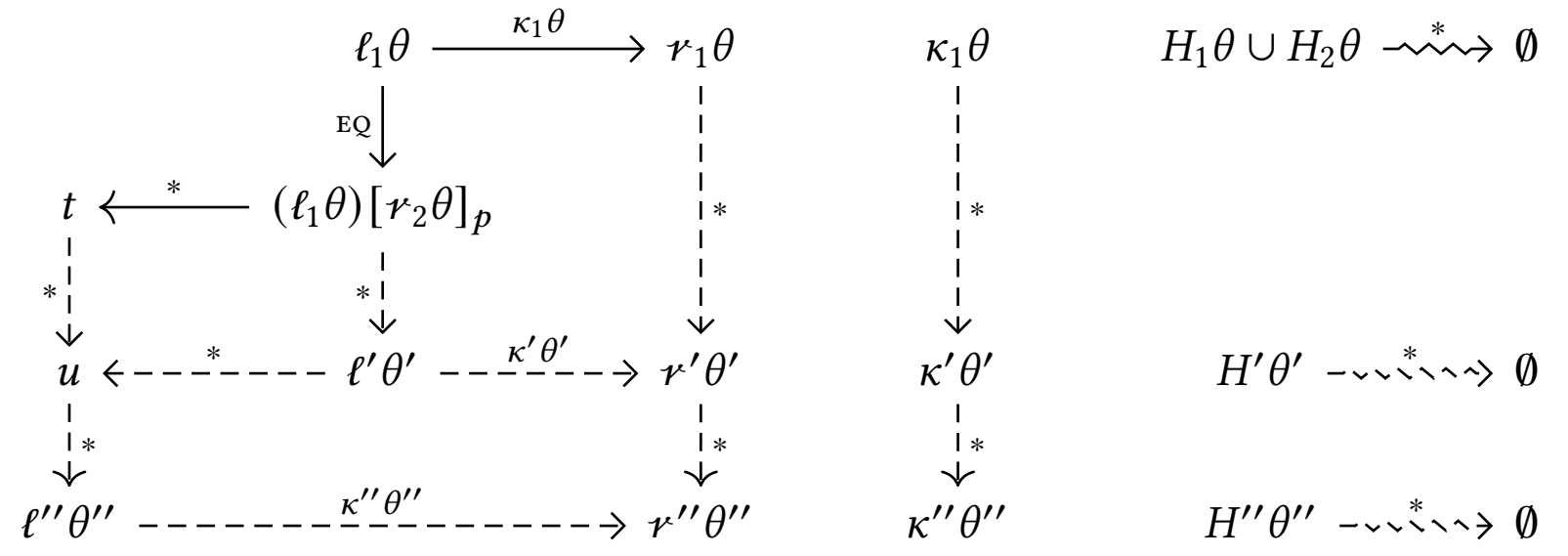
\begin{center}
\begin{tikzcd}[>=Straight Barb]
& \ell_1\theta \ar[d,->,"\eq",swap] \ar[r,->,"\kappa_1\theta"] & 
\err_1\theta  \ar[dd,->,"*",dashed] & 
\kappa_1\theta \ar[dd,->,dashed,"*"] & 
H_1\theta\cup H_2\theta \ar[r,->,"*",squiggly] & \emptyset\\
t \ar[d,->,"*"{left},dashed] & (\ell_1\theta)[\err_2\theta]_p \ar[d,->,"*"{left},dashed] \ar[l,"*"{above}] & & & \\
u \ar[d,dashed,"*"]& \ell'\theta' \ar[r,->,"\kappa'\theta'",dashed] \ar[l,->,"*"{above},dashed] & \err'\theta' \ar[d,dashed,"*"] & 
\kappa'\theta' \ar[d,dashed,"*"] &
H'\theta' \ar[r,->,"*",dashed,squiggly]& \emptyset \\
\ell''\theta'' \ar[rr,->,"\kappa''\theta''",dashed] &  & \err''\theta'' & \kappa''\theta'' & 
H''\theta'' \ar[r,->,"*",dashed,squiggly]& \emptyset
\end{tikzcd}
\end{center}
By ground confluence, $t\rtmrew{*} u \ltmrew{*} \ell'\theta'$ for some term $u\in T_\Sigma$.
By the induction hypothesis applied to $\ell'\theta'$, 
there exist a transition rule $\Forall{Y''}\kappa''(\ell'',\err'') \lif H''$ and a substitution $\theta'':Y''\to \emptyset$ such that 
\begin{enumerate*}[label=(\alph*)]
\item $u \rtmrew{*}\ell''\theta''$,
\item $\err'\theta'\rtmrew{*} \err''\theta''$,
\item $\kappa'\theta'\rtmrew{*}\kappa''\theta''$ and 
\item $H''\theta''\glrew{*}\emptyset$.
\end{enumerate*}
\item[Non-critical overlap.]
In this case,
we assume 
a ground term $t\in T_\Sigma$,
a transition rule $\Forall{Y_1}\kappa_1(\ell_1,\err_1)\lif H_1\in \Gamma$ and 
an equation rule $\Forall{Y_2}\ell_2=\err_2\lif H_2\in \Gamma$ such that
$t\ltmrew{*}(\ell_1\theta_1)[\err_2\theta_2]_{p_1 q_1} \ltmrew{\eq} \ell_1\theta_1 \rtmrew{\kappa_1\theta_1} \err_1\theta_1$, 
where 
\begin{enumerate*}[label=(\alph*)]
\item $p_1\in \mathit{Pos}(\ell_1)$ such that $\ell_1|_{p_1}=x$ for some variable $x\in var(\ell_1)$,
\item $\theta_1:Y_1\to \emptyset$ such that $H_1\theta_1\glrew{*}\emptyset$,
\item $\theta_2:Y_2\to \emptyset$ such that $H_2\theta_2\glrew{*}\emptyset$, and
\item $q_1\in \mathit{Pos}(\theta_1(x))$ such that $\theta_1(x)|_{q_1}$ is not a variable and $\theta_1(x)|_{q_1}=\ell_2\theta_2$.
\end{enumerate*}
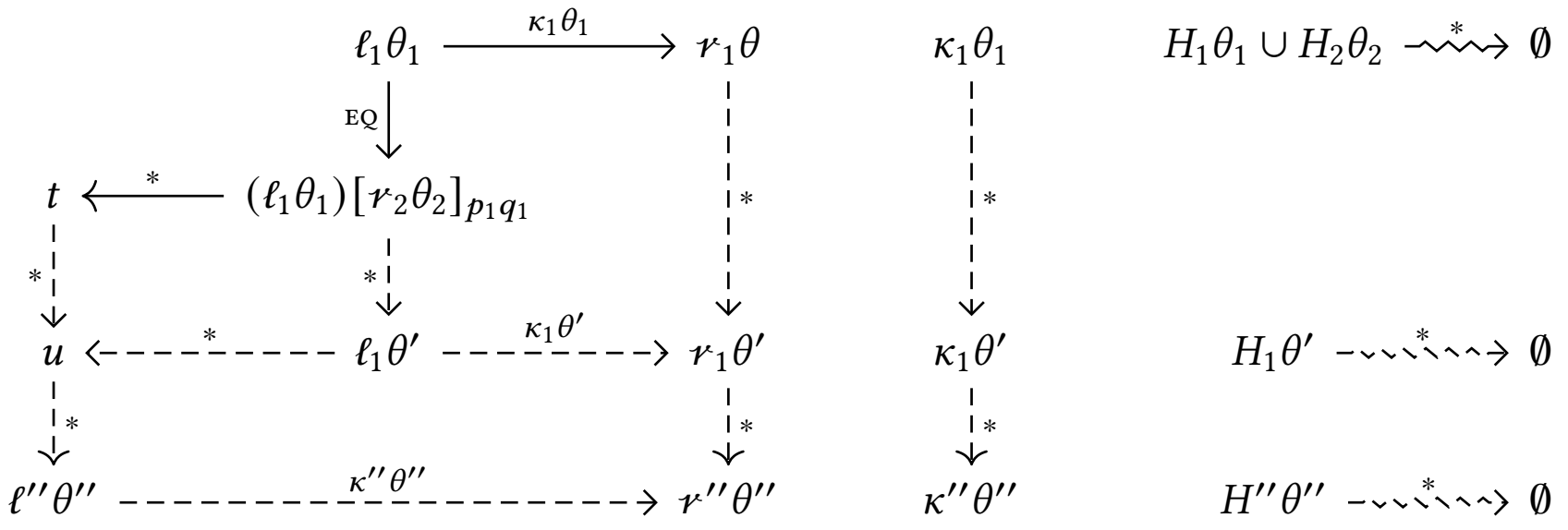
\begin{center}
\begin{tikzcd}[>=Straight Barb]
& \ell_1\theta_1 \ar[d,->,"\eq",swap] \ar[r,->,"\kappa_1\theta_1"] & 
\err_1\theta  \ar[dd,->,"*",dashed] & 
\kappa_1\theta_1 \ar[dd,->,dashed,"*"] & 
H_1\theta_1\cup H_2\theta_2 \ar[r,->,"*",squiggly] & \emptyset\\
t \ar[d,->,"*"{left},dashed] & 
(\ell_1\theta_1)[\err_2\theta_2]_{p_1 q_1} \ar[d,->,"*"{left},dashed] \ar[l,"*"{above}] & & & \\
u \ar[d,dashed,"*"] & 
\ell_1\theta' \ar[r,->,"\kappa_1\theta'",dashed] \ar[l,->,"*"{above},dashed] & \err_1\theta' \ar[d,dashed,"*"] & 
\kappa_1\theta' \ar[d,dashed,"*"] &
H_1\theta' \ar[r,->,"*",dashed,squiggly] & \emptyset \\
\ell''\theta'' \ar[rr,->,"\kappa''\theta''",dashed] &  & \err''\theta'' & \kappa''\theta'' & 
H''\theta'' \ar[r,->,"*",dashed,squiggly]& \emptyset
\end{tikzcd}
\end{center}
Let $\theta':Y_1\to\emptyset$ be the substitution defined by $\theta'(y)=\nf{\theta_1(y)}$ for all variables $y\in Y_1$.
Since $H_1\theta_1\glrew{*}\emptyset$ and $H_1$ is preserved under rewriting,
$H_1\theta'\glrew{*}\emptyset$.
By ground confluence, $t\rtmrew{*} u \ltmrew{*} \ell_1\theta'$ for some term $u\in T_\Sigma$.
By the induction hypothesis applied to $\ell_1\theta'$, 
there exist a transition rule $\Forall{Y''}\kappa''(\ell'',\err'') \lif H''$ and a substitution $\theta'':Y''\to \emptyset$ such that 
\begin{enumerate*}[label=(\alph*)]
\item $u \rtmrew{*}\ell''\theta''$,
\item $\err'\theta'\rtmrew{*} \err''\theta''$,
\item $\kappa'\theta'\rtmrew{*}\kappa''\theta''$ and 
\item $H''\theta''\glrew{*}\emptyset$.
\end{enumerate*}
\end{description}
Hence, $(\Sigma,\Gamma)$ is coherent.
\end{proof}
Observe that the second condition of Lemma~\ref{lemma:r-coherence} holds in any unconditional term rewriting system.

To further investigate coherence in rewriting modulo associativity, commutativity, and/or the identity element, the interested reader may consult \cite{DBLP:conf/concur/Noll99}.
It is straightforward to verify that $\mathtt{BUBBLE}$ is ground terminating and ground confluent.
Critical pairs arise from the interaction of $\mathtt{idi}$ with $\mathtt{trj}$, where $i\in\{0,1\}$ and $j\in\{0,1,2\}$.
We show that all critical pairs are joinable:
\begin{itemize}
\item [] 
Observe that for all natural numbers $\mathtt{n,m}\in\mathbb{N}$ we have 
$\mathtt{(s^n 0 <= s^m 0)= true}$ iff $\mathtt{n\leq m}$ .
It follows that for any substitution $\mathtt{\theta: \{x,y\}\to \emptyset}$, 
we have that $\mathtt{(\theta(y) <= \theta(x)) =true }$  iff $\mathtt{\theta(x)=s^n 0}$ and $\mathtt{\theta(y)=s^m 0}$ for some $\mathtt{n,m}\in\mathbb{N}$ such that $\mathtt{m\leq n}$.
For all $i\in\{0,1\}$ and all $j\in\{0,1,2\}$,
from $\mathtt{idi}$ and $\mathtt{trj}$,
we get the following critical pairs:
\begin{enumerate}
\item $\mathtt{i(s^n 0) ; (i(s^m 0) ; list)\ltmrew{\eq} empty;(i(s^n 0);(i(s^m 0); list)) \rtmrew{true} empty; (i(s^m 0);(i(s^n 0);list))}$
\item $\mathtt{list ; (i(s^n 0); i(s^m 0))\ltmrew{\eq} list;(i(s^n 0);(i(s^m 0); empty)) \rtmrew{true} list; (i(s^m 0);(i(s^n 0);empty))}$
\item $\mathtt{i(s^n 0) ; i(s^m 0) \ltmrew{\eq} i(s^n 0);(i(s^m 0); empty) \rtmrew{true} i(s^m 0);(i(s^n 0); empty)}$
\item $\mathtt{i(s^n 0) ; i(s^m 0)\ltmrew{\eq} empty;(i(s^n 0);i(s^m 0)) \rtmrew{true} empty ; (i(s^m 0) ; i(s^n 0))}$
\end{enumerate}
where $\mathtt{list}$ is any list of natural numbers. 
All critical pairs are joinable.
Take, for example, the first one:
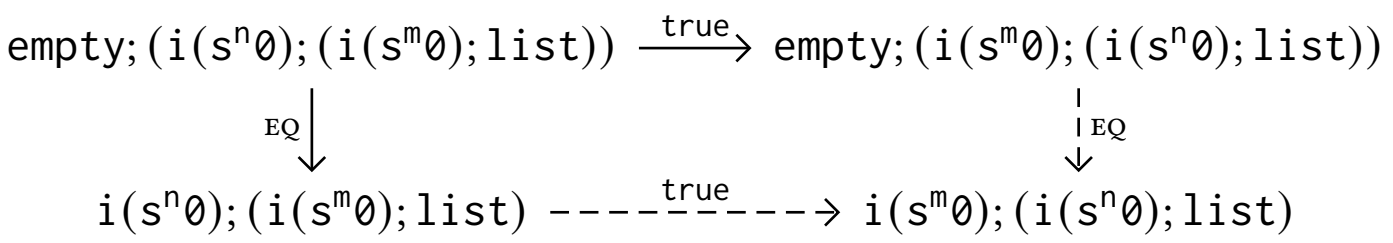
\begin{center}
\begin{tikzcd}[>=Straight Barb]
\mathtt{empty;(i(s^n 0);(i(s^m 0); list))} \ar[r,->,"\mathtt{true}"] \ar[d,->,"\eq",swap]& \mathtt{empty; (i(s^m 0);(i(s^n 0);list))} \ar[d,->,"\eq",dashed] \\
\mathtt{i(s^n 0);(i(s^m 0); list)} \ar[r,->,"\mathtt{true}",dashed]& \mathtt{i(s^m 0);(i(s^n 0);list)}
\end{tikzcd}
\end{center}
\end{itemize}
Hence, $\mathtt{BUBBLE}$ is coherent.
Moreover, note that if any of the transition rules $\mathtt{tr1}$, $\mathtt{tr2}$, or $\mathtt{tr3}$ is removed, the resulting term rewriting system is no longer coherent.

The main result of this section establishes conditions that
\begin{enumerate*}[label=(\alph*)] 
\item ensure a compatibility of transitions with equations and
\item allow goal rewriting to be carried out by means of term rewriting.
\end{enumerate*}
\begin{proposition} \label{prop:coherence}
Let $(\Sigma,\Gamma)$ be a ground confluent and coherent term rewriting system.
\begin{enumerate}
\item \label{prop:coherence-1} If $t_1\rtmrew{\act}t_2$ and $t_1\rtmrew{*}u_1$, then there exists a term $u_2$ such that 
$u_1\rtmrew{\act}u_2$ and $t_2\rtmrew{*}u_2$.
\item \label{prop:coherence-2} If $\{\act(t_1,t_2)\}\glrew{*} \emptyset$, then $t_1\rtmrew{\act}u \ltmrew{*} t_2$\, for some $\Sigma$-term $u$.
\end{enumerate}
\end{proposition}
\begin{proof}
For the first statement,  we proceed by induction on the structure of $\act$:
\begin{description}[font=\normalfont]
\item[\tacase{\tr}]
  Suppose $\act$ is an atomic action $\tau$.
  By the definition of $\rtmrew{\tau}$, there exist a rule $\Forall{Y} \kappa(\ell,\err) \lif H \in \Gamma$ and a substitution $\theta:Y\to\emptyset$ such that $t_1\rtmrew{*}\ell\theta\rtmrew{\kappa\theta}\err\theta\rtmrew{*}t_2$, $\tau\rtmrew{*}\kappa\theta$ and $H\theta\glrew{*}\emptyset$.
  Since $u_1\ltmrew{*}t_1\rtmrew{*} \ell\theta$,
  by ground confluence, 
  there exists a term $v\in T_\Sigma$ such that $\ell\theta\rtmrew{*} v\ltmrew{*}u_1$.
  By coherence, there exists a rewrite rule 
  $\Forall{Y'} \kappa'(\ell',\err') \lif H' \in \Gamma$ and a substitution $\theta':Y'\to \emptyset$ such that 
  $v\rtmrew{*}\ell'\theta'\rtmrew{\kappa'\theta'}\err'\theta'\ltmrew{*}\err\theta$,
  $\kappa\theta \rtmrew{*} \kappa'\theta'$ and 
  $H'\theta'\rtmrew{*}\emptyset$.
  Since $\err'\theta'\ltmrew{*}\err\theta\rtmrew{*}t_2$, 
  by ground confluence, 
  $\err'\theta'\rtmrew{*} u_2\ltmrew{*} t_2$ 
  for some $\Sigma$-term $u_2$.
  
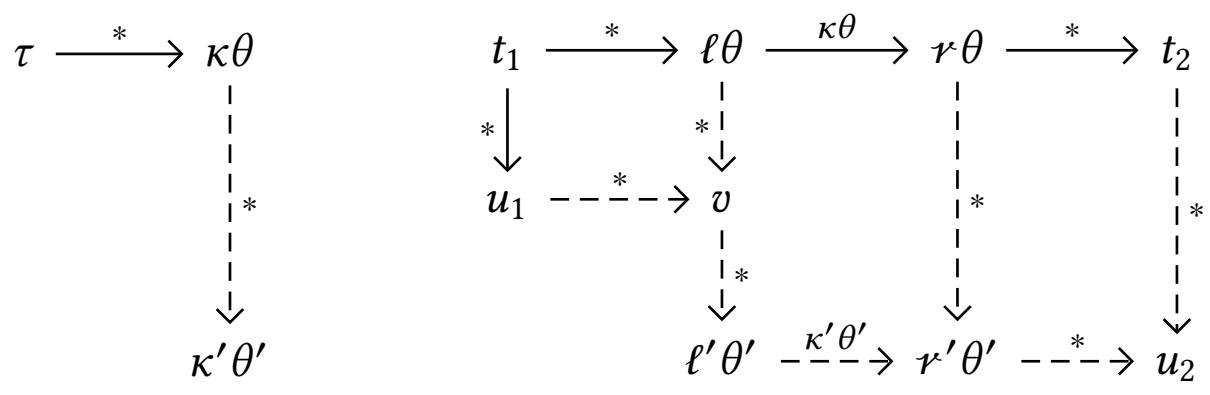
\begin{center}
    \begin{tikzcd}[>=Straight Barb]
      \tau \ar[r,->,"*"] & \kappa\theta \ar[dd,->,dashed,"*"]\\
      & \phantom{\tau''} \\
      &  \kappa'\theta'
    \end{tikzcd}
    \hspace{1cm}
    \begin{tikzcd}[>=Straight Barb]
      t_1 \ar[d,->,"*",swap] \ar[r,->,"*"] & 
      \ell\theta \ar[r,->,"\kappa\theta"] \ar[d,->,dashed,"*",swap] & 
      \err\theta  \ar[r,->,"*"] \ar[dd,->,dashed,"*"]& 
      t_2 \ar[dd,->,dashed,"*"] \\
      u_1 \ar[r,->,dashed,"*"]& 
      v \ar[d,->,dashed, "*"] & &  \\
      & \ell'\theta' \ar[r,->,dashed,"\kappa'\theta'"]& \err'\theta' \ar[r,->,dashed,"*"]& 
      u_2 
    \end{tikzcd}
  \end{center}
  It follows that $u_{1} \rtmrew{*} \ell'\theta'$, $\tau \rtmrew{*} \kappa'\theta'$ and $\err'\theta' \rtmrew{*} u_{2}$.
  Therefore, by the definition of term transitions, $u_{1} \rtmrew{\tau} u_{2}$.
\item[\tacase{\comp}]
Suppose $t_1\rtmrew{\act_1\comp\act_2}t_2$.
By definition, it follows that
$t_1\rtmrew{\act_1}t$ and $t\rtmrew{\act_2}t_2$ for some term $t\in T_\Sigma$.
By the induction hypothesis for $\act_{1}$, there exists a term $u$ such that 
$u_1\rtmrew{\act_1}u$ and $t\rtmrew{*}u$.
Applying the induction hypothesis once more, for $\act_{2}$, we get
$u\rtmrew{\act_2}u_2$ and $t_2\rtmrew{*}u_2$ 
for some $\Sigma$-term $u_2$.
\begin{center}
\begin{tikzcd}[>=Straight Barb]
t_1 \ar[r,->,"\act_1"] \ar[d,->,"*",swap]& 
t \ar[d,->,"*",dashed] \ar[r,->,"\act_2"]& 
t_2 \ar[d,->,dashed,"*"]\\
u_1 \ar[r,->,"\act_1",dashed]& 
u \ar[r,->,dashed,"\act_2"] & u_2
\end{tikzcd}
\end{center}
Hence, by the definition of term transitions,
$u_1\rtmrew{\act_1\comp\act_2}u_2$.
\item[\tacase{\cup}]
  Suppose $t_1\rtmrew{\act_1\cup\act_2}t_2$.
  This means that $t_1\rtmrew{\act_i}t_2$ for some $i\in\{1,2\}$.
  So, by induction hypothesis for $a_{i}$,
 $u_1\rtmrew{\act_i}u_2$ and $t_2\rtmrew{*}u_2$
for some $\Sigma$-term $u_2$.
It follows that
$u_1\rtmrew{\act_1\cup\act_2}u_2$ and $t_2\rtmrew{*}u_2$.
\item[\tacase{\circlearrowright}]
Suppose $t_1\rtmrew{\act^*}t_2$.
This means that $t_1\rtmrew{\act^n}t_2$ for some $n\in\mathbb{N}$.
That is, there is a sequence of terms $v_{1}, v_{2}, \dotsc, v_{n+1}$ such that
$v_{1} = t_{1}$, $v_{n+1} = t_{2}$, and $v_{i} \rtmrew{\act} v_{i+1}$ for all $1 \leq i \leq n$.
By the induction hypothesis for $\act$, there exists a sequence of terms $w_{1}, w_{2}, \dotsc, w_{n+1}$ such that
$w_{1} = u_{1}$ and $w_{i} \rtmrew{\act} w_{i+1} \ltmrew{*} v_{i+1}$ for all $1 \leq i \leq n$.
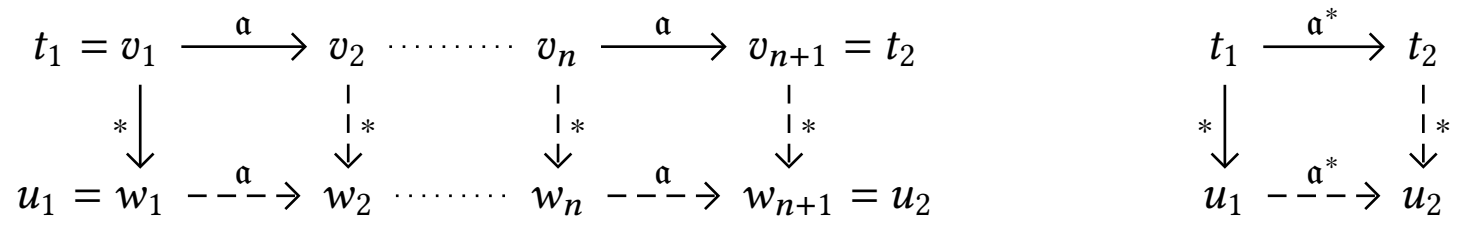
\begin{center}
\begin{tikzcd}[>=Straight Barb]
  \mathllap{t_1 = {}} v_1 \ar[r,->,"\act"] \ar[d,->,"*",swap]
  & v_2 \ar[r,dash,dotted] \ar[d,->,"*",dashed]
  & v_n \ar[r,->,"\act"] \ar[d,->,"*",dashed]
  & v_{n+1}\mathrlap{{} = t_2} \ar[d,->,"*",dashed]\\
  \mathllap{u_1 = {}} w_1 \ar[r,->,"\act",dashed]
  & w_2 \ar[r,dash,dotted]
  & w_n \ar[r,->,"\act",dashed]
  & w_{n+1}\mathrlap{{} = u_2}
\end{tikzcd}
\hspace{2cm}
\begin{tikzcd}[>=Straight Barb]
t_1 \ar[r,->,"\act^*"] \ar[d,->,"*",swap]& 
t_2 \ar[d,->,"*",dashed]  \\
u_1 \ar[r,->,"\act^*",dashed]& u_2
\end{tikzcd}
\end{center}
Now let $u_{2}$ be the term $w_{n+1}$.
It follows that  $u_1\rtmrew{\act^*}u_2$ and $t_2\rtmrew{*}u_2$.
\qedhere
\end{description}

 For the second statement, suppose $\{\act(t_1,t_2)\} \glrew{\textit{lb}} G \glrew{n} \emptyset$ for some label $\textit{lb} \in \{\eq, \tr, \comp, \cup, \circlearrowright\}$ (we exclude the label `$=$' since that kind of rewriting step does not apply to transitions).
  We proceed by case analysis on $\textit{lb}$:
  \begin{description}[font=\normalfont]
  \item[\tacase{\eq}]
    Depending on the choice of extended context, the rewriting takes place either within $\act$, $t_1$, or $t_2$.
    Assume it does so within $\act$, since that is the most complex scenario; the other two are similar.
    This means that there exist
    actions $\bact$ and $\cact$ over the signatures $\Sigma$ and $\Sigma \cup \{\gap\}$, respectively,
    and also
    a rule $\Forall{Y} (\ell=\err) \lif H \in \Gamma$,
    and a substitution $\theta:Y\to \emptyset$ such that
    $\act = \cact(\gap \mapsto \ell\theta)$,
    $\bact = \cact(\gap \mapsto \err\theta)$,
    and
    $G = \{\bact(t_{1}, t_{2})\} \cup H\theta$.
    Since $\{\bact(t_{1}, t_{2})\} \cup H\theta \glrew{n} \emptyset$,
    we have $\{\bact(t_{1}, t_{2})\}  \glrew{n_1} \emptyset$ and $H\theta \glrew{n_2} \emptyset$
    for some natural numbers $n_1$ and $n_2$ such that $n=n_1+n_2$.
    Since $H\theta \glrew{n_2} \emptyset$, we have $\act \rtmrew{*} \bact$.
    Since $\{\bact(t_{1}, t_{2})\}  \glrew{n_1} \emptyset$, by induction hypothesis, there is a $\Sigma$-term $u$ such that
    $t_{1} \rtmrew{\bact} u \ltmrew{*} t_{2}$.
    Finally, by Lemma~\ref{lemma:tr+tt}(\ref{lemma:tr+tt-1}), we get $t_{1} \rtmrew{\act} u \ltmrew{*} t_{2}$.

  \item[\tacase{\tr}]
    Suppose $\act$ is an atomic action $\tau$ and that there exist a rule $\Forall{Y}\kappa(\ell,\err) \lif H \in \Gamma$ and a substitution $\theta:Y\to\emptyset$ such that $\tau=\kappa\theta$ and $t_1=\ell\theta$.
    This means that $G = \{\err\theta = t_{2}\} \cup H\theta$.
    Therefore, since $G \glrew{n} \emptyset$, we get both $\{\err\theta = t_{2}\} \glrew{*} \emptyset$ and $H\theta \glrew{*} \emptyset$.
    The former rewriting implies that there is a term $u$ such that $\err\theta \rtmrew{*} u \ltmrew{*} t_{2}$, whereas the latter entails $t_{1} \rtmrew{\tau} \err\theta$.
    We conclude, by the definition of term transition, that $t_{1} \rtmrew{\tau} u \ltmrew{*} t_{2}$.

  \item[\tacase{\comp}]
    In this case, $G = \{\act_1(t_1,t), \act_2(t,t_2)\}$ for any $\Sigma$-term $t$.
    It follows that $\{\act_1(t_1,t)\} \glrew{n_{1}} \emptyset$ and $\{\act_2(t,t_2)\} \glrew{n_{2}} \emptyset$ for some $n_{1}, n_{2} \leq n$.
    By the induction hypothesis, there exist two $\Sigma$-terms $u_{1}$ and $u_{2}$ such that
    $t_1\rtmrew{\act_1}u_1\ltmrew{*}t$ and 
    $t\rtmrew{\act_2}u_2\ltmrew{*}t_2$, as depicted below.
    \vspace{-\baselineskip}
    \begin{center}
      \begin{tikzcd}[>=Straight Barb]
        & 
        t \ar[d,->,"*"] \ar[r,->,"\act_2"]& 
        u_2 \ar[d,->,dashed,"*"] &
        t_2 \ar[l,->,"*"{above}]\\
        t_1 \ar[r,->,"\act_1"]& 
        u_1 \ar[r,->,dashed,"\act_2"] & 
        v
      \end{tikzcd}
    \end{center}
    Since $(\Sigma, \Gamma)$ is ground confluent and coherent, by the first statement, we know there is a $\Sigma$-term $u$ such that $u_1\rtmrew{\act_2}v\ltmrew{*}u_2$.
    Since $t_1\rtmrew{\act_1}u_1\rtmrew{\act_2}u$ and $t_2\rtmrew{*}u_2\rtmrew{*}u$,
    we get $t_1\rtmrew{\act_1\comp\act_2}u \ltmrew{*}t_2$.

  \item[\tacase{\cup}]
    Suppose $\act = \act_{1} \cup \act_{2}$.
    It follows that $G = \{\act_{i}(t_{1}, t_{2})\}$ for some $i \in \{1, 2\}$.
    By the induction hypothesis, there exists a $\Sigma$-term $u$ such that $t_1\rtmrew{\act_i}u\ltmrew{*}t_2$
    Therefore, by the definition of term transitions, $t_1\rtmrew{\act_1\cup\act_2}u\ltmrew{*}t_2$.
    
  \item[\tacase{\circlearrowright}]
    In this case, $G = \{\act^{m}(t_{1}, t_{2})\}$ for some $m \in \mathbb{N}$.
    By the  induction hypothesis, $t_1\rtmrew{\act^m}u\ltmrew{*}t_2$ for some term $u$, which entails, by the definition of term transitions, that $t_1\rtmrew{\act^*}u\ltmrew{*}t_2$.
    \qedhere
  \end{description}
\end{proof}
%

\section{Strict rewriting} \label{sec:strict}
Under coherence and canonicity, transitions from a given term can be computed in a concise and systematic way: the term is first reduced to its normal form, and a transition rule is then applied to the resulting term. 
To this end, we introduce a stricter form of rewriting goals:

\begin{definition}[Strict goal rewriting] \label{def:strict-goal-rew}
  We say that a goal $G$ \emph{rewrites strictly} to $G'$, and write $G \glrew{*!} G'$, when there exists a sequence of rewriting steps\/ $G \glrew{} G_{1} \glrew{} G_{2} \glrew{} \dotsb \glrew{} G'$ where all the occurrences of the `$\comp$' step are instances of the following stricter decomposition of actions:
\begin{description}[font=\normalfont]
\item[\tacase{\comp!}] $\{(\act_1\comp\act_2)(t_1,t_2)\} \cup G \glrew{\comp!} \{\act_1(t_1,t), \act_2(t,t_2)\} \cup G$,
for any \emph{irreducible} $\Sigma$-term $t$, that is, there is no $\Sigma$-term $u$ such that $t\rtmrew{\eq}u$.
\end{description}
That is, ${\glrew{*!}} = \bigcup \{{\glrew{n!}} \mid n \in \mathbb{N}\}$, where:
  \begin{enumerate*}[label=(\alph*)]
  \item $G \glrew{0!} G$, and
  \item if $G_1\glrew{\textit{lb}} G_2$ and $G_2\glrew{n!} G_3$,
    then 
    $G_1 \glrew{(n+1)!} G_3$, for each label $\textit{lb}\in \{\eq,=,\tr,\comp!,\cup,\circlearrowright\}$.
  \end{enumerate*}
\end{definition}
Strict goal rewriting induces a more constrained variant of term rewriting, which we call \emph{strict term rewriting}, where conditions are rewritten according to Definition~\ref{def:strict-goal-rew}.
\begin{definition}[Strict term rewriting]
Let $(\Sigma,\Gamma)$ be a term rewriting system.
For any context $c$, we have 
\begin{description}[font=\normalfont]
\item[\tacase{\eq!}]  $c[\ell\theta]\rtmrew{\eq!} c[\err\theta]$ if there exist an equational rule $\Forall{Y} (\ell=\err) \lif H \in \Gamma$ and a substitution $\theta \colon Y \to \emptyset$ such that $H\theta \glrew{*!} \emptyset$.
\end{description}
The family of \emph{$n$-step strict rewriting} relations $\{{\rtmrew{n!}} \subseteq T_{\Sigma} \times T_{\Sigma} \mid n \in \mathbb{N}\}$ is defined inductively as follows:
\begin{enumerate*}[label=(\alph*)]
\item $t\rtmrew{0!} t$, and 
\item if $t_{1}\rtmrew{\eq!} t_{2}$ and $t_{2}\rtmrew{n!} t_{3}$, then $t_{1} \rtmrew{(n+1)!} t_{3}$.
\end{enumerate*}
We write $t\rtmrew{*!}t'$ whenever $t\rtmrew{n!}t'$ for some $n\in\mathbb{N}$.
\end{definition}
Observe that that for any unconditional term rewriting system, term rewriting coincides with strict term rewriting.
Based on strict term rewriting, we define a constrained version of term transition, which we call \emph{strict term transition}.
\begin{definition}[Strict term transition]
We say that a term $t$ \emph{transitions strictly} to $t'$ via an action $\act$, and write $t \rtmrew{\act!} t'$, when $t \rtmrew{\act} t'$ such that all the occurrences of `\tr' and `$\comp$' steps are instances of the following stricter rules:
\begin{description}[topsep=.25em, itemsep=.5em, font=\normalfont]
\item [\tacase{\tr!}]
    $t_1\rtmrew{\tau!} t_2$
    if there exist
    a rewrite rule $\Forall{Y}\kappa(\ell,\err) \lif H \in \Gamma$ and a substitution $\theta:Y\to \emptyset$ 
    such that
    \begin{enumerate*}[label=(\alph*)]
    \item  $t_1\rtmrew{*!} \ell\theta$,
    \item $\tau\rtmrew{*!}\kappa\theta$,
    \item $\err\theta\rtmrew{*!}t_2$, and
    \item $H\theta\glrew{*!}\emptyset$;
    \end{enumerate*}
\item[\tacase{\comp!}] 
$t_1\rtmrew{(\act_1\comp\act_2)!}t_2$ if $t_1\rtmrew{\act_1!}t$ and $t\rtmrew{\act_2!}t_2$ for some irreducible term $t\in T_\Sigma$;
 \item[\tacase{\cup!}]
    $t_1\rtmrew{(\act_1\cup\act_2)!}t_2$ 
    if $t_1\rtmrew{\act_i!}t$ for some $i\in\{1,2\}$; and
  \item[\tacase{\circlearrowright!}]
    $t_1\rtmrew{\act^*!}t_2$
    if $t_1\rtmrew{\act^n!}t_2$ for some $n\in\mathbb{N}$.

\end{description}
\end{definition}
\begin{proposition}\label{prop:str+stt}
Let $(\Sigma,\Gamma)$ be a term rewriting system.
\begin{enumerate}
\item \label{prop:str+stt-1} If $\act\rtmrew{*!}\bact$ and $t_1\rtmrew{\bact!}t_2$, then $t_1\rtmrew{\act!}t_2$.
\item \label{prop:str+stt-2} 
$t_1 \rtmrew{*!} u \ltmrew{*!} t_2$ for some term $u$ \quad iff \quad 
$\{t_1=t_2\} \glrew{*!} \emptyset$.
\item \label{prop:str+stt-3}  
$t_1\rtmrew{\act!} u \ltmrew{*!} t_2$ for some term $u$ \quad iff \quad
$\{\act(t_1,t_2)\}\glrew{*!} \emptyset$.
\end{enumerate}
\end{proposition}
\begin{proof}
The proof of the first statement is straightforward, by induction on the structure of the action $\act$. \smallskip

For the second statement, the forward implication, one can straightforwardly prove by induction on the number of strict rewrite steps that
$t_1 \rtmrew{n!} u$ iff $\{t_1=x\}\rtmrew{n!}\{u=x\}$ for any natural number $n$ and any new variable $x$ for $\Sigma$.
Then from $t_1 \rtmrew{*!} u \ltmrew{*!} t_2$ we get $\{t_1=x\}\glrew{*!}\{u=x\}$ and $\{t_2=x\}\glrew{*!}\{u=x\}$ for some new variable $x$ for $\Sigma$.
It follows that $\{t_1=u\}\glrew{*!}\{u=u\}$ and $\{t_2=u\}\glrew{*!}\{u=u\}$.
Hence, $\{t_1=t_2\}\glrew{*!} \{ u=t_2 \}\glrew{*!} \{u=u\}\glrew{0!}\emptyset$. 

The backward implication of the second statement follows easily by induction on the number of rewrite steps.
\smallskip

For the third statement, the forward implication, one can straightforwardly prove by induction on the number of strict rewrite steps that:
\begin{enumerate*}[label=(\alph*)]
\item if $\act\rtmrew{*!}\bact$, then $\act(t_1,t_2)\glrew{*!}\bact(t_1,t_2)$, 
\item if $t_1\rtmrew{*!}u_1$, then $\act(t_1,t_2)\glrew{*!}\act(u_1,t_2)$, and
\item if $t_2\rtmrew{*!}u_2$, then $\act(t_1,t_2)\glrew{*!}\act(t_1,u_2)$.
\end{enumerate*}
Second, apply induction on the structure of $\act$ to prove that $t_1\rtmrew{\act!}u$ implies $\act(t_1,u)\glrew{*!}\emptyset$.
Third, we get $t_1\rtmrew{\act!} u \ltmrew{*!} t_2$ implies $\{\act(t_1,t_2)\}\glrew{*!}\{\act(t_1,u)\} \glrew{*!}\emptyset$.

For the third statement, the backward implication, 
suppose we have $\{\act(t_1,t_2)\} \glrew{\textit{lb}} G \glrew{n!} \emptyset$ for some label $\textit{lb} \in \{\eq, \tr, \comp!, \cup, \circlearrowright\}$ (we exclude the label `$=$' since that kind of rewriting step does not apply to transitions).
 We proceed by case analysis on $\textit{lb}$:
  \begin{description}[font=\normalfont]
  \item[\tacase{\eq}]
    Depending on the choice of extended context, the rewriting takes place either within $\act$, $t_1$, or $t_2$.
    Assume it does so within $\act$, since that is the most complex scenario; the other two are similar.
    This means that there exist
    action $\bact$ over the signature $\Sigma \cup \{\gap\}$,
    a rule $\Forall{Y} (\ell=\err) \lif H \in \Gamma$,
    and a substitution $\theta:Y\to \emptyset$ such that
    $\act = \bact[\ell\theta]$ and $G=\{\bact[\err\theta](t_1,t_2)\}\cup H\theta$. 
    Let us denote $\bact[\err\theta]$ by $\act'$, and we have
    $\{\act(t_1,t_2)\}\glrew{\textit{\eq}} \{\act'(t_1,t_2)\}\cup H\theta  \glrew{n!} \emptyset$.
    It follows that $\{\act'(t_1,t_2)\} \glrew{n_1!} \emptyset$ and $H\theta  \glrew{n_2!} \emptyset$ for some natural numbers $n_1$ and $n_2$ with $n=n_1+n_2$.
    Since $H\theta  \glrew{n_2!} \emptyset$, we have $\act \glrew{\eq!}\act'$.
    Since $\{\act'(t_1,t_2)\} \glrew{n_1!} \emptyset$, by induction hypothesis, $t_1\rtmrew{\act'!}u \ltmrew{*!} t_2$\, for some $\Sigma$-term $u$.
    Finally, from the first statement, we get $t_{1} \rtmrew{\act!} u \ltmrew{*!} t_{2}$.
  \item[\tacase{\tr!}]
    Suppose $\act$ is an atomic action $\tau$ and that there exist a rule $\Forall{Y}\kappa(\ell,\err) \lif H \in \Gamma$ and a substitution $\theta:Y\to\emptyset$ such that $\tau=\kappa\theta$ and $t_1=\ell\theta$.
    This means that $G = \{\err\theta = t_{2}\} \cup H\theta$.
    Therefore, since $G \glrew{n!} \emptyset$, we get both $\{\err\theta = t_{2}\} \glrew{*!} \emptyset$ and $H\theta \glrew{*!} \emptyset$.
    The former rewriting implies that there is a term $u$ such that $\err\theta \rtmrew{*!} u \ltmrew{*!} t_{2}$, whereas the latter entails $t_{1} \rtmrew{\tau!} \err\theta$.
    We conclude, by the definition of term transition, that $t_{1} \rtmrew{\tau!} u \ltmrew{*!} t_{2}$.

  \item[\tacase{\comp}]
  Suppose $\act = \act_{1} \comp \act_{2}$.
  In this case, $G = \{\act_1(t_1,t), \act_2(t,t_2)\}$,  where $t$ is an irreducible $\Sigma$-term.
  We get $\{\act_1(t_1,t)\} \glrew{n_1!} \emptyset$ and $\{\act_2(t,t_2)\} \glrew{n_2!} \emptyset$ for some natural numbers $n_{1}$ and $n_{2}$ with $n=n_1+n_2$.
 By the induction hypothesis, there exist two $\Sigma$-terms $u_{1}$ and $u_{2}$ such that
 $t_1\rtmrew{\act_1!}u_1\ltmrew{*!}t$ and 
 $t\rtmrew{\act_2!}u_2\ltmrew{*!}t_2$.
 Since $t$ is irreducible, $u_1=t$.
 Hence, $t_1\rtmrew{(\act_1\comp\act_2)!}u_2 \ltmrew{*!}t_2$.
  \item[\tacase{\cup}]
    Suppose $\act = \act_{1} \cup \act_{2}$.
    It follows that $G = \{\act_{i}(t_{1}, t_{2})\}$ for some $i \in \{1, 2\}$.
    By the induction hypothesis, there exists a $\Sigma$-term $u$ such that $t_1\rtmrew{\act_i!}u\ltmrew{*!}t_2$
    Therefore, by the definition of strict term transition, $t_1\rtmrew{\act_1\cup\act_2!}u\ltmrew{*!}t_2$.
    
  \item[\tacase{\circlearrowright}]
    Suppose $\act = \act^*$.
    In this case, $G = \{\act^{m}(t_{1}, t_{2})\}$ for some $m \in \mathbb{N}$.
    By the  induction hypothesis, $t_1\rtmrew{\act^m!}u\ltmrew{*!}t_2$ for some term $u$, which entails, by the definition of strict term transition, that $t_1\rtmrew{\act^*!}u\ltmrew{*!}t_2$.
    \qedhere
  \end{description}
\end{proof}
Obviously, strict rewriting and strict transition are special cases of rewriting and transition, respectively. This naturally leads us to ask under which conditions
\begin{enumerate*}[label=(\alph*)]
\item strict rewriting is equivalent to ordinary rewriting, and
\item every sequence of transitions can be realized as a sequence of strict transitions.
\end{enumerate*}

We address the first question and we define a notion of \emph{modular} term rewriting system.
\begin{definition}[Modular term rewriting system] \label{def:modular}
A term rewriting system $(\Sigma,\Gamma)$ is \emph{modular} if $\Gamma$ is a disjoint union of rules, that is, $\Gamma=\Gamma_\eq\uplus \Gamma_\tr$ such that 
\begin{enumerate}
\item $\Gamma_\eq$ is free of composition `$\comp$' and iteration `$*$'.
\item $\Gamma_\eq$ contains all conditional equations from $\Gamma$, that is, $\Forall{X} \ell=\err \lif H\in \Gamma_\eq$ whenever $\Forall{X}\ell=\err \lif H\in \Gamma$.
\item For all $\Forall{X} \varphi\lif \{ \kappa_1(\ell_1,\err_1)\} \cup H\in\Gamma_\eq$ and  $\Forall{Y} \kappa_2(\ell_2,\err_2) \lif Q\in\Gamma_\tr$,
the sort of  $\kappa_1$ is different from the sort of $\kappa_2$.
\end{enumerate}
\end{definition}
The last condition in Definition~\ref{def:modular} is a closure property ensuring that, when evaluating conditions from $\Gamma_{\eq}$, no transition rules from $\Gamma_{\tr}$ are used; in particular, $(\Sigma,\Gamma_\eq)$ is a self-contained term rewriting system.
Together with the first condition in Definition~\ref{def:modular}, this ensures that, for any modular term rewriting system, ordinary term rewriting coincides with strict term rewriting.
\begin{lemma}\label{lemma:modular}
Let $(\Sigma,\Gamma)$ be a modular term rewriting system, 
where $\Gamma=\Gamma_\eq\uplus \Gamma_\tr$ and $\Gamma_\eq$ and $\Gamma_\tr$ satisfy the properties from Definition~\ref{def:modular}.
For all $\Sigma$-terms $t_1$ and $t_2$ of the same sort, the following are equivalent:
\begin{center}
\begin{enumerate*}[label=(\alph*)]
\item $t_1\rtmrew{*}t_2$ over $(\Sigma,\Gamma)$,
\item $t_1\rtmrew{*}t_2$ over $(\Sigma,\Gamma_\eq)$,
\item $t_1\rtmrew{*!}t_2$ over $(\Sigma,\Gamma_\eq)$.
\end{enumerate*}
\end{center}
\end{lemma}
\begin{proof}
By the second and third condition of Definition~\ref{def:modular},
$t_1\rtmrew{*}t_2$ over $(\Sigma,\Gamma)$ if and only if $t_1\rtmrew{*}t_2$ over $(\Sigma,\Gamma_\eq)$.
By the first condition of Definition~\ref{def:modular},
$t_1\rtmrew{*}t_2$ over $(\Sigma,\Gamma_\eq)$ if and only if $t_1\rtmrew{*!}t_2$ over $(\Sigma,\Gamma_\eq)$.
\end{proof}
Maude specifications are modular by default in the sense of Definition~\ref{def:modular}, 
since the condition $H$ of an equation rule $\Forall{Y}\ell=\err\lif H$ is not permitted to contain transitions. 
The rest of the section provides conditions ensuring that every transition sequence can be realized by a sequence of strict transitions.
\begin{definition}[Strict coherence]
  A term rewriting system $(\Sigma,\Gamma)$ is strictly coherent if for 
\begin{enumerate}  
\item all transition rules $\Forall{Y} \kappa(\ell,\err) \lif H \in \Gamma$,
\item all substitutions $\theta:Y\to \emptyset$ such that $H\theta\glrew{*}\emptyset$, and
\item all normal forms $\nf{\ell\theta}$ of $\ell\theta$,
\end{enumerate}
there exist a transition rule $\Forall{Y'} \kappa'(\ell',r') \lif H' \in \Gamma$ and 
a substitution $\theta':Y'\to \emptyset$ such that
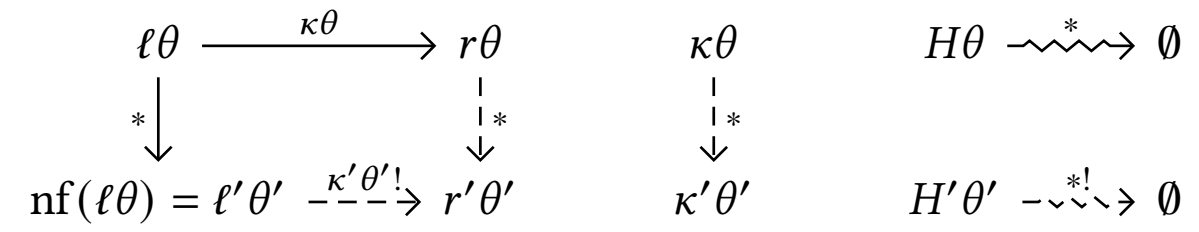
\begin{center}
\begin{enumerate*}[label=(\alph*)]
\item $H'\theta'\glrew{*!}\emptyset$,
\item $\ell'\theta'=\nf{\ell\theta}$, 
\item $\kappa\theta \rtmrew{*} \kappa'\theta'$ and 
\item $r\theta\rtmrew{*}r'\theta'$. 
\end{enumerate*}
\end{center}
  \begin{center}
    \begin{tikzcd}[>=Straight Barb]
      \ell\theta \ar[d,->,"*",swap] \ar[r,->,"\kappa\theta"] & 
      r\theta  \ar[d,->,"*",dashed] & 
      \kappa\theta \ar[d,dashed,->,"*"] & 
      H\theta \ar[r,->,"*",squiggly] & \emptyset\\
      \nf{\ell\theta}=\ell'\theta' \ar[r,->,"\kappa'\theta'!",dashed] & r'\theta' & 
      \kappa'\theta'  &
      H'\theta' \ar[r,->,"*!",dashed,squiggly]& \emptyset
    \end{tikzcd}
  \end{center}
\end{definition}
If the underlying term rewriting system $(\Sigma,\Gamma)$ is ground terminating then strict coherence implies coherence.
However, this implication does not hold in general.
\begin{example}\label{ex:collapse}
  The term rewriting system $\mathtt{COLLAPSE}$ is obtained by putting together two smaller term rewriting systems $\mathtt{NAT}$ and $\mathtt{BOOL}$  (from Examples~\ref{ex:nat} and~\ref{ex:bubble}) and then adding on top four more axioms.
  Note that, in this case, we are \texttt{including} $\mathtt{NAT}$ instead of \texttt{protecting} it because the last two of the added axioms identify `old' elements (natural numbers).
{\small
\begin{verbatim}
spec COLLAPSE is
  including NAT .
  protecting BOOL .
  var x : Nat .
  ax 0 =(true)> 0 .
  ax s x =(false)> s 0 .
  ax 0 = s 0 .
  ax s x = s s x .
endspec
\end{verbatim}}
\end{example}
Observe that the equation $\mathtt{\Forall{x}s~x=  s~s~x}$ in $\mathtt{COLLAPSE}$ causes a failure of termination.
As a consequence, no term of sort $\mathtt{Nat}$ has a normal form. 
This implies that $\mathtt{COLLAPSE}$ is strictly coherent.
However, $\mathtt{COLLAPSE}$ is not coherent.
Indeed, for any $n>0$, we have
$\mathtt{s^n 0  \ltmrew{*} s~0 \ltmrew{\eq} 0  \rtmrew{true} 0 \rtmrew{\eq} s~0}$ and
$\mathtt{s^n 0 \rtmrew{\mathit{false}} s~0}$, while
$\mathtt{true\rtmrew{*} false}$ does not hold.
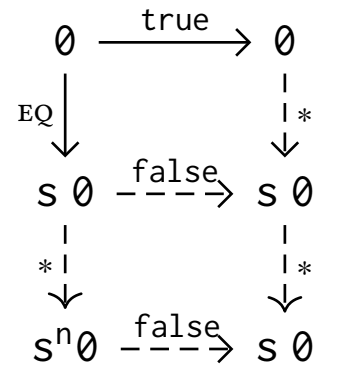
\begin{center}
    \begin{tikzcd}[>=Straight Barb]
      \mathtt{0} \ar[d,->,"\eq",swap] \ar[r,->,"\mathtt{true}"] & 
      \mathtt{0}  \ar[d,->,"*",dashed] \\
      \mathtt{s~0} \ar[d,dashed,"*"{left}] \ar[r,->,"\mathtt{false}",dashed] & \mathtt{s~0} \ar[d,dashed,"*"]\\
      \mathtt{s^n0} \ar[r,->,"\mathtt{false}",dashed] & \mathtt{s~0} 
    \end{tikzcd}
  \end{center}
The main result of this section provides conditions ensuring that any sequence of transition steps can be achieved by a sequence of strict transitions.
\begin{proposition}\label{prop:ordinary-strict}
Consider a term rewriting system $(\Sigma,\Gamma)$ that is 
\begin{enumerate*}[label=(\alph*)]
\item ground confluent, 
\item strictly coherent, and 
\item modular in the sense of Definition~\ref{def:modular} such that $(\Sigma,\Gamma_\eq)$ is ground terminating.
\end{enumerate*}
If $t_1\rtmrew{\act}t_2$ 
then $\nf{t_1}\rtmrew{\act!}\nf{t_2}$,
where $\nf{t_i}$ denote the (unique) normal form of $t_i$, for each $i \in \{1,2\}$.
\end{proposition}
\begin{proof}
We proceed by induction on the structure of $\act$:
\begin{description}[font=\normalfont]
\item[\tacase{\tr}]
  Suppose $\act$ is an atomic action $\tau$.
  By the definition of $\rtmrew{\tau}$, there exist a transition rule $\Forall{Y} \kappa(\ell,\err) \lif H \in \Gamma$ and a substitution $\theta:Y\to\emptyset$ such that 
\begin{center}
\begin{enumerate*}[label=(\alph*)]
\item $t_1\rtmrew{*}\ell\theta\rtmrew{\kappa\theta}\err\theta\rtmrew{*}t_2$, 
\item $\tau\rtmrew{*}\kappa\theta$ and 
\item $H\theta\glrew{*}\emptyset$.
\end{enumerate*}
\end{center}
By strict coherence, there exist 
a rule $\Forall{Y'} \kappa'(\ell',\err') \lif H' \in \Gamma$ and 
a substitution $\theta':Y'\to \emptyset$ such that 
\begin{center}
\begin{enumerate*}[label=(\alph*)]
\item $\nf{\ell\theta}=\ell'\theta'\rtmrew{\kappa'\theta'}\err'\theta'$,
\item $\err\theta \rtmrew{*} \err'\theta'$,
\item $\kappa\theta \rtmrew{*} \kappa'\theta'$ and
\item $H'\theta'\rtmrew{*!}\emptyset$.
\end{enumerate*}
\end{center}
Since $\err'\theta'\ltmrew{*}\err\theta\rtmrew{*}t_2$, 
by ground confluence, 
$\err'\theta'\rtmrew{*} \nf{t_2}\ltmrew{*} t_2$.
\begin{center}
\begin{tikzcd}[>=Straight Barb]
\tau \ar[r,->,"*"] & \kappa\theta \ar[d,->,dashed,"*"]\\
&  \kappa'\theta'
\end{tikzcd}
    \hspace{1cm}
    \begin{tikzcd}[>=Straight Barb]
      t_1 \ar[r,->,"*"] & 
      \ell\theta \ar[r,->,"\kappa\theta"] \ar[d,->,dashed,"*",swap] & 
      \err\theta  \ar[r,->,"*"] \ar[d,->,dashed,"*"]& 
      t_2 \ar[d,->,dashed,"*"] \\
      &\nf{\ell\theta}= \ell'\theta' \ar[r,->,dashed,"\kappa'\theta'!"]& \err'\theta' \ar[r,->,dashed,"*"]& 
      \nf{t_2}
    \end{tikzcd}
  \end{center}
It follows that $\nf{t_1} =\nf{\ell\theta}= \ell'\theta'$, 
$\err'\theta' \rtmrew{*} \nf{t_2}$ and
$\tau \rtmrew{*} \kappa'\theta'$.
Since $(\Sigma,\Gamma)$ is modular, by Lemma~\ref{lemma:modular}, 
$t_1\rtmrew{*!}\nf{\ell\theta}$ and $\err'\theta' \rtmrew{*!} \nf{t_2}$.
By the definition of strict term transition, $\nf{t_1} \rtmrew{\tau!} \nf{t_2}$.
\item[\tacase{\comp}]
Suppose $t_1\rtmrew{\act_1\comp\act_2}t_2$.
By definition, it follows that
$t_1\rtmrew{\act_1}t$ and $t\rtmrew{\act_2}t_2$ for some term $t\in T_\Sigma$.
Since $(\Sigma,\Gamma_\eq)$ is ground terminating,
$t$ has a normal form $\nf{t}$.
By the induction hypothesis for $\act_{1}$, 
$\nf{t_1}\rtmrew{\act_1!}\nf{t}$.
Applying the induction hypothesis once more, for $\act_{2}$, we get $\nf{t}\rtmrew{\act_2!}\nf{t_2}$.
By definition, $\nf{t_1}\rtmrew{(\act_1\comp\act_2)!}\nf{t_2}$.
\item[\tacase{\cup}]
  Suppose $t_1\rtmrew{\act_1\cup\act_2}t_2$.
  This means that $t_1\rtmrew{\act_i}t_2$ for some $i\in\{1,2\}$.
  So, by induction hypothesis for $a_{i}$,
 $\nf{t_1}\rtmrew{\act_i!}\nf{t_2}$.
 By definition, $\nf{t_1}\rtmrew{(\act_1\cup\act_2)!}\nf{t_2}$.
\item[\tacase{\circlearrowright}]
Suppose $t_1\rtmrew{\act^*}t_2$.
This means that $t_1\rtmrew{\act^n}t_2$ for some $n\in\mathbb{N}$.
By the induction hypothesis for $\act^n$, we have
$\nf{t_1}\rtmrew{\act^n!}\nf{t_2}$.
By definition,
$\nf{t_1}\rtmrew{\act^*!}\nf{t_2}$.
\qedhere
\end{description}
\end{proof}

\section{Completeness of operational semantics} \label{sec:op-complete}

The completeness of rewriting and narrowing is much more complex than soundness and it holds only under certain conditions.
All the results that we develop in this section are based on confluence and coherence assumptions.

\subsection{Completeness of rewriting}
In this section, we prove the completeness of both rewriting and strict rewriting.
\begin{theorem} [Completeness rewriting and strict rewriting] \label{th:CTR}
Assume a ground confluent term rewriting system $(\Sigma,\Gamma)$ and a goal $G \subseteq \Sen(\Sigma)$.
\begin{enumerate}
\item \label{th:CTR-1} If $(\Sigma,\Gamma)$ is coherent then following are equivalent:
\begin{enumerate*}[label=(\alph*)]
\item \label{rew-comp-a} $G\glrew{*}\emptyset$, ~~
\item \label{rew-comp-b}$\Gamma\models G$, ~~ and ~~
\item \label{rew-comp-c} $\A^\Gamma\models G$.
\end{enumerate*}
\item \label{th:CTR-2} 
If $(\Sigma,\Gamma)$ is strictly coherent and modular in the sense of Definition~\ref{def:modular} such that $(\Sigma,\Gamma_\eq)$ is ground terminating then the following are equivalent:
\begin{enumerate*}[label=(\roman*)]
\item \label{strict-rew-comp-a} $G\glrew{*!}\emptyset$, ~~
\item \label{strict-rew-comp-b} $\Gamma\models G$, ~~ and ~~
\item \label{strict-rew-comp-c} $\A^\Gamma\models G$.
\end{enumerate*}
\end{enumerate}
\end{theorem}
\begin{proof}
  Let $\conv$ be the usual convergence, or joinability, relation on terms.
  That is, $t_{1} \conv t_{2}$ if and only if $t_{1} \rtmrew{*} u \ltmrew{*} t_{2}$ for some term $u$.
Just as in equational logic, it is easy to see that $\conv$ is a congruence on the term algebra $T_{\Sigma}$.
Let $\A^\Gamma$  be the transition algebra obtained from $T_\Sigma/_{\conv}$ by interpreting each atomic action $[\tau]_s$ as follows:
$$
([t_1],[t_2]) \in [\tau]_s
\quad\text{iff}\quad
t_{1} \rtmrew{\tau} u \ltmrew{*} t_{2}\ \text{for some term}\ u
$$
where, for each term $t \in T_{\Sigma}$, $[t]$ is its congruence class with respect to $\conv$.

We show that $[\tau]_s$ is well-defined.
Suppose $t_{1} \rtmrew{\tau} u \ltmrew{*} t_{2}$ for some term $u$ and let $[\tau] = [\tau']$, $[t_{1}] = [t_{1}']$, and $[t_{2}] = [t_{2}']$.
This means that there exist terms $\lambda$, $u_{1}$, and $u_{2}$ such that
$\tau \rtmrew{*} \lambda \ltmrew{*} \tau'$,
$t_{1} \rtmrew{*} u_{1} \ltmrew{*} t_{1}'$, and
$t_{2} \rtmrew{*} u_{2} \ltmrew{*} t_{2}'$ as depicted in the diagram below.
By the definition of term transitions, there also exist a rule $\Forall{Y}\kappa(\ell,\err) \lif H \in \Gamma$ and a substitution $\theta:Y\to \emptyset$ such that
$\tau \rtmrew{*} \kappa\theta$, $t_{1} \rtmrew{*} \ell\theta \rtmrew{\kappa\theta} \err\theta \rtmrew{*} u$, and $H\theta \glrew{*} \emptyset$.
The ground confluence property gives us the terms $v_{1}$ and $\varepsilon$ in the diagram below.
Then, by coherence, we know there exist a rule $\Forall{Y'}\kappa'(\ell',\err') \lif H' \in \Gamma$ and a substitution $\theta':Y'\to \emptyset$ such that
$\varepsilon \rtmrew{*} \kappa'\theta'$, $v_{1} \rtmrew{*} \ell'\theta' \rtmrew{\kappa'\theta'} \err'\theta' \ltmrew{*} \err\theta$, and $H'\theta' \glrew{*} \emptyset$.
By ground confluence (applied twice) we also obtain the terms $v_{2}$ and $u'$ together with the four rewrites depicted in the diagram below.
$$
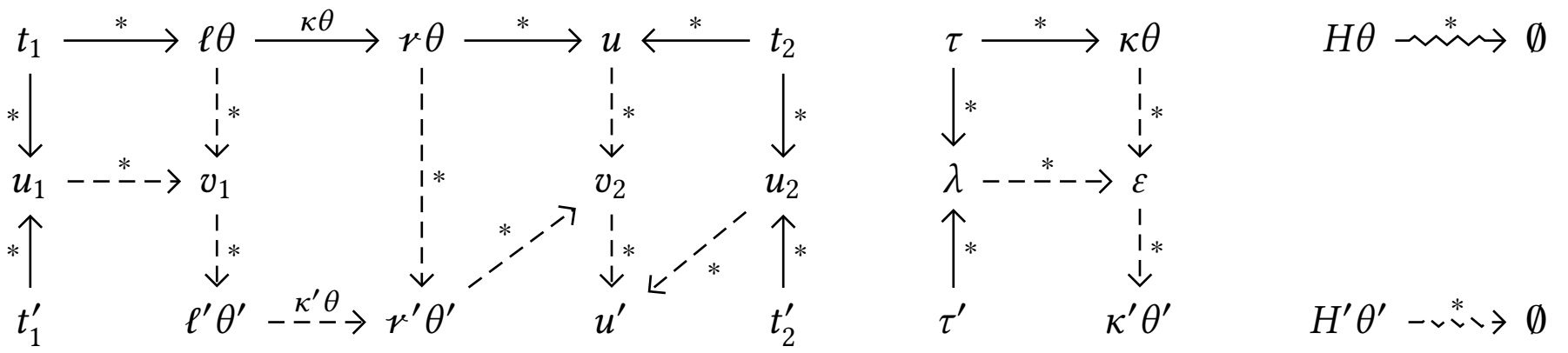
\begin{tikzcd}[>=Straight Barb]
  t_{1}
  \ar [->] {r} {*}
  \ar [->] {d} [left]{*}
  & \ell\theta
  \ar [->] {r} {\kappa\theta}
  \ar [->, dashed] {d} {*}
  & \err\theta
  \ar [->] {r} {*}
  \ar [->, dashed] {dd} {*}
  & u
  \ar [->, dashed] {d} {*}
  & t_{2}
  \ar [->] {l} [above]{*}
  \ar [->] {d} {*}
  & \tau
  \ar [->] {r} {*}
  \ar [->] {d} {*}
  & \kappa\theta
  \ar [->, dashed] {d} {*}
  & H\theta
  \ar [->, squiggly] {r} {*}
  & \emptyset
  \\
  u_{1}
  \ar [->, dashed] {r} {*}
  & v_{1}
  \ar [->, dashed] {d} {*}
  & & v_{2}
  \ar [->, dashed] {d} {*}
  & u_{2}
  \ar [->, dashed] {dl} {*}
  & \lambda
  \ar [->, dashed] {r} {*}
  & \varepsilon
  \ar [->, dashed] {d} {*}
  \\
  t_{1}'
  \ar [->] {u} {*}
  & \ell'\theta'
  \ar [->, dashed] {r} {\kappa'\theta}
  & \err'\theta'
  \ar [->, dashed] {ur} {*}
  & u'
  & t_{2}'
  \ar [->] {u} [right]{*}
  & \tau'
  \ar [->] {u} [right]{*}
  & \kappa'\theta'
  &
  H'\theta'
  \ar [->, dashed, squiggly] {r} {*}
  & \emptyset
\end{tikzcd}
$$
We have thus obtained $t_{1}' \rtmrew{*} \ell'\theta' \rtmrew{\kappa'\theta'} \err'\theta' \rtmrew{*} u' \ltmrew{*} t_{2}'$, $\tau' \rtmrew{*} \kappa'\theta'$, and $H'\theta' \glrew{*} \emptyset$.
By the definition of term transitions, it follows that $t_{1}' \rtmrew{\tau'} u' \ltmrew{*} t_{2}'$.

Next, we show two key properties of $\A^{\Gamma}$.
The first one is that, for all goals $G \subseteq \Sen(\Sigma)$,
\begin{equation} \tag{$\dagger$} \label{eq:rw}
\A^\Gamma \models G
\quad\text{implies}\quad
G \glrew{*} \emptyset.
\end{equation}
We prove it by induction on the structure of the sentences in $G$:
\begin{description}[font=\normalfont]
\item[\tacase{\eq}]
  Consider an equality $t_{1} = t_{2}$.
  By the construction of $\A^{\Gamma}$, we know that $\A^{\Gamma} \models t_{1} = t_{2}$ if and only if there exists a term $u$ such that $t_{1} \rtmrew{*} u \ltmrew{*} t_{2}$.
  By Lemma~\ref{lemma:tr+tt}(\ref{lemma:tr+tt-2}), the latter property entails $\{t_{1} = t_{2}\} \glrew{*} \emptyset$.

\item[\tacase{\tr}]
  Now consider an atomic transition $\tau(t_{1}, t_{2})$.
  By definition, we have $\A^{\Gamma} \models \tau(t_1,t_2)$ if and only if $([t_{1}],[t_{2}]) \in [\tau]_s$, which holds if and only if $t_{1} \rtmrew{\tau} u \ltmrew{*} t_{2}$ for some term $u$.
  By Lemma~\ref{lemma:tr+tt}(\ref{lemma:tr+tt-3}), the latter implies $\{\tau(t_{1}, t_{2})\} \glrew{*} \emptyset$.

\item[\tacase{\comp}]
For the composition of actions, 
since $\A^\Gamma$ is reachable,
$\A^{\Gamma} \models (\act_{1} \comp \act_{2})(t_{1}, t_{2})$ iff there exists a term $t$ such that $\A^{\Gamma} \models \act_{1}(t_{1}, t)$ and $\A^{\Gamma} \models \act_{2}(t, t_{2})$.
By the induction hypothesis, $\{\act_{1}(t_{1}, t)\} \glrew{*} \emptyset$ and $\{\act_{2}(t, t_{2})\} \glrew{*} \emptyset$.
  Therefore, by the definition of goal rewriting, we get
  $\{(\act_{1} \comp \act_{2})(t_{1}, t_{2})\} \glrew{\comp} \{\act_{1}(t_{1}, t), \act_{2}(t, t_{2})\} \glrew{*} \emptyset$.

\end{description}
The remaining cases, for unions and iterations, can be treated similarly to compositions, hence we leave out the details.
The second property of interest of $\A^{\Gamma}$ is that
\begin{equation}\tag{$*$} \label{eq:init}
\A^{\Gamma} \models \Gamma
\end{equation}
To see why that is the case, consider a clause $\Forall{Y} \varphi \lif H \in \Gamma$.
Since $\A^{\Gamma}$ is reachable, it follows that $\A^{\Gamma} \models \Forall{Y} \varphi \lif H$ if and only if
$\A^{\Gamma} \models H\theta$ implies $\A^{\Gamma} \models \varphi\theta$ for all substitutions $\theta \colon Y \to \emptyset$.
Now let $\theta \colon Y \to \emptyset$ be a substitution such that $\A^{\Gamma} \models H\theta$.
By \eqref{eq:rw}, we know that $H\theta \glrew{*} \emptyset$.
Hence we have two cases to consider depending on the nature of $\varphi$:
\begin{enumerate}
\item If $\varphi$ is an equality $\ell = \err$, then
  $\{\ell\theta = \err\theta\} \glrew{\eq} \{\err\theta = \err\theta\} \cup H\theta \glrew{=} H\theta \glrew{*} \emptyset$.
  Therefore, by Lemma~\ref{lemma:tr+tt}(\ref{lemma:tr+tt-2}), $\ell\theta \rtmrew{*} u \ltmrew{*} \err\theta$ for some term $u$.
  By the definition of $\A^{\Gamma}$, this means that $\A^{\Gamma} \models \ell\theta = \err\theta$.

\item If $\varphi$ is an atomic transition $\kappa(\ell, \err)$, then
  $\{\kappa\theta(\ell\theta, \err\theta)\} \glrew{\tr} \{\err\theta = \err\theta\} \cup H\theta \glrew{=} H\theta \glrew{*} \emptyset$.
  So, by Proposition~\ref{prop:coherence}(\ref{prop:coherence-2}), $\ell\theta \rtmrew{k\theta} u \ltmrew{*} \err\theta$ for some term $u$, which is equivalent to $\A^{\Gamma} \models \kappa\theta(\ell\theta,\err\theta)$.
\end{enumerate}
By soundness, \ref{rew-comp-a} $\implies$ \ref{rew-comp-b}.
By~\eqref{eq:init}, \ref{rew-comp-b} $\implies$ \ref{rew-comp-c}.
By~\eqref{eq:rw}, \ref{rew-comp-c} $\implies$ \ref{rew-comp-a}.
\smallskip

For the second statement, notice that:
\begin{enumerate}
\item Since $(\Sigma,\Gamma_\eq)$ is ground terminating, strict coherence implies coherence, which means 
$G\glrew{*!}\emptyset$ implies $G\glrew{*}\emptyset$.
\item By Proposition~\ref{prop:coherence}(~\ref{prop:coherence-2}), 
Proposition~\ref{prop:ordinary-strict} and Proposition~\ref{prop:str+stt}(\ref{prop:str+stt-3}),
$G\glrew{*}\emptyset$ implies $G\glrew{*!}\emptyset$.
\end{enumerate}
Using the first statement, we get \ref{strict-rew-comp-a} $\iff$ \ref{strict-rew-comp-b} $\iff$ \ref{strict-rew-comp-c}.
\end{proof}
\subsection{Completeness of narrowing}

As usual for narrowing-based procedures for solving queries (e.g.,~\cite{MiddeldorpH94}), we cannot expect to compute arbitrary solutions to queries.
Instead, we are looking only for those solutions $\theta \colon X \to X'$ that are \emph{normal}, meaning that $x\theta$ is a normal/irreducible term for all variables $x \in X$.

The final major tool used to prove the completeness of narrowing is a Lifting Lemma that allows us to retrace strict goal rewrites through narrowing.

\begin{lemma}[Lifting Lemma]\label{lemma:lifting}
  Let $(\Sigma,\Gamma)$ be a term rewriting system such that:
  \begin{enumerate}
  \item for each equational rule $\Forall{Y} (\ell=\err) \lif H \in \Gamma$, $Y = \var(\ell)$, and
  \item for each transition rule $\Forall{Y} \kappa(\ell,\err) \lif H \in \Gamma$, $Y = \var(\kappa) \cup \var(\ell)$.
  \end{enumerate}
  For every query $\Exists{X} Q$, normal substitution $\theta \colon X \to Z$, and goal $G$ over $\Sigma[Z]$ such that $Q\theta \glrew{n!} G$, there exist a narrowing derivation $\Exists{X} Q \narrow{n}_{\psi} \Exists{X'} Q'$ and a normal substitution $\delta \colon X' \to Z$ such that $Q'\delta = G$, and $\psi \comp \delta = \theta$.
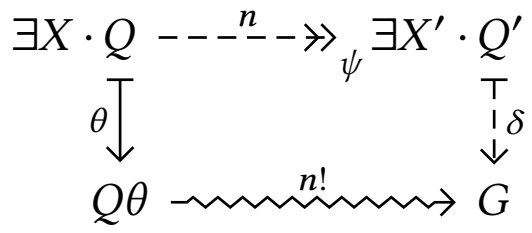
    $$
    \newdimen\qwidth
    \setbox0=\hbox{~\(\Exists{X'} {}\)}
    \qwidth=\wd0
    \begin{tikzcd}[column sep=6em, >=Straight Barb]
      {\mathllap{~\Exists{X} {}} Q}
      \ar [|->]{d} [left]{\theta}
      \ar [->>, shorten >=\qwidth, dashed]{r} [above, pos=.3]{n} [below, pos=.65]{\psi}
      & {\mathllap{\Exists{X'} {}} Q\mathrlap{'}}
      \ar [|->, dashed]{d} {\delta}
      \\
      {Q\theta}
      \ar [->, squiggly]{r} {n!}
      & {G}
    \end{tikzcd}
  $$
\end{lemma}
\begin{proof}
  We prove the statement by induction on the number of steps.
  Since the base case is trivial, we focus only on the induction step.
  We proceed by case analysis on the first step of the rewriting.
  The only interesting cases are $\eq$, `$=$', and `$\comp$'; $\tr$ is analogous to $\eq$, hence we omit it, and the remaining ones are straightforward.
\begin{description}[font=\normalfont]
\item[\tacase{\eq}]
  Let $\theta^{\gap} : X \uplus \{\gap\} \to Z \uplus \{\gap\}$ be the extension of $\theta$ that maps $\gap$ to $\gap$, and
  suppose $C\theta^{\gap}[t\theta] \glrew{\eq} C\theta^{\gap}[\err\mu] \cup H\mu \glrew{n!} G$
  for some rule $\Forall{Y} (\ell=\err) \lif H \in \Gamma$ and substitution $\mu \colon Y \to Z$ such that $t\theta = \ell\mu$.
  Since the substitution $\theta$ is normal, and because $t\theta = \ell\mu$ is rewritten to $\err\mu$, it follows that $t$ is not a variable, hence we can narrow it.

  Without loss of generality, assume that the sets of variables $X$ and $Y$ are disjoint.
  Now take $\zeta \colon X \uplus Y \to Z$ as the unique substitution that agrees with $\theta$ and $\mu$ on $X$ and $Y$, respectively.
  It follows that $\zeta$ unifies $t$ and $\ell$.
Therefore, $t$ and $\ell$ admit a most general unifier, 
say $\xi_{1} \colon X \uplus Y \to X_{1}$, 
with $X_{1} = \bigcup \{\var(z\xi_{1}) \mid z \in X \uplus Y\}$.
Hence, there exists a substitution $\delta_{1} \colon X_{1} \to T_{\Sigma}(Z)$ that $\zeta = \xi_{1} \comp \delta_{1}$.
And because $\theta$ is normal, it follows that $\delta_{1}$ is normal too.
To see why that holds, consider the following cases, for every $x_{1} \in X_{1}$:
  \begin{enumerate}
  \item If $x_{1} \in \var(x\xi_{1})$ for some variable $x \in X$, then $x_{1}\delta_{1}$ is subterm of $x\xi_{1}\delta_{1} = x\theta$.
    Since $x\theta$ is normal, it follows that $x_{1}\delta_{1}$ must be normal as well.

  \item If $x_{1} \in \var(y\xi_{1})$ for some variable $y \in Y$, then $x_{1} \in \var(\ell\xi_{1})$ because $Y = \var(\ell)$.
    Since $\ell\xi_{1} = t\xi_{1}$, we deduce that $x_{1} \in \var(t\xi_{1})$, so there exists a variable $x \in \var(t) \subseteq X$ such that $x_{1} \in \var(x\xi_{1})$.
    It follows that $x_{1}\delta_{1}$ is a subterm of $x\theta$, and thus a normal subterm.
  \end{enumerate}

  Finally, let us choose $\psi_{1} = \xi_{1} \red_{X}$.
  It is straightforward to check that $\Exists{X} C[t] \narrow{\eq}_{\psi_{1}} \Exists{X_1} (C[\err] \cup H)\xi_{1}$.
  Moreover, we have $(C[\err] \cup H)\xi_{1}\delta_{1} = (C[\err] \cup H)\zeta = C\theta^{\gap}[\err\mu] \cup H\mu$,
  and $\psi_{1} \comp \delta_{1} =  (\xi_{1} \comp \delta_{1}) \red_{X} = \zeta \red_{X} = \theta$.
\item[\tacase{=}]
  Suppose $Q\theta = \{t_{1}\theta = t_{2}\theta\} \cup O\theta \glrew{=} O\theta \glrew{n!} G$.
  It follows that $\theta$ unifies $t_{1}$ and $t_{2}$, so there is a most general unifier $\psi_{1} \colon X \to X_{1}$ of $t_{1}$ and $t_{2}$, with $X_{1} = \bigcup \{\var(x\psi_{1}) \mid x \in X\}$, and a substitution $\delta_{1} \colon X_{1} \to Z$ such that $\theta = \psi_{1} \comp \delta_{1}$.
  Therefore, $\Exists{X} \{t_{1} = t_{2}\} \cup O \narrow{=}_{\psi_{1}} \Exists{X_{1}} O\psi_{1}$.
  Moreover, for every variable $x_{1} \in X_{1}$ there exists $x \in X$ such that $x_{1}\delta_{1}$ is a subterm of $x\theta$.
  Since $x\theta$ is normal, we deduce that $x_{1}\delta_{1}$ must de normal too.
\item[\tacase{\comp}]
  Suppose there exists a normal term $t$ such that $\{(\act_1\theta\comp\act_2\theta)(t_1\theta,t_2\theta)\} \cup O\theta \glrew{\comp!} \{\act_1\theta(t_1\theta,t), \act_2\theta(t,t_2\theta)\} \cup O\theta \glrew{n!} G$.
  By the definition of narrowing, it follows that
  $\Exists{X} \{(\act_1\comp\act_2)(t_1,t_2)\} \cup O \narrow{\comp}_{\psi_{x}} \Exists{X \cup \{x\}} \{\act_1(t_1,x) ,\act_2(x,t_2)\} \cup O$ for some fresh variable $x$ and the inclusion $\psi_{x} \colon X \hookrightarrow X \cup \{x\}$.
  Let $\delta_{1} \colon X \cup \{x\} \to Z$ be the substitution that agrees with $\theta$ and $X$ and maps $x$ to $t$.
  It follows that $(\{\act_1(t_1,x) ,\act_2(x,t_2)\} \cup O)\delta_{1} = \{\act_1\theta(t_1\theta,t), \act_2\theta(t,t_2\theta)\} \cup O\theta$ and $\psi_{x} \comp \delta_{1} = \theta$.
  Moreover, since the substitution $\theta$ and the term $t$ are normal, $\delta_{1}$ is normal too.
  
\end{description}
We have thus shown that, given $Q\theta \glrew{1!} G_{1} \glrew{n!} G$, there exists a narrowing step $\Exists{X} Q \narrow{1!}_{\psi_{1}} \Exists{X_{1}} Q_{1}$ and a normal substitution $\delta_{1} \colon X_{1} \to Z$ such that $Q_{1}\delta_{1} = G_{1}$ and $\psi_{1} \comp \delta_{1} = \theta$.
By the induction hypothesis, we also have $\Exists{X_{1}} Q_{1} \narrow{n!}_{\psi_{2}} \Exists{X'} Q'$ and a normal substitution $\delta \colon X' \to Z$ such that $Q'\delta = G$ and $\psi_{2} \comp \delta = \delta_{1}$.
To conclude, let $\psi = \psi_{1} \comp \psi_{2}$.
It follows that $\Exists{X} Q \narrow{n+1!}_{\psi} \Exists{X'} Q'$ and $\psi \comp \delta = \psi_{1} \comp \delta_{1} = \theta$.
\end{proof}

The completeness of narrowing holds under much stricter conditions than for rewriting, as it requires confluence, termination, coherence, and also syntactical constraints on the use of variables within rules.

\begin{theorem} \label{th:CN}
  Let $(\Sigma,\Gamma)$ be a term rewriting system that is confluent, strictly coherent and modular in the sense of Definition~\ref{def:modular} such that $\Gamma=\Gamma_\eq\uplus\Gamma_\tr$ and $(\Sigma,\Gamma_\eq)$ is terminating.
Additionally, we assume that 

  \begin{enumerate}
  \item for each equational rule $\Forall{Y} (\ell=\err) \lif H \in \Gamma$, $Y = \var(\ell)$,
  \item for each transition rule $\Forall{Y} \kappa(\ell,\err) \lif H \in \Gamma$, $Y = \var(\kappa) \cup \var(\ell)$, and
  \item all equation rules in $\Gamma_\eq$ are free of composition $(\comp)$ and iteration $(*)$.
  \end{enumerate}
  For every normal solution $\theta \colon X \to Z$ to a query $\Exists{X} Q$ there exists a narrowing derivation $\Exists{X} Q \narrow{*}_{\psi} \Exists{X'} \emptyset$ and a substitution $\delta \colon X' \to Z$ such that $\theta = \psi \comp \delta$.
\end{theorem}
\begin{proof}
  By definition, $\theta:X\to Z$ is a solution to $\Exists{X} Q$ if and only if $\Gamma \models_{\Sigma[Z]} Q\theta$.
  Since $(\Sigma,\Gamma)$ is confluent, $(\Sigma[Z],\Gamma)$ is ground confluent and $(\Sigma[Z],\Gamma_\eq)$ is ground terminating.
  By the completeness of strict rewriting, it follows that $Q\theta \glrew{*!} \emptyset$.
  The latter gets lifted, by Lemma~\ref{lemma:lifting}, to a derivation $\Exists{X} Q \narrow{*}_{\psi} \Exists{X'} Q'$ and a substitution $\delta \colon X' \to Z$ such that $Q'\delta = \emptyset$ (entailing $Q' = \emptyset$) and $\theta = \psi \comp \delta$.
\end{proof}

\begin{corollary}[Completeness of narrowing]
Under the conditions of Theorem~\ref{th:CN}, the following are equivalent:
\begin{center}
\begin{enumerate*}[label=(\alph*)]
\item \label{cor:narrow-a} $\Exists{X} Q \narrow{*}_{\psi} \Exists{X'} \emptyset$, \quad
\item \label{cor:narrow-b} $\Gamma\models \Exists{X}Q$, \quad and \quad
\item \label{cor:narrow-c}$\A^\Gamma\models\Exists{X} Q$. 
\end{enumerate*}
\end{center}
\end{corollary}
\begin{proof} 
By Theorem~\ref{th:narrow-sound}, we get
\ref{cor:narrow-a}$\implies$\ref{cor:narrow-b}.
Since $\A^\Gamma\models\Gamma$, we obtain
 \ref{cor:narrow-b}$\implies$\ref{cor:narrow-c}.
For \ref{cor:narrow-c}$\implies$\ref{cor:narrow-a},
assume that $\Gamma\models \Exists{X}Q$.
Let  $\A^\Gamma$ be a reachable initial model of $\Gamma$ as in Theorem~\ref{th:init}.
It follows that $\A^\Gamma\models \Exists{X}Q$. 
By Proposition~\ref{prop:reach}, $\A^\Gamma\models Q\theta$ for some substitution $\theta:X\to\emptyset$. 
Since $(\Sigma,\Gamma_\eq)$ is canonical, by soundness of term rewriting, we can assume without loss of generality that $\theta$ is normal.
By Theorem~\ref{th:init} (\ref{th:init-2}), $\Gamma\models_\Sigma Q\theta$, hence $\theta$ is a normal solution to $\Exists{X}Q$.
By Theorem~\ref{th:CN}, there exists a narrowing derivation $\Exists{X} Q \narrow{*}_{\psi} \Exists{X'} \emptyset$.
\end{proof}



\section{Extending completeness} \label{sec:birkhoff}
%
\begin{table}\centering
\begin{tabular}{|l l|}
\hline
&\\
$(\mathit{Comp_I})~\Frac{\Gamma\cup\{ \act_1(t_1,x), \act_2(x, t_2)  \}\vdash_{\Sigma[x]} \gamma }{\Gamma\cup\{ (\act_1\comp\act_2)(t_1, t_2) \}\vdash \gamma }$ & 
$(\mathit{Union_I})~\Frac{\Gamma\cup \{ \act_i(t_1, t_2) \} \vdash \gamma \text{ for all } i\in\{1,2\} }{\Gamma\cup \{ (\act_1\cup\act_2)(t_1, t_2) \} \vdash \gamma}$ \\
& \\
$(\mathit{Star_I})~\Frac{\Gamma\cup \{ \act^n(t_1, t_2) \} \vdash \gamma \text{ for all } n\in\mathbb{N}}{\Gamma\cup \{ \act^*(t_1, t_2) \} \vdash \gamma }$ & 
$(\mathit{Star_0})~\Frac{ \Gamma\cup \{ t_2 = t_1 \} \vdash \gamma }{ \Gamma\cup \{ \act^0(t_1,t_2) \} \vdash \gamma}$ \Space
$(\mathit{Imp})~\Frac{\Gamma\cup \{\varphi_2\} \vdash \varphi_1}{\Gamma\vdash \varphi_1 \lif \varphi_2 }$ \\
& \\
$(\mathit{EQ})~\Frac{\Gamma\cup Q \vdash_{\Sigma[X]} \gamma}{\Gamma\cup \{\Exists{X} Q\}\vdash_{\Sigma} \gamma }$ & 
$(\mathit{UQ})~\Frac{\Gamma\vdash_{\Sigma[X]} \gamma}{\Gamma\vdash_{\Sigma} \Forall{X}\gamma }$ \\
& \\
\hline
\end{tabular}
\caption{Actions, implication and quantification}
\label{tab:birkhoff}
\end{table}
A program can be formalized as a set of clauses $\Gamma$ over a signature $\Sigma$.
However, the properties we aim to express about programs often extend beyond queries.
To capture such properties, we introduce the notion of \emph{extended clause}, defined as a sentence of the form $\Forall{X}(\varphi_1 \lif \varphi_2)$, where $\varphi_1$ and $\varphi_2$ are queries.
Evidently, every clause is a special instance of an extended clause in which no existential quantifiers occur.
In this section, we are interested in proving $\Gamma\models \gamma$, where $\Gamma$ is a set of clauses describing a program, while $\gamma$ is an extended clause formalizing a property of that program. 
See Table~\ref{tab:birkhoff} for the proof rules used to reason about extended clauses starting from a given set of (ordinary) clauses.
\begin{lemma} [Soundness]
  The proof rules defined in Tables~\ref{tab:basic},~\ref{tab:init}, and~\ref{tab:birkhoff} are sound.
  That is, the least entailment system closed under the proof rules defined in Tables~\ref{tab:basic},~\ref{tab:init}, and~\ref{tab:birkhoff} is sound.
\end{lemma}
Since the premises of the clauses may include arbitrary transitions, we must go beyond the realm of clauses and introduce proof rules that handle both clauses and transitions.
For instance, to deduce a clause of the form $(\varphi_1 \lif \varphi_2)$ using rule $(\mathit{Imp})$, we must add $\varphi_2$ as a hypothesis to the clause set $\Gamma$. However, since $\varphi_2$ is not itself a set of clauses, this requires reasoning in a hybrid setting that incorporates both clauses and queries.
Consequently, the completeness established in this section necessitates a more structured methodology than the ones introduced in \cite{CodescuG08,gai-cbl}.

Moreover, if $T_\Sigma$ is finite for some sort $s$, then, since the initial model $\A^\Gamma$ is reachable, the set $\A^\Gamma_s$ is finite as well.
Based on this observation, if we work over $\FTA$ signatures $(S\supseteq S^f,F)$ such that $T_\Sigma(X)_s$ is finite for all sorts $s\in S^f$ and all finite blocks of variables $X$, then we can extend the completeness results from $\TA$ to $\FTA$.

\begin{theorem}[Completeness] \label{th:birkhoff}
Let $\vdash$ be any sound entailment system closed under the proof rules defined in Tables~\ref{tab:basic},~\ref{tab:init}, and~\ref{tab:birkhoff}.
Assume a set of clauses $\Gamma$ and an extended clause $\gamma$ defined over the same signature $(S\supseteq S^f,F)$ such that
$T_\Sigma(X)_s$ is finite  for all finite blocks of variables $X$ and all sorts $s\in S^f$.
Then the following are equivalent:
\begin{center}
\begin{enumerate*}[label=(\alph*)]
\item \label{th:birkhoff-a} $\Gamma\models \gamma$,
\item \label{th:birkhoff-b} $\Gamma\models^f \gamma$, and 
\item \label{th:birkhoff-c} $\Gamma\vdash\gamma$.
\end{enumerate*}
\end{center}
\end{theorem}
\begin{proof}
First, we prove the following property for 
all sets of ground equations and transitions $H$ and 
all queries $\varphi$, the following are equivalent:
\begin{center}
\begin{enumerate*}[label=(\alph*)]
\item $\Gamma\cup H\models \varphi$,
\item $\Gamma\cup H\models^f \varphi$, and 
\item $\Gamma\cup H\vdash\varphi$.
\end{enumerate*}
\end{center}
We proceed by well-founded induction on $(m,n)$, 
where $m$ is the total number of occurrences of iteration operator in $H$, and 
$n$ is the total number of occurrences of the composition or union operators in $H$.
The ordering we consider on these pairs is the usual lexicographic ordering.

For the \emph{base case}, we assume that $H$ consists of atomic sentences.
That is, $m=0$ and $n=0$.
This case is covered by Theorem~\ref{th:init}(\ref{th:init-3}) and the fact that $\A^\Gamma_s$ is finite for all $s\in S^f$.

For the \emph{induction step}, there are three cases to consider:
\begin{description}[leftmargin=4.5em, itemsep=.5\baselineskip]
\item [{\case[3em]{\act_1\cup \act_2}}] 
Assume that $\Gamma\cup H\cup\{(\act_1\cup \act_2)(t_1, t_2) \}\models \varphi$.
Let $m$ be the total number of occurrences of the iteration operator in $H\cup\{(\act_1\cup \act_2)(t_1, t_2) \}$, and 
$n$ be the total number of occurrences of the composition or union operators in $H\cup\{(\act_1\cup \act_2)(t_1, t_2) \}$.
\begin{description}
\item[$(a)\iff(b)$] $\Gamma\cup H\cup\{(\act_1\cup \act_2)(t_1, t_2) \}\models \varphi \xLeftrightarrow{\mathit{Semantics}}
\Gamma\cup H\cup\{\act_i(t_1, t_2) \}\models \varphi \text{ for all } i\in\{1,2\} \xLeftrightarrow{(m,n-1)}$
$\Gamma\cup H\cup\{\act_i(t_1, t_2) \}\models^f \varphi \text{ for all } i\in\{1,2\} \xLeftrightarrow{\mathit{Semantics}} 
\Gamma\cup H\cup\{(\act_1\cup \act_2)(t_1, t_2) \}\models^f \varphi$.
Notice that $\xLeftrightarrow{(m,n-1)}$ indicates that the equivalence follows from applying the induction hypothesis for $(m,n-1)$.
\item [$(b)\implies(c)$] 
$\Gamma\cup H\cup\{(\act_1\cup \act_2)(t_1, t_2) \}\models \varphi \xLeftrightarrow{\mathit{Semantics}} 
\Gamma\cup H\cup\{\act_i(t_1, t_2) \}\models\varphi \text{ for all } i\in\{1,2\} 
 \xLeftrightarrow{(m,n-1)}$  
$\Gamma\cup H\cup\{(\act_1\cup \act_2)(t_1, t_2) \}\vdash \varphi \xRightarrow{(\mathit{Union_I})} 
\Gamma\cup H\cup\{(\act_1\cup \act_2)(t_1, t_2) \}\vdash \varphi$.
\item [$(c)\implies(a)$] By soundness.
\end{description}
\item [{\case[3em]{\act_1\comp \act_2}}] 
Assume that  $\Gamma\cup H\cup\{(\act_1\comp \act_2)(t_1, t_2) \}\models_\Sigma \gamma$. 
Let $m$ be the total number of occurrences of the iteration operator in $H\cup\{(\act_1\comp \act_2)(t_1, t_2) \}$, and 
$n$ be the total number of occurrences of the composition or union operators in $H\cup\{(\act_1\comp \act_2)(t_1, t_2) \}$.
\begin{description}
\item[$(a)\iff(b)$]
$\Gamma\cup H\cup\{(\act_1\comp \act_2)(t_1, t_2) \}\models_\Sigma \varphi \xLeftrightarrow{\mathit{Semantics}}
\Gamma\cup H\cup\{\act_1(t_1,z),\act_2(z,t_2) \}\models_{\Sigma[z]} \varphi \xLeftrightarrow{(m,n-1)}$
$\Gamma\cup H\cup\{\act_1(t_1,z),\act_2(z,t_2) \}\models_{\Sigma[z]}^f \varphi \xLeftrightarrow{\mathit{Semantics}} 
\Gamma\cup H\cup\{(\act_1\comp \act_2)(t_1, t_2) \}\models_\Sigma^f \varphi $.  
\item[$(b)\implies(c)$]  
$\Gamma\cup H\cup\{(\act_1\comp \act_2)(t_1, t_2) \}\models_\Sigma \varphi \xLeftrightarrow{\mathit{Semantics}}
\Gamma\cup H\cup\{\act_1(t_1,z),\act_2(z,t_2) \}\models_{\Sigma[z]} \varphi \xLeftrightarrow{(m,n-1)}$
$\Gamma\cup H\cup\{\act_1(t_1,z),\act_2(z,t_2) \}\vdash_{\Sigma[z]} \varphi \xRightarrow{(Comp_I)} 
\Gamma\cup H\cup\{(\act_1\comp \act_2)(t_1, t_2) \}\vdash_\Sigma \varphi $.
\item[$(c)\implies(a)$] By soundness.
\end{description}
\item [{\case[3em]{\act^*}}] Assume that  $\Gamma\cup H\cup \{\act^*(t_1, t_2) \} \models \varphi$.
Let $m$ be the total number of occurrences of the iteration operator in $H\cup \{\act^*(t_1, t_2) \}$, and 
$n$ be the total number of occurrences of the composition or union operators in $H\cup \{\act^*(t_1, t_2) \}$.
\begin{description}
\item[$(a)\iff(b)$]
$\Gamma\cup H\cup \{\act^*(t_1, t_2) \} \models \varphi \xLeftrightarrow{\mathit{Semantics}}
\Gamma\cup H\cup \{\act^n(t_1, t_2) \} \models \varphi \text{ for all } n\in \mathbb{N} \xLeftrightarrow{(m-1,n)}$
 $\Gamma\cup H\cup \{\act^n(t_1, t_2) \} \models^f\varphi\text{ for all }n\in\mathbb{N} \xLeftrightarrow{\mathit{Semantics}} \Gamma\cup H\cup \{\act^*(t_1, t_2) \} \models^f \varphi $.
\item[$(b)\implies(c)$] 
$\Gamma\cup H\cup \{\act^*(t_1, t_2) \} \models \varphi \xLeftrightarrow{\mathit{Semantics}}
\Gamma\cup H\cup \{\act^n(t_1, t_2) \} \models \varphi \text{ for all } n\in \mathbb{N} \xLeftrightarrow{(m-1,n)}$ 
$\Gamma\cup H\cup \{\act^n(t_1, t_2) \} \vdash \varphi \text{ for all } n\in \mathbb{N} \xRightarrow{(\mathit{Star_I})}
\Gamma\cup H\cup \{\act^*(t_1, t_2) \} \vdash \varphi$.
\item[$(c)\implies(a)$] By soundness.
\end{description}
\end{description}
Secondly, from first part of the proof, by $\mathit{(EQ)}$, 
for all queries $\varphi_1$ and $\varphi_2$, we have: 
\begin{center}
\begin{enumerate*}[label=(\alph*)]
\item $\Gamma \cup \{\varphi_1\} \models \varphi_2$,
\item $\Gamma \cup \{\varphi_1\} \models^f \varphi_2$ and 
\item $\Gamma \cup \{\varphi_1\} \vdash \varphi_2$.
\end{enumerate*}
\end{center}
Thirdly, we prove the statement of the theorem by induction on the structure of clauses.

\begin{description}[leftmargin=4.5em, itemsep=.75\baselineskip, font=\normalfont]
\item[{\case[3em]{\text{base}}}] By Theorem~\ref{th:init} (\ref{th:init-3}).
\item [{\case[3em]{\varphi_1 \lif \varphi_2}}]
In this case, $\varphi_1$ and $\varphi_2$ are queries.
\begin{description}[itemsep=.5em]
\item[$(a)\iff(b)$]
$\Gamma\models \varphi_1 \lif\varphi_2 \xLeftrightarrow{\mathit{Semantics}}   \Gamma \cup \{\varphi_2\}\models \varphi_1 \xLeftrightarrow{IH}
\Gamma \cup \{\varphi_2\}\models^f \varphi_1 \xLeftrightarrow{\mathit{Semantics}}  \Gamma\models^f \varphi_1 \lif\varphi_2$.
\item[$(b)\implies(c)$] 
$\Gamma\models \varphi_1 \lif\varphi_2 \xLeftrightarrow{\mathit{Semantics}}  
\Gamma \cup \{\varphi_2\}\models \varphi_1 \xLeftrightarrow{IH} 
\Gamma \cup \{\varphi_2\}\vdash \varphi_1 \xRightarrow{(Imp)} 
\Gamma \vdash \varphi_1 \lif \varphi_2$.
\item[$(c)\implies(a)$] By soundness.
\end{description}
\item [{\case[3em]{\Forall{X}\gamma}}]
Let  $\Forall{X}\gamma$ be an extended clause.

\begin{description}[itemsep=.5em]
\item[$(a)\iff(b)$]
$\Gamma\models_\Sigma \Forall{X}\gamma \xLeftrightarrow{\mathit{Semantics}} 
\Gamma \models_{\Sigma[X]} \gamma  \xLeftrightarrow{IH} 
\Gamma \models_{\Sigma[X]}^f \gamma  \xLeftrightarrow{\mathit{Semantics}}
 \Gamma\models_\Sigma^f \Forall{X}\gamma$.
\item[$(b)\implies(c)$] 
$\Gamma\models_\Sigma \Forall{X}\gamma \xLeftrightarrow{\mathit{Semantics}} 
\Gamma \models_{\Sigma[X]} \gamma  \xLeftrightarrow{IH} 
\Gamma \vdash_{\Sigma[X]} \gamma \xRightarrow{(UQ_I)}
\Gamma\vdash_\Sigma \Forall{X}\gamma$.
\item[$(c)\implies(a)$] By soundness.
\qedhere
\end{description}
\end{description}
\end{proof}
Observe that we can take $S^f=\emptyset$ to obtain the completeness of $\TA$.
Moreover, the conditions of Theorem~\ref{th:birkhoff} concerning the signature can be replaced with a more general assumption:
for all finite blocks of variables $X$, the set of clauses $\Gamma$ has an initial model $\A^{\Gamma_X}$ over $(S,F[X])$ such that $\A^{\Gamma_X}_s$ is finite for all $s\in S^f$.
Take, for example, the signature of natural numbers $\Sigma_\texttt{NAT}=(\mathtt{\{Nat\},\{0:\to Nat, s:Nat\to Nat\}})$  and the equation $\gamma=\{\Forall{x} \mathtt{s}^n(x)= x\}$, where $n\in \mathbb{N}\setminus\{0\}$. 
Then for all finite blocks of variables $X$, the equation $\gamma$ has an initial model over $\Sigma_\texttt{NAT}[X]$ which is finite.

In practical case studies, when proving properties formalized as extended clauses free of existential quantification, the proof rules from Tables~\ref{tab:basic} and \ref{tab:init} are replaced with a single rewriting rule:
$$(\mathit{Rewrite})\Frac{G \glrew{*} \emptyset}{\Gamma\vdash G}$$
The completeness of extended clauses is preserved, provided that the underlying set of clauses forms a term rewriting system that satisfies the conditions of Theorem~\ref{th:CTR}(~\ref{th:CTR-1}).
Similarly, when proving properties formalized as extended clauses, the proof rules from Tables~\ref{tab:basic} and \ref{tab:init} are replaced by a single narrowing rule:
$$(\mathit{Narrow})\Frac{\Exists{X} Q \narrow{*}_{\psi} \Exists{X'} \emptyset}{\Gamma\vdash \Exists{X}Q}$$
The completeness stated in Theorem~\ref{th:birkhoff} is preserved, provided that the underlying set of clauses forms a term rewriting system satisfying the conditions of Theorem~\ref{th:CN}.


\section{Reachability, induction, and finite models} \label{sec:rifm}
%
By defining an operational semantics for sets of clauses, we have taken significant steps toward establishing a methodology for verifying systems modeled with $\TA$ clauses. However, in practical case studies, the properties of interest often go far beyond individual clauses, and developing modeling techniques and formal proof methods with a high degree of automation remains challenging.
The foundations laid in this paper enable the construction of several formal methods on top of them. 
In this section, we highlight the expressive power of $\TA$ by demonstrating its ability to model reachability and finiteness, and to prove properties of systems that are reachable or finite. 
In particular, this approach directly addresses the issue that $\mathit{Star_I}$ from Table~\ref{tab:birkhoff} contains an infinite number of proof rules.
A proof rule is useful in practice only if its premise can be checked mechanically by a software tool. Clearly, an infinite number of premises can be handled only through inductive reasoning.

We return to the specification of natural numbers from Example~\ref{ex:nat}, but we modify it, first by labelling the two axioms that define the successor relation, and then by dropping the conditions used in the two axioms defining the addition of natural numbers. In the specification below, those axioms of the addition operation, labelled \texttt{add-base} and \texttt{add-ind}, apply to all elements of sort \texttt{Nat}, not only to those that are reachable along the successor relation.
{\small \begin{verbatim}
spec NAT-ADD is
  protecting NAT .
  protecting SUCC .
  op _+_ : Nat Nat -> Nat .
  vars x, y : Nat .
  ax x =(succ)> s x [label: succ-a] .
  ax y = s x if x =(succ)> y [label: succ-b] .
  ax x + 0 = x [label: add-base] .
  ax x + s y = s (x + y) [label: add-ind] .
endspec
\end{verbatim}}
Therefore, the set of clauses $\Gamma_\texttt{NAT-ADD}$ is  $\{\texttt{succ-a,succ-b, add-base, add-ind}\}$.
It is not difficult to check that \texttt{NAT-ADD} does not entail the left-identity property of 
$0$; that is, $\Gamma_\texttt{NAT-ADD}\not\models \Forall{x} 0+x=x$.
\begin{itemize}
\item [] Indeed, consider the model of natural numbers $\mathfrak{N}$, which interprets all symbols in the usual way:
\begin{enumerate*}[label=(\alph*)]
\item $\mathfrak{N}_\texttt{Nat}$ is the set of natural numbers $\mathbb{N}$, 
\item $\mathfrak{N}_\texttt{Label}=\{\texttt{succ}\}$,
\item $0^\mathfrak{N}=0$, 
\item $\texttt{s}^\mathfrak{N}:\mathbb{N}\to\mathbb{N}$ is the successor function, 
\item $+^\mathfrak{N}:\mathbb{N}\times\mathbb{N}\to\mathbb{N}$ is the addition, 
\item $\mathtt{succ_{Nat}} \subseteq \mathbb{N}\times \mathbb{N}$ is the successor relation, and 
\item $\mathtt{succ_{Label}}$ is the empty set.
\end{enumerate*}
Obviously, $\mathfrak{N}\models \Gamma_\texttt{NAT-ADD}$.
We construct another model $\mathfrak{M}$ from $\mathfrak{N}$ by adding a new element $0'$ such that 
\begin{enumerate*}[label=(\alph*)]
\item $\mathtt{s}^\mathfrak{M} 0'=1$, 
\item $0' +^\mathfrak{M} n =  n +^\mathfrak{M} 0'= n$ for all $n\in \mathbb{N}\setminus\{0\}$,
\item $0'+^\mathfrak{M}0=0'$,
\item $0+^\mathfrak{M}0'=0$, and
\item $\texttt{succ}(0',1)$.
\end{enumerate*}
We have that $\mathfrak{M}\models\Gamma_\texttt{NAT-ADD}$ but $\mathfrak{M}\not\models\Forall{x}0 + x = x$, which means $\Gamma_\texttt{NAT-ADD}\not\models\Forall{x}0 + x = x$.
\end{itemize}
However, 
$\Gamma_\texttt{NAT-ADD}$ satisfies the left identity property restricted to the elements reachable from $0$ via the successor relation, that is,
$\Gamma_\texttt{NAT-ADD}\models \Forall{x} 0 + x = x \lif \mathtt{succ}^*(0,x)$, which is equivalent with
$\Gamma_\texttt{NAT-ADD}\models \Forall{x} 0 + x = x \lif \mathtt{succ}^n(0,x)$ for all $n\in \mathbb{N}$.
We apply induction.
\begin{description}[leftmargin=3.5em, font=\normalfont]
\item[{\case[2em]{base}}]
$\Gamma_\texttt{NAT-ADD}\vdash \Forall{x}0+  x=x \lif \mathtt{succ}^0(0,x)$.

{\small\begin{tabular}{l l l}
1 & $\Gamma_\texttt{NAT-ADD}\vdash_{\Sigma_\texttt{NAT-ADD}} \Forall{x} 0+x= x \lif \mathtt{succ}^0(0,x)$ & 
by $(\mathit{UQ})$ and $(\mathit{Imp})$\\
2 &  $\Gamma_\texttt{NAT-ADD}\cup\{\mathtt{succ}^0(0,x) \} \vdash_{\Sigma_\texttt{NAT-ADD}[x]} 0 + x=x $ & by $(\mathit{Star_0})$ \\
3 &  $\Gamma_\texttt{NAT-ADD}\cup\{ x = 0\} \vdash_{\Sigma_\texttt{NAT-ADD}[x]} 0+ x=x$ & 
by $(\mathit{Rewrite})$ \\
4 &  $\{0+x=x \}\glrew{*}\emptyset$ over $(\Sigma_\texttt{NAT-ADD}[x],\Gamma_\texttt{NAT-ADD}\cup\{ x = 0\} )$ & automatically checked 
\end{tabular}}
\item[{\case[2em]{\mathit{step}}}]
Assuming 
$\Gamma_\texttt{NAT-ADD}\vdash \Forall{x} 0+x=x \lif \mathtt{succ}^n(0,x)$, 
we prove 
$\Gamma_\texttt{NAT-ADD}\vdash \Forall{x,y} 0 +x=x \lif \mathtt{succ}^{n+1}(0,x)$,
which is equivalent to
$\Gamma_\texttt{NAT-ADD}\cup \mathit{IH} \vdash \Forall{x} 0+x=x \lif \mathtt{succ}^{n+1}(0,x)$, 
where $\mathit{IH} \coloneqq \Forall{x} 0+x=x \lif \mathtt{succ}^n(0,x)$.

{\small\begin{tabular}{l l l}
1 & $\Gamma_\texttt{NAT-ADD}\cup \mathit{IH} \vdash \Forall{x} 0+x=x \lif \mathtt{succ}^{n+1}(0,x)$ & 
by $(\mathit{UQ})$ and $(\mathit{Imp})$ \\
2 & $\Gamma_\texttt{NAT-ADD}\cup \mathit{IH}\cup\{\mathtt{succ}^{n+1}(0,x)\}\vdash_{\Sigma_\texttt{NAT-ADD}[x]}  0+x=x$ & 
by $(\mathit{Comp_I})$\\
3 & $\Gamma_\texttt{NAT-ADD}\cup \mathit{IH}\cup\{\mathtt{succ}^n (0,z), \mathtt{succ}(z,x)\} \vdash_{\Sigma_\texttt{NAT-ADD}[x,z]}  0+x=x$ &
by $\texttt{succ-b}\{\mathtt{N}\leftarrow x, \mathtt{M}\leftarrow z\}$ \\
4 & $\Gamma_\texttt{NAT-ADD}\cup \mathit{IH}\cup\{ \mathtt{succ}^n(0,z), \mathtt{succ}(z,x), x=\mathtt{s}~z\} \vdash_{\Sigma_\texttt{NAT-ADD}[x,z]}  0+x=x$ &
by $(\mathit{Rewrite})$ \\
5 & $\{0+x=x\}\glrew{*}\emptyset$ over $(\Sigma_\texttt{NAT-ADD}[x,z], \Gamma_\texttt{NAT-ADD}\cup \mathit{IH}\cup\{\mathtt{succ}^n(0,z), \mathtt{succ}(z,x), x=\mathtt{s}~z\})$ & automatically checked 
\end{tabular}}
\end{description}



The verification methodology advanced in this paper
focuses on simplifying the formula to prove by eliminating the sentence operators used in its construction.
The first target of this process is the iteration operator $*$ that appears in the conditions of the extended clauses. To handle it, we apply induction schemes, as illustrated in the case study above.
However, the use of induction schemes---and the corresponding addition of induction hypotheses to the clause set---comes at a cost. 
We must rely on lemma discovery, that is, on the $(\mathit{Cut})$ rule, which we were able to avoid during the completeness proof.
For example, in order to prove
$\Gamma_\texttt{NAT-ADD}\models \Forall{x,y}x+y=y+x\lif \mathtt{succ}^*(0,x)\land\mathtt{succ}^*(0,y)$ 
we must first establish the following lemmas:
$\Gamma_\texttt{NAT-ADD}\models \Forall{x}0+x=x\lif \mathtt{succ}^*(0,x)$ and
$\Gamma_\texttt{NAT-ADD}\models \Forall{x,y}\mathtt{s}~x+y= \mathtt{s}(x+y)\lif \mathtt{succ}^*(0,x)\land\mathtt{succ}^*(0,y)$.

Another potential source of problems is that we may start from a specification that is terminating and confluent, but during the verification process the underlying specification undergoes transformations that may lead to the loss of these desirable properties, depending on how the property to be proved is formalized.
In our example, the initial specification is $(\Sigma_\texttt{NAT-ADD},\Gamma_\texttt{NAT-ADD})$, whereas the specifications at the leaves of the proof tree are $(\Sigma_\texttt{NAT-ADD}[x],\Gamma_\texttt{NAT-ADD}\cup\{x = 0\})$ and $(\Sigma_\texttt{NAT-ADD}[x,z], \Gamma_\texttt{NAT-ADD}\cup \mathit{IH}\cup\{\mathtt{succ}^n(0,z), \mathtt{succ}(z,x), x=\mathtt{s}~z\})$.

Suppose we are interested in properties that hold only for finite reachable models such as $\mathbb{Z}_n$, where $n>0$.
Notice that $\mathbb{Z}_n\models \Forall{y}\Exists{x}x+y=0$ while  $\mathbb{N}\not\models\Forall{y}\Exists{x}x+y=0$.
In this case, we can add $\mathtt{succ}^+(0,0)$ to $\Gamma_\texttt{NAT-ADD}$ and prove 
\begin{equation}\label{eq:zn}
\Gamma_\texttt{NAT-ADD}\cup\{\mathtt{succ}^+(0,0)\}\models \Forall{y}\Exists{x}x+y=0 \text{ if } \mathtt{succ}^*(0,y).
\end{equation} 
The action $\mathtt{succ}^+(0,0)$ restricts the models to algebras which have a finite number of reachable elements, while
the condition $\mathtt{succ}^*(0,y)$ asserts that $y$ is reachable from $0$ by applying the successor function a finite number of times.
\footnote{$\mathtt{succ}^+$ denotes the transitive closure of $\mathtt{succ}$.}
\begin{description}
\item[(goal)] The goal (\ref{eq:zn}) is equivalent with 
$\Gamma_\texttt{NAT-ADD}\cup\{\mathtt{succ}^m(0,0)\}\models \Forall{y}\Exists{x}x+y=0 \text{ if } \mathtt{succ}^*(0,y)$ for all $m>0$.
\item[(lemma1)] 
$\Gamma_\texttt{NAT-ADD}\cup\{\mathtt{succ}^m(0,0)\}\vdash s^m 0=0$, 
since $\Gamma_\texttt{NAT-ADD} \vdash \Forall{x,y} y = s^n x  \text{ if } \mathtt{succ}^n(x,y)$ for all $n\in\mathbb{N}$.
\item[(lemma2)] $\Gamma_\texttt{NAT-ADD}\cup\{s^m 0 = 0\} \vdash  s^n 0 = s^{(n~\mathtt{mod}~m)} 0$ for all $m>0$ and all $n\geqslant 0$.

\item[(lemma3)] $\Gamma_\texttt{NAT-ADD}\cup\{s^m 0 = 0\} \vdash  s^{(n~\mathtt{mod}~m)} s^{m-(n~\mathtt{mod}~m)} 0=0$ for all $m>0$ and all $n\geqslant 0$.

\item[(induction)] 
Using the above lemmas, we prove by induction on $n\in\mathbb{N}$ that for all $m>0$,
\begin{equation}
\Gamma_\texttt{NAT-ADD}\cup\{\mathtt{succ}^m(0,0)\} \vdash \Forall{y}\Exists{x}x+y=0 \text{ if } \mathtt{succ}^n(0,y).
\end{equation}

For the base case, one can easily prove $\Gamma_\texttt{NAT-ADD}\cup\{\mathtt{succ}^m(0,0)\} \vdash \Forall{y}\Exists{x}x+y=0 \text{ if } \mathtt{succ}^0(0,y)$.

For the induction step, we show 
$\Gamma_\texttt{NAT-ADD}\cup\{\mathtt{succ}^m(0,0),IH\} \vdash \Forall{y}\Exists{x}x+y=0 \text{ if } \mathtt{succ}^{n+1}(0,y)$, 
where $IH \coloneqq \Forall{y}\Exists{x} x+y =0 \text{ if } \mathtt{succ}^n(0,y)$.
This is equivalent to 
\begin{equation}
\Gamma_\texttt{NAT-ADD}\cup\{\mathtt{succ}^m(0,0), IH\} \cup lemmas  \vdash \Forall{x}\Exists{y}x+y=0 \text{ if } \mathtt{succ}^{n+1}(0,y)
\end{equation}
where $lemmas\coloneqq \{(s^m 0=0),(s^n 0 = s^{(n~\mathtt{mod}~m)} 0), s^{((n+1)~\mathtt{mod}~m)} s^{m-((n+1)~\mathtt{mod}~m)}0=0 \}$.

\noindent
{\small\begin{tabular}{ l l l}
1 & $\Gamma_\texttt{NAT-ADD}\cup\{\mathtt{succ}^m(0,0), IH\}\cup lemmas  \vdash \Forall{y}\Exists{x}x+y=0 \text{ if } \mathtt{succ}^{n+1}(0,y)$ & \\
2 & $\Gamma_\texttt{NAT-ADD}\cup\{\mathtt{succ}^m(0,0), IH, \mathtt{succ}^{n+1}(0,y)\} \cup lemmas \vdash_{\Sigma_\texttt{NAT-ADD}[y]} \Exists{x}x+y=0$ & 
by $(\mathit{UQ})$ and $(\mathit{Imp})$\\
3 & $\Gamma_\texttt{NAT-ADD}\cup\{\mathtt{succ}^m(0,0), IH, \mathtt{succ}^n(0,z),\mathtt{succ}(z,y)\}\cup lemmas  \vdash_{\Sigma_\texttt{NAT-ADD}[y,z]} \Exists{x}x+y=0$ & 
by $(\mathit{Comp}_I)$\\
4 & $\Gamma_\texttt{NAT-ADD}\cup \dots \cup lemmas \vdash_{\Sigma_\texttt{NAT-ADD}[y,z]} \Exists{x}x + s^{n+1}0 =0$ & 
by $(\mathit{Rewrite})$ \\
5 & $\Gamma_\texttt{NAT-ADD}\cup\dots  \cup lemmas\vdash_{\Sigma_\texttt{NAT-ADD}[y,z]} \Exists{x}s^{(n+1)~\mathtt{mod}~ m} x=0$ & 
by $(\mathit{Rewrite})$\\
6 & $\Gamma_\texttt{NAT-ADD}\cup \dots  \cup lemmas \vdash_{\Sigma_\texttt{NAT-ADD}[y,z]} s^{((n+1)~\mathtt{mod}~m)} s^{m-((n+1)~\mathtt{mod}~m)}0=0$ & 
by $(\mathit{Narrow})$\\
7 & $\Gamma_\texttt{NAT-ADD}\cup \dots  \cup lemmas \vdash_{\Sigma_\texttt{NAT-ADD}[y,z]} 0=0$ & 
by $(\mathit{Rewrite})$\\
\end{tabular}}
\end{description}


\section{Tool support}\label{section:imp}

To showcase the usefulness of Transition Algebra, we have developed a dedicated tool that supports writing specifications, reducing terms, and checking properties both by rewriting and by narrowing.
In this section, we outline the main aspects of our tool. First, we recall and also expand, by means of a case study, the syntax already introduced in Examples~\ref{ex:nat}, \ref{ex:bubble}, and~\ref{ex:collapse}.
More precisely, we demonstrate how the $\TA$ specification language can be used to formalize the CCS process calculus and the MapReduce programming model.
In addition, we present some of the commands available in the $\TA$ tool that can be used for verification purposes.
Then, in the last part of the section, we outline the main aspects of the implementation.
The tool is developed on top of SpeX~\cite{Tutu22}, a rewriting-based environment written in Maude~\cite{conf/maude/2007} that facilitates the experimental development of formal-specification languages.
Its source code together with illustrative examples is available at \url{https://github.com/Transition-Algebra/TATP}.
\subsection{Process Algebra}
\label{subsection:CCS}
Building on work reported in~\cite{go-icalp24}, we use Robin Milner's \emph{calculus of communicating systems} (CCS)~\cite{Milner80,Milner89} for illustrating how to specify a complex system.
CCS is a well know process calculus used for reasoning about concurrency by means of inter-process interactions that take place along certain communication channels.
Its basic building blocks are a set of so-called process identifiers together with a set of channel names, which model interactions between processes.
We start our specification by describing \emph{channels} and \emph{CCS actions}, as follows:

{\small
\begin{verbatim}
spec CCS is
  sorts Channel, Action .
  op rcv : Channel -> Action .
  op snd : Channel -> Action .
  op tau : () -> Action .  
\end{verbatim}
}

The specification \texttt{CCS} simply declares two sorts: \texttt{Channel} and \texttt{Action}.
Particular channel names will be defined later on for a concrete example.
We use the operation symbols \texttt{rcv} and \texttt{snd} to make explicit the \emph{actions} of receiving or sending information along a given channel.
The silent action $\tau$ is specified as a constant \texttt{tau}.
In CCS terminology, the terms \texttt{tau}, \texttt{rcv(c)} and \texttt{rcv(c)}, where \texttt{c} denotes a channel name, are all regarded as actions. 
To avoid confusion with the actions of Transition Algebra, however, we will refer to them throughout this paper as \emph{CCS actions}.

Processes are defined recursively as follows.
First, let a constant \texttt{0} stand for an inactive process:

{\small
\begin{verbatim}
  sort Process .
  op 0 : () -> Process .
\end{verbatim}
}

\noindent Then, a CCS action $a$ can be used as a prefix for a process $p$ (for which we employ the infix syntax \verb"_._").
This builds a new process, $a\verb"."p$, whose execution is meant to start with $a$ and then, once $a$ is performed, to continue with $p$:
  
{\small
\begin{verbatim}
  op _._ : Action Process -> Process .
  ax (A . P) =(A)> P [label: act] .
\end{verbatim}
}

\noindent In the axiom above, \texttt{A} and \texttt{P} are variables of sort \texttt{Action} and \texttt{Process}, respectively (besides \texttt{P}, we will also use \texttt{P'}, \texttt{Q}, \texttt{Q'}, and \texttt{R} for processes in what follows, when discussing other operators).
The syntax \texttt{(A . P) =(A)> P} specifies a transition between \texttt{A . P} and \texttt{P} using the CCS action \texttt{A}.
We label this axiom \texttt{act} in order to be able to reference it later on, when checking properties -- and we will do the same for some of the other axioms.

The operator \verb"+" denotes a non-deterministic choice between processes.
We declare it associative and commutative, which allows us to specify its operational semantics using just one axiom.
That axiom states that, if any of two processes advances with a CCS action \texttt{A} (in the condition), then the composition also advances using that action:

{\small
\begin{verbatim}
  op _+_ : Process Process -> Process [assoc comm] .
  ax (P + Q) =(A)> P' if P =(A)> P' [label: sum] .
\end{verbatim}
}

\noindent We also specify that the non-deterministic choice is idempotent and with \texttt{0} as an identity using equations:
{\small
\begin{verbatim}
  ax P + P = P .
  ax P + 0 = P .
\end{verbatim}
}
These axioms are not labelled since they are simple enough to be applied automatically when checking properties.

The operator \verb"|" stands for the parallel composition of processes.
In this case, the axiomatization requires two clauses.
The first one (labelled \texttt{ucom} for unsynchronized composition) allows processes to advance independently from one another, while the second (labelled \texttt{scom}) is in charge of synchronization (when one of the processes sends information through a channel and the other one is waiting to receive information across the same channel).
Note that, when synchronizing processes, the system advances silently.

{\small
\begin{verbatim}
  op _|_ : Process Process -> Process [assoc comm] .
  ax (P | Q) =(A)> R if P =(A)> P' and R = P' | Q [label: ucom] .
  ax (P | Q) =(tau)> R if P =(snd(C))> P' and Q =(rcv(C))> Q' and R = P' | Q' [label: scom] .
\end{verbatim}
}

\noindent Siminarly to non-deterministic choice, we specify that \texttt{0} is also an identity for the parallel composition operator using an equation:
{\small
\begin{verbatim}
  ax P | 0 = P .
\end{verbatim}
}

We limit the presentation of CCS to these three process-building operators, namely action prefixing, non-deterministic choice, and parallel composition, because they suffice for the case study that we will discuss next.
However, it is worth noting that CCS is a richer language that also supports relabelling and label restriction, both of which can be specified in $\TA$ in a straightforward manner -- see, for instance,~\cite{go-icalp24} and the `institute' example from that paper (which is now fully formalized in the repository of the $\TA$ tool) for a way to define and use label restriction.



Finally, to support recursive processes, whose behaviour may be infinite, CCS provides \emph{process definitions} which are specified by transitions rules  of the form  \texttt{P =(def)> Q}, where
\begin{enumerate*}[label=(\alph*)]
\item \texttt{P} is a constant of sort \texttt{Process}, 
\item \texttt{Q} is a process built from constants (possibly even \texttt{P}) using action prefixing, composition, etc, and
\item \texttt{def} is a purposely defined label.
\end{enumerate*}

{\small
\begin{verbatim}
  sort ProcDef .
  op def : () -> ProcDef .
\end{verbatim}
}

\noindent We use a fresh sort for \texttt{def} in order to separate it from the CCS actions that define the behaviour of processes.
Process definitions integrate into the semantics of CCS using the following rule for constants, which is standard for CCS:

{\small
\begin{verbatim}
  ax P =(A)> Q' if P =(def)> Q and Q =(A)> Q' [label: con] .
\end{verbatim}
}

\noindent The specification concludes with two special clauses for discarding processes defined as \texttt{0}.
These occur naturally when writing inductive definitions of parameterized processes, as we will see in the next case study on MapReduce.

{\small
\begin{verbatim}
  ax P + Q = Q if P =(def)> 0 .
  ax P | Q = Q if P =(def)> 0 .
endspec
\end{verbatim}
}

\subsection{Parallel and distributed data processing}

We employ \texttt{CCS} as an intermediate language in order to specify and analyse MapReduce~\cite{DeanG08}, a widely used programming framework for processing large data sets in a parallel and distributed manner.
To set the stage, suppose we need to compute the outcome of a `global' function $g$ for some fixed keys $k$ and a large input $d$ that is distributed across various `nodes' (e.g., computers, file systems, databases).
That is, $g$ is of type $K \times \prod_{1 \leq i \leq n} D_{i} \to V$, where $K$ is a finite set of so-called keys, $n$ is the number of nodes, $D_{i}$ is the set of all possible data available at node $i$, and $V$ is the set of output values.
MapReduce enables the computation of $g$ through the composition of two kinds of processing units:
\begin{itemize}

\item \emph{mappers}, which are in one-to-one correspondence with the nodes; each mapper computes the local output of $g$ for all keys $k \in K$ and for the local data stored at its corresponding node; that is, a mapper for node $i$ computes a function of the form $f_{i} \colon K \times D_{i} \to V$;

\item \emph{reducers}, which are in one-to-one correspondence with the keys; each reducer aggregates all the local outputs for its corresponding key.

\end{itemize}

Therefore, MapReduce builds on the assumption that the local outputs of $g$ are computable and can be aggregated (which we will denote using addition); moreover, aggregation needs to be associative and commutative because the local outputs can arrive at reducers in any order.
In symbols, this means that MapReduce computes
\begin{equation}
  \label{eq:MapReduce}
  g(k, d_{1}, \dots d_{n}) = \sum_{1 \leq i \leq n} f_{i}(k, d_{i})
\end{equation}
where $f_{i}(k, d_{i})$ is the local output of $g$ at node $i$ for the key $k$.
Without loss of generality, to simplify notations, we assume $D_{i} = D_{j} = D$ and $f_{i} = f_{j} = f$ for all nodes $i$, $j$.
Alternatively, consider $D = \biguplus_{1 \leq i \leq n} D_{i} = \{\langle i : d \rangle \mid 1 \leq i \leq n\ \text{and}\ d \in D_{i}\}$ and $f \colon K \times D \to V$ given by $f(k, \langle i : d \rangle) = f_{i}(k, d)$ for all keys $k \in K$, nodes $1 \leq i \leq n$, and local data $d \in D_{i}$.

The actual computation is performed in five stages, as follows:
\begin{enumerate*}

\item first, we initialize the mappers and the reducers;

\item next, we run the mappers, which yield multiple local outputs, one for each node and key;

\item then, we shuffle the local outputs so that, for each key $k$, all local outputs with key $k$ are delivered to the $k^{\text{th}}$ reducer;

\item we run the reducers, thus aggregating all local outputs; and, finally,

\item we collect the global outputs that are generated by reducers.

\end{enumerate*}
Some of these stages may overlap.
For instance, reducers can start aggregating local outputs as soon as they become available, without having to wait for all the mappers to finish their computations.

\begin{example}[Counting occurrences]
  To illustrate the process, consider a simple distributed binary database consisting of three nodes, $N_{1}$, $N_{2}$, and $N_{3}$ with local data consisting of the following bit strings $100$, $10$, and $101$, respectively.
  Suppose we aim to count the occurrences of the bits $0$ and $1$ in the database.
  For that purpose, we let the set $K$ of keys be $\{0, 1\}$, the set $V$ of values be $\mathbb{N}$, we define aggregation as addition, and $f$ as the function that counts the occurrences of a key (bit) in a bit string. Hence $f(0, 100) = 2$, $f(1, 100) = 1$, and so on.
  The diagram below depicts the corresponding MapReduce computation.
  On the left, we have the three nodes of the database.
  Each of them has an associated mapper; we use dashed lines to indicate what local data is available to each mapper.
  Next, we use solid arrows to represent the communication between mappers and reducers.
  Each arrow, say between a mapper $M_{i}$ and a reducer $R_{k}$, is labelled with a local output, which is a key-value pair of the form $\langle k : f(k, d_{i}) \rangle$.
  Lastly, on the right, we have the aggregated outcomes, which are obtained by summing all the local outputs that arrive at each reducer; again, we use dashed lines to indicate which outcome corresponds to each reducer.
  At the bottom of the diagram, we mark the three main stages of the MapReduce process: map, shuffle, and reduce.
\end{example}
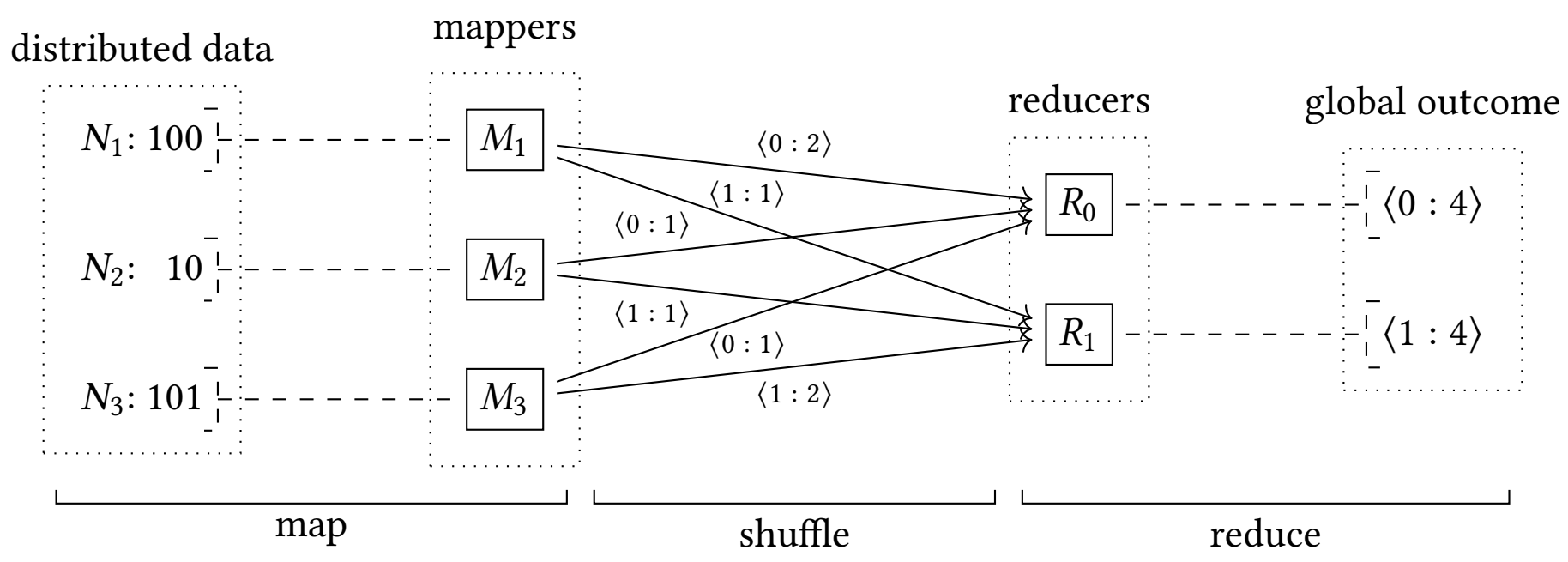
\begin{center}
    \begin{tikzpicture}[x=7em]
      \begin{scope}[anchor=east]
        \node (N1) at (0,1) {$N_{1}$: $100$};
        \node (N2) at (0,0) {$N_{2}$: \phantom{$0$}$10$};
        \node (N3) at (0,-1) {$N_{3}$: $101$};
      \end{scope}

      \begin{scope}[every node/.style={draw, black, outer sep=3pt}]
        \node (M1) at (1,1) {$M_{1}$};
        \node (M2) at (1,0) {$M_{2}$};
        \node (M3) at (1,-1) {$M_{3}$};

        \node (R0) at (3,.5) {$R_{0}$};
        \node (R1) at (3,-.5) {$R_{1}$};
      \end{scope}

      \begin{scope}[anchor=west]
        \node (O0) at (4,.5) {$\langle 0 : 4 \rangle$};
        \node (O1) at (4,-.5) {$\langle 1 : 4 \rangle$};
      \end{scope}

      \begin{scope}[dashed]
        \foreach \i in {1,2,3} {
          \draw ($(N\i.north east)-(3pt,0)$) -- (N\i.north east) -- (N\i.south east) -- ($(N\i.south east)-(3pt,0)$);
          \draw (N\i.east) -- (M\i.west);
        }
        \foreach \i in {0,1} {
          \draw (R\i.east) -- (O\i.west);
          \draw ($(O\i.north west)+(3pt,0)$) -- (O\i.north west) -- (O\i.south west) -- ($(O\i.south west)+(3pt,0)$);
        }
      \end{scope}

      \begin{scope}[->, every node/.style={font=\scriptsize}]
        \draw (M1) -- node [above] {$\langle 0 : 2 \rangle$} (R0);
        \draw (M1) -- node [above, pos=.4] {$\langle 1 : 1 \rangle$} (R1);
        \draw (M2) -- node [above, pos=.2] {$\langle 0 : 1 \rangle$} (R0);
        \draw (M2) -- node [below, pos=.2] {$\langle 1 : 1 \rangle$} (R1);
        \draw (M3) -- node [below, pos=.4] {$\langle 0 : 1 \rangle$} (R0);
        \draw (M3) -- node [below] {$\langle 1 : 2 \rangle$} (R1);
      \end{scope}

      \begin{scope}[every node/.style={draw, dotted, black, inner sep=5pt}]
        \node (DD) [fit={(N1) (N2) (N3)}, label=above: distributed data] {};
        \node (MS) [fit={(M1) (M2) (M3)}, label=above: mappers] {};
        \node (RS) [fit={(R0) (R1)}, label=above: reducers] {};
        \node (OS) [fit={(O0) (O1)}, label=above: global outcome] {};
      \end{scope}

      \node (S) at ($(MS.south)-(0,8pt)$) {};

      \draw ($(DD.south west |- S)+(3pt,3pt)$) -- ($(DD.south west |- S)+(3pt,0pt)$) -- node [below] {map} ($(S -| MS.south east)+(-3pt,0pt)$) -- ($(S -| MS.south east)+(-3pt,3pt)$);
      \draw ($(MS.south east |- S)+(3pt,3pt)$) -- ($(MS.south east |- S)+(3pt,0pt)$) -- node [below] {shuffle} ($(S -| RS.south west)+(-3pt,0pt)$) -- ($(S -| RS.south west)+(-3pt,3pt)$);
      \draw ($(RS.south west |- S)+(3pt,3pt)$) -- ($(RS.south west |- S)+(3pt,0pt)$) -- node [below] {reduce} ($(S -| OS.south east)+(-3pt,0pt)$) -- ($(S -| OS.south east)+(-3pt,3pt)$);
    \end{tikzpicture}
  \end{center}
To model MapReduce in $\TA$, we start by extending the specification \texttt{CCS} with declarations of the basic entities involved in MapReduce: node identifiers, keys, finite sets of keys, data, values, key-value pairs, and the underlying function $f$.\footnote{We use the keyword \texttt{extending} to indicate that we add new elements on `old' sorts, notably \texttt{Process} and \texttt{Channel}, imported from the specification \texttt{CCS}.}
We write \texttt{S1 S2} for the union of two sets \texttt{S1} and \texttt{S2} (of node identifiers, keys, or key-value pairs), \texttt{none} for the empty set, \texttt{Id(X)} for the singleton set consisting of a node identifier \texttt{X}, \texttt{Key(K)} for the singleton set consisting of a key \texttt{K}, \texttt{V1 + V2} for the aggregation of values \texttt{V1} and \texttt{V2}, \texttt{0} for the identity element of \texttt{+} (which we use as a starting value during reduction), and \verb"< K : V >" for the singleton set consisting of the key-value pair with key \texttt{K} and value \texttt{V}.

{\small\begin{verbatim}
spec MAP-REDUCE is
  extending CCS .
  sorts Id, IdSet .
  op none : () -> IdSet .
  op Id : Id -> IdSet .
  op __ : IdSet IdSet -> IdSet [assoc comm id: (none)] .
  sort Data .
  sorts Key, KeySet .
  op none : () -> KeySet .
  op Key : Key -> KeySet .
  op __ : KeySet KeySet -> KeySet [assoc comm id: (none)] .
  sort Value .
  op 0 : () -> Value .
  op _+_ : Value Value -> Value [assoc comm id: (0)] .
  sorts KVSet .
  op none : () -> KVSet .
  op <_:_> : Key Value -> KVSet .
  op __ : KVSet KVSet -> KVSet [assoc comm id: (none)] .
  op f : Key Data -> Value .
\end{verbatim}}  

Next, we define the CCS processes used in modelling MapReduce.
The first two are \texttt{JobManager} and \texttt{Node}.
Both are parameterized -- as will be all other CCS processes that we declare.
\texttt{JobManager} serves as a seed for MapReduce; its parameters are a set of keys and a set of node identifiers, which will be used to determine which mappers and which reducers need to be initialized.
The \texttt{Node} process has two parameters as well; the first one is its identifier, and the second is its underlying local data.
Hence we consider the following $\TA$ declarations:

{\small\begin{verbatim}
  op JobManager : KeySet IdSet -> Process .
  op Node : Id Data -> Process .
\end{verbatim}}

\noindent These enable us to capture the starting point of a MapReduce computation as a composed process of the form
\begin{center}
  \ttfamily
  JobManager(KS, Id($\mathtt{X}_{1}$) \dotso Id($\mathtt{X}_{n}$)) | Node($\mathtt{X}_{1}$, $\mathtt{D}_{1}$) | \dotso | Node($\mathtt{X}_{n}$, $\mathtt{D}_{n}$)
\end{center}
where \texttt{KS} is a set of keys, $\mathtt{X}_{i}$ is a node identifier for each $1 \leq i \leq n$, and $\mathtt{D}_{i}$ is the local data at node $\mathtt{X}_{i}$.

The \texttt{JobManager} spawns four other processes that either initialize, monitor, or conclude the MapReduce computation.
Its definition is as follows:

{\small\begin{verbatim}
  ax JobManager(KS, IdS)
     =(def)> TaskDistributor(KS, IdS)
           | TaskMonitor(KS, IdS)
           | Reducers(KS)
           | Collector(KS, none)
     [label: def-JobManager] .
\end{verbatim}}
where
\begin{itemize}

\item \texttt{TaskDistributor} is a process that triggers the map stage of MapReduce, initializing mappers;

\item \texttt{TaskMonitor} (as the name implies) tracks the mapper processes and, when they terminate, notifies reducers in order to finalize the aggregation of the generated local outputs;

\item \texttt{Reducers} initializes the reducer processes; and

\item \texttt{Collector} gathers all global outputs compiled by reducers; its second parameter is an increasing subset (initially empty) of the set of key-value pairs that we aim to compute through MapReduce.
\end{itemize}
The signatures of all theses processes are declared similarly to those of \texttt{JobManager} and \texttt{Node}.

The \texttt{TaskDistributor} and \texttt{Node} processes work together to initialize mappers.
To that end, they communicate along a parameterized channel \texttt{map(KS, X)}, where \texttt{KS} is a set of keys and \texttt{X} is a node identifier.
The $\TA$ declaration of \texttt{map} and the definitions of \texttt{TaskDistributor} and \texttt{Node} are as follows:

{\small\begin{verbatim}
  op map : KeySet Id -> Channel .
  ax TaskDistributor(KS, Id(X) IdS)
     =(def)> (snd(map(KS, X)) . TaskDistributor(KS, IdS))
     [label: def-TaskDistributor] .
  ax TaskDistributor(KS, none)
     =(def)> 0
     [label: def-TaskDistributor] .
  ax Node(X, D)
     =(def)> (rcv(map(KS, X)) . Mapper(X, KS, D))
     [label: def-Node] .
\end{verbatim}}

It is worth noting that \texttt{TaskDistributor} sends information through \texttt{map} channels sequentially, but the choice of recipient (node) at each step is non-deterministic.
That is because the set-building operator \verb"__" is associative and commutative.
Also, what we are seeing here is actually an instance of value-passing in CCS: \texttt{TaskDistributor} provides the set \texttt{KS} of keys (through the \texttt{map} channel), which are unknown to nodes and which get coupled (through process synchronization) with the local data \texttt{D} of the node \texttt{X} in order to initialize the corresponding mapper for \texttt{X}.

Clauses such as \texttt{def-Node} having fresh variables in the right-hand side of transitions---an essential aspect for value-passing---are generally problematic from an operational perspective and fall outside the scope of the transition rules considered in Section~\ref{section:op-semantics}.
However, because the specification of MapReduce is based on \texttt{CCS}, we can mitigate this issue through narrowing applications of the axiom \texttt{scom} (from the specification \texttt{CCS}) that defines the operational semantics for the synchronous composition of processes.
Take, for instance, a transition of the form $\tau(\texttt{TaskDistributor(ks, Id(x) ids)} \mid \texttt{Node(x, d)}, \texttt{TaskDistributor(ks, ids)} \mid \texttt{Mapper(x, ks, d)})$, where \texttt{ks}, \texttt{x}, \texttt{ids}, and \texttt{d} are all constant terms.
That transition can be derived through narrowing, using \texttt{scom}, by solving the following joint queries, where \texttt{C} is the variable introduced by \texttt{scom}:
\begin{itemize}

\item $\texttt{snd(C)}(\texttt{TaskDistributor(ks, Id(x) ids)}, \texttt{TaskDistributor(ks, ids)})$
  
\item $\texttt{rcv(C)}(\texttt{Node(x, d)}, \texttt{Mapper(x, ks, d)})$

\end{itemize}
By continuing to apply narrowing for the first query using the axioms \texttt{con} (from \texttt{CCS}), \texttt{def-TaskDistributor} (from \texttt{MAP-REDUCE}), and \texttt{act} (from \texttt{CCS}), we get $\texttt{C} = \texttt{map(ks, x)}$.
Similarly, for the second query, using the axioms \texttt{con}, \texttt{def-Node}, and \texttt{act}, we get $\texttt{C} = \texttt{map(KS, x)}$, where \texttt{KS} is the fresh variable introduced by \texttt{def-Node}.
We put these steps together (in order to conclude the application of \texttt{scom}) by unifying the terms \texttt{map(ks, x)} and \texttt{map(KS, x)}.
This yields a substitution for the variable \texttt{KS}, namely $\texttt{KS} \mapsto \texttt{ks}$, thus avoiding the problem of having a fresh variable in the right-hand side of \texttt{def-Node}.
We will use this line of reasoning repeatedly in the final part of the section (but assisted by the $\TA$ tool in that case) in order to check the correctness of MapReduce.

The reducers get initialized in a similar (non-deterministic) manner, by applying the following definitions:

{\small\begin{verbatim}
  ax Reducers(Key(K) KS)
     =(def)> Reducer(K, 0)
           | Reducers(KS)
     [label: def-Reducers] .
  ax Reducers(none)
     =(def)> 0
     [label: def-Reducers] .
\end{verbatim}}

This concludes the initialization stage.
The map, shuffle, and reduce stages are all condensed in the synchronization of mappers and reducers.
First, mappers send local outputs for all the keys whose values they have been instructed to compute, followed by a special \texttt{eom} (end of mapping) message once all the local outputs have been delivered.
That \texttt{eom} message is meant to indicate to \texttt{TaskMonitor} that the map stage has ended for the current node.
In turn, reducers receive either a local output (through synchronization, from one of the mappers) or a special \texttt{eor} (end of reduction) message (from \texttt{TaskMonitor}).
In the first case, they add the local output received to their tally and continue the computation; and in the second case, they inform the \texttt{Collector} of the global output computed for the current key and then terminate their execution.
The formal CCS definitions are as follows:

{\small\begin{verbatim}
  ax Mapper(X, Key(K) KS, D)
     =(def)> (snd(local-output(K, f(K, D))) . Mapper(X, KS, D))
     [label: def-Mapper] .
  ax Mapper(X, none, D)
     =(def)> (snd(eom(X)) . Node(X, D))
     [label: def-Mapper] .
  ax Reducer(K, R)
     =(def)> ((rcv(local-output(K, V)) . Reducer(K, R + V))
            + (rcv(eor(K)) . snd(global-output(K, R)) . 0))
     [label: def-Reducer] .
\end{verbatim}}

We note, once more, that the specification does not commit to any ordering of sending and receiving local outputs.
Instead, all possible synchronizations are explored in parallel, non-deterministically.
That is due to two factors: first, because the process-composition operator \texttt{|} is associative and commutative, at each step, we can (try to) synchronize each available mapper and reducer; second, because the set-building operator \verb"__" is associative and commutative, the key \texttt{K} whose value is computed next by the selected mapper is also chosen non-deterministically.

The only remaining processes to describe are \texttt{TaskMonitor} and \texttt{Collector}.
As mentioned above, the first one acts as a barrier: it waits for all mappers to end their execution, then notifies the reducers that they can conclude too.

{\small\begin{verbatim}
  ax TaskMonitor(KS, Id(X) IdS)
     =(def)> (rcv(eom(X)) . TaskMonitor(KS, IdS))
     [label: def-TaskMonitor] .
  ax TaskMonitor(Key(K) KS, none)
     =(def)> (snd(eor(K)) . TaskMonitor(KS, none))
     [label: def-TaskMonitor] .
  ax TaskMonitor(none, none)
     =(def)> 0
     [label: def-TaskMonitor] .
\end{verbatim}}

Lastly, the \texttt{Collector} synchronizes with the terminating reducers in order to record their compiled global outputs.
This brings to an end the final stage of MapReduce.

{\small\begin{verbatim}
  ax Collector(Key(K) KS, KVS)
     =(def)> (rcv(global-output(K, V)) . Collector(KS, < K, V > KVS))
     [label: def-Collector] .
endspec
\end{verbatim}}

\noindent The full specifications of CCS and MapReduce, as well as the proof score for the next verification, are available as examples in the source-code repository of the tool at \url{https://github.com/Transition-Algebra/TATP}.

In the remaining part of this section, we look into the correctness of MapReduce.
More specifically, we check that the CCS definitions given above allow for the computation of the function described in equation~\eqref{eq:MapReduce}.
To simplify the presentation, we fix the number of nodes and the number of keys: two in each case.
But this is not a significant limitation, because the proof scales without difficulty to arbitrarily many nodes and keys simply by repeating some of the proof steps that we will discuss next.
We can even consider a general proof by simultaneous structural induction on the parameters of \texttt{JobManager} (i.e., on the sets of nodes and keys used to initialize mappers and reducers), but that is beyond the scope of this paper.
So what we want to check is that the specification \texttt{MAP-REDUCE} entails a quantified transition of the form
$$
\Forall{Z}\Exists{R}
\mathtt{tau}^{*}(S(Z), T(Z, R))
$$
where
\begin{itemize}
  
\item $Z = \{\mathtt{K}_{1}, \mathtt{K}_{2} \colon \mathtt{Key}; \mathtt{X}_{1}, \mathtt{X}_{2} \colon \mathtt{Id}; \mathtt{D}_{1}, \mathtt{D}_{2} \colon \mathtt{Data}\}$,

\item $R$ is a variable of sort \texttt{Process},

\item $S(Z) = \mathtt{JobManager}(\mathtt{Key}(\mathtt{K}_{1})\;\mathtt{Key}(\mathtt{K}_{2}), \mathtt{Id}(\mathtt{X}_{1})\;\mathtt{Id}(\mathtt{X}_{2}))
  \mid \mathtt{Node}(\mathtt{X}_{1}, \mathtt{D}_{1})
  \mid \mathtt{Node}(\mathtt{X}_{2}, \mathtt{D}_{2})$, and

\item $T(Z, R) = \mathtt{Collector}(\mathtt{none}, \langle \mathtt{K}_{1}, \mathtt{f}(\mathtt{K}_{1}, \mathtt{D}_{1}) + \mathtt{f}(\mathtt{K}_{1}, \mathtt{D}_{2})  \rangle\;\langle \mathtt{K}_{2}, \mathtt{f}(\mathtt{K}_{2}, \mathtt{D}_{1}) + \mathtt{f}(\mathtt{K}_{2}, \mathtt{D}_{2}) \rangle)
  \mid R$.
  
\end{itemize}

By $(\mathit{UQ})$, this is equivalent to checking that the axioms of \texttt{MAP-REDUCE} entail the query $\Exists{R} \mathtt{tau}^{*}(S(Z), T(Z, R))$ with respect to the extended signature $\Sigma_{\texttt{MAP-REDUCE}}[Z]$.
We can confirm that this entailment holds using narrowing, which the $\TA$ tool supports through a dedicated \texttt{check} command.
But the general narrowing procedure is impractical for this property because the search space is vast.
We can limit the number of narrowing steps used (via the \texttt{set narrowing bound} command), but that is still not enough.
The transition that we are verifying can be witnessed by composing 12 atomic \texttt{tau} transitions: two to initialize the mappers, two to initialize the reducers and to deliver the first local outputs to them, two more to deliver the remaining local outputs, two to inform the \texttt{TaskMonitor} that the mappers have completed their execution, two to also complete the execution of the reducers, and lastly, two to collect the global outputs.
All in all, these transitions require more than 100 narrowing steps.
To make narrowing effective, we use an explicit-check command (introduced by the \texttt{echeck} keyword) that, in addition to a query, takes a list of axiom labels indicating which equations or rules are meant to be used in narrowing steps, and in which order.
The list may include both labels from the specifications above, such as \texttt{def-JobManager}, \texttt{def-Node}, \texttt{scom}, or \texttt{act}, and labels that are intrinsic to Transition Algebra such as \texttt{comp} for `$\comp$' steps (as presented in Definition~\ref{definition:narrowing-step}), \texttt{choice} for `$\cup$' steps, and \texttt{star} for `$\circlearrowright$' steps.
We also write comments, introduced by \verb"``", to indicate which narrowing steps are applied for each transition.

The full proof is written in an open-close environment reminiscent of the OBJ family of languages~\cite{Goguen2000}.
Within such an environment, we first indicate which specification is being used for verification purposes, then issue a series of commands that mechanize the proof.
In our case, we use the specification \texttt{MAP-REDUCE}, then issue three \texttt{let} commands to extend the signature, and conclude with a verification by narrowing through the \texttt{echeck} command.

{\small\begin{verbatim}
open MAP-REDUCE
  let X1, X2 : Id
  let D1, D2 : Data
  let K1, K2 : Key
  echeck
    JobManager(Key(K1) Key(K2), Id(X1) Id(X2))
    | Node(X1, D1)
    | Node(X2, D2)
  =((tau)*)>
    Collector(none, < K1 : f(K1, D1) + f(K1, D2) > < K2 : f(K2, D1) + f(K2, D2) >)
    | R:Process
  using (star comp
         `` 1st transition: initialize the 1st mapper by synchronizing TaskDistributor and one of the nodes
         scom con def-JobManager ucom con def-TaskDistributor act ucom con def-Node act
         star comp
         `` 2nd transition: initialize the 2nd mapper by synchronizing TaskDistributor and the remaining node
         ucom scom con def-TaskDistributor act con def-Node act
         star comp
         `` 3rd transition: initialize the 1st reducer and deliver the first local output to it
         ucom scom con def-Mapper act con def-Reducers ucom con def-Reducer sum act
         star comp
         `` 4th transition: initialize the 2nd reducer and deliver the next local output
         ucom scom con def-Mapper act con def-Reducers con def-Reducer sum act
         star comp
         `` 5th and 6th transitions: deliver the remaining local output to the reducers
         ucom scom con def-Mapper act con def-Reducer sum act
         star comp
         ucom scom con def-Mapper act con def-Reducer sum act
         star comp
         `` 7th and 8th transitions: notify the TaskMonitor that the mappers have completed their execution
         ucom scom con def-Mapper act con def-TaskMonitor act
         star comp
         ucom scom con def-Mapper act con def-TaskMonitor act
         star comp
         `` 9th and 10th transitions: inform the reducers that there are no more local outputs to aggregate
         ucom scom con def-TaskMonitor act con def-Reducer sum act
         star comp
         ucom scom con def-TaskMonitor act con def-Reducer sum act
         star comp
         `` 11th and 12th transitions: collect the global outputs
         ucom scom act con def-Collector act
         star comp
         ucom scom act con def-Collector act
         star)
close
\end{verbatim}}


\subsection{Implementation notes}

We summarize here how the main aspects of the implementation have been developed.

\paragraph{Reduction} We can take advantage of the meta-level capabilities of Maude for designing a resolution
procedure for reductions. 
As with Maude itself, the resulting implementation is far from straightforward, involving numerous 
subtle interactions whose correct handling requires considerable care.
In particular, given a term $t$, we look for an equation $\ell = \err \lif H$, such that
$\ell$ matches $t$ (possibly not at the top), obtaining a substitution $\theta$ and a context $c$.
Then, a new substitution $\theta'$ extending $\theta$ is required for satisfying the condition $H$; $\theta'$
will be used for instantiating $\err$, replacing $\ell$ in the context $c$. The whole process starts
with the \texttt{reduce T} command, where \texttt{T} is the term to be reduced, and is repeated
until no equations can be applied to the current term.

As explained in the previous sections, the condition $H$ may contain both equalities and transitions. For equalities,
the current substitution is applied to both sides of the equation and they are reduced independently
as described. Once they cannot be further reduced, equality modulo axioms is checked.

For transitions, which are not expected to be confluent, we do not compute a unique substitution but a set
of them, so all reachable terms can be obtained. Moreover, the process is slightly different to the
one before. First, the matching is twofold: we need the term to match the left-hand side and the actions
to match the labels. Then, the implementation includes built-in transitions implementing the rules for
composition, choice, and iteration, which are used when required. Finally, the matching with the right-hand
side might extend the substitution.

The substitution $\theta'$ is computed by using backtracking following a depth-first strategy
and adding an extra condition requiring matching with the right-hand side.
In particular, for each transition in the
condition (equalities do not extend the substitution), as many child nodes can be generated as 
there are possible substitutions. It is important to
note that, because we work modulo equational axioms such as associativity and commutativity,
several different substitutions can be generated when matching a term and a pattern. 
Also note that equalities, even though they do not extend the
substitution, might need to be recomputed in general when the substitution changes.
Finally, fresh variables are required during the process because substitutions are propagated across
transition applications, and the same variable may appear in several of them. Therefore, clashes must 
first be avoided. Once the process is completed, a substitution containing only user-defined variables 
is reconstructed.

Termination is ensured by supporting two bounds: in the number of steps (including both reductions
and transitions) and in the number of conditions, as explained above.
It is interesting to note that the bounds are
considered differently: while the number of steps is computed globally, so each
application of an equation or a transition reduces the bound (whether in the current reduction or in
the following ones), the number of conditions is used locally for each reduction step and restarted 
again for the next one.
Moreover, our implementation stores the terms already tried to reduce in the conditions, avoiding
loops that often occur when dealing with some operators, in particular iteration.

\paragraph{Narrowing} For solving queries, we adapt and use a general logic-programming procedure that is already integrated into SpeX~\cite{Tutu22} and matches closely the definition of narrowing in Section~\ref{section:op-semantics}.
The implementation of that procedure relies on the Maude strategy language~\cite{EkerMMRV23} and involves the rewriting of three kinds of `configurations': \emph{initial}, \emph{open}, and \emph{solved}.
An initial configuration is merely an encoding of a logic-programming problem.
That is, a pair consisting of a specification (set of Horn clauses) and a query that is meant to be checked.
Optionally, for \texttt{echeck}, an initial configuration may also provide a list of (sets of) clauses (identified by labels) to be used in subsequent narrowing steps.
An open configuration includes, besides the specification and a query, information about the partial solution computed up to that point.
This means that, for any chain of narrowing steps $\Exists{X_{0}} Q_{0} \narrow{}_{\psi_{1}} \Exists{X_{1}} \emptyset_{1}  \narrow{}_{\psi_{2}} \Exists{X_{2}} \emptyset_{2}  \narrow{}_{\psi_{3}} \dotsb$ we have a corresponding chain of configuration rewrites such that, for each index $n$, the $n$th open configuration includes an encoding of the substitution $\psi_{1} \comp \psi_{2} \comp \dotsb \comp \psi_{n}$.
Moreover, open configurations may also carry lists of (sets of) clauses for \texttt{echeck} and information about the variables in use.
The latter information is passed to built-in Maude functions for computing unifiers.
Finally, a solved configuration is a configuration whose query is $\ltrue$.

The rewriting of configurations is controlled using Maude strategies.
Every change of configurations consists of several sub-steps:
\begin{enumerate*}[label=(\alph*)]
\item first, we non-deterministically select a context or formula within the current query where narrowing is applied;
\item then we select (also non-deterministically) a suitable clause,
\item we compute all possible unifiers (which may not be unique if the computations are performed modulo associativity and commutativity),
\item apply substitutions (according to Definition~\ref{definition:narrowing-step}),
\item compile new partial solutions, and
\item update the information on the variable names that are currently in use and on the list of (sets of) clauses meant to guide the following narrowing steps.
\end{enumerate*}

Similarly to term rewriting, termination can be enforced by specifying a bound on the number of narrowing steps that can be applied.
Still, the search space for solved configurations can easily become infeasibly large even for low bounds.
Because of this, the explicit search (implemented for \texttt{echeck}) is often more convenient as it drastically limits the extent of the search space by specifying which clauses are admissible at each step.
Its downside is that the search procedure is no longer fully automatic, and completeness cannot be guaranteed.
To counterbalance this limitation, if no solved configuration is found (given a sequence \texttt{HCL} of Horn-clause labels), then the explicit search returns a final list of open configurations that may be rewritten to a solved one (by modifying or extending \texttt{HCL}).




\section{Conclusions and ongoing work}\label{section:conc}

Several logics have been proposed for modelling and reasoning about the behaviour of concurrent systems, giving freedom to researchers
to choose among them depending of their particular needs.
In this paper, we have examined Transition Algebra, an extension of many-sorted first-order logic that includes features from
dynamic logic, where actions are used as labels of transitions, thus allowing us to capture and analyse complex behaviours.
This logic is useful for modelling and studying systems where the actions are as important as the system states upon which they produce an effect.
Unlike previous definitions, and for greater flexibility in the use of actions, we have extended (and, at the same time, simplified) the framework introduced in~\cite{go-icalp24} in order to support arbitrary terms as actions.
We have devised a sound and complete set of proof rules for reasoning about Horn clauses.
Moreover, we have developed an operational semantics for Transition Algebra based on term rewriting, transition checking, and narrowing, a powerful mechanism for solving queries.
We have shown that rewriting and narrowing are both sound and complete, thus advocating for Transition Algebra as an appropriate logical setting for system verification.
To further demonstrate its applicability, we have conducted a comprehensive case study on the analysis of the MapReduce programming framework.

The features above provide advantages from both theoretical and practical points of view. On the one hand, systems
can be more easily and accurately described than with conventional equational logics.
Furthermore, Transition Algebra allows for integrating specification and verification within a single framework, which is useful especially when considering properties that cannot be easily expressed equationally.
On the other hand, its simple but powerful operational semantics is very well suited for axiomatization. We have implemented
a Maude prototype that supports specifications in Transition Algebra and solving queries on them.

This work lays the foundation for a series of future developments.
On the theoretical side, we want to explore additional action operators, such as tests and parallel composition.
Also, we aim to continue the study of narrowing.
Though sound and complete, in its current form, narrowing is still rather inefficient from a computational perspective.
We intend to look further into different forms of narrowing, as can be found in the equational-logic literature.
Moreover, we aim to explore the use of explicit narrowing, shown in Section~\ref{section:imp}, as a useful tool in interactive theorem proving.
On the practical side, we are interested in continuing the development of the tool, and also in specifying and analysing a variety of systems to asses the approach and to design strategies to ease the verification process.



\bibliographystyle{ACM-Reference-Format}
\bibliography{ta}

@techreport{Dijkstra1970StructuredProgramming,
  author      = {Edsger W. Dijkstra},
  title       = {Notes on Structured Programming},
  institution = {Technological University Eindhoven},
  year        = {1970},
  address     = {Eindhoven, The Netherlands},
  type        = {Technical Report}
}

@inproceedings{bugs,
author = {Cotroneo, Domenico and De Simone, Luigi and Liguori, Pietro and Natella, Roberto and Bidokhti, Nematollah},
title = {How bad can a bug get? an empirical analysis of software failures in the {OpenStack} cloud computing platform},
year = {2019},
isbn = {9781450355728},
publisher = {Association for Computing Machinery},
address = {New York, NY, USA},
url = {https://doi.org/10.1145/3338906.3338916},
doi = {10.1145/3338906.3338916},
booktitle = {Proceedings of the 2019 27th ACM Joint Meeting on European Software Engineering Conference and Symposium on the Foundations of Software Engineering},
pages = {200–211},
numpages = {12},
location = {Tallinn, Estonia},
series = {ESEC/FSE 2019}
}

@article{DBLP:journals/tcs/ClavelMP07,
  author       = {Manuel Clavel and Jos{\'{e}} Meseguer and Miguel Palomino},
  title        = {Reflection in membership equational logic, many-sorted equational logic, Horn logic with equality, and rewriting logic},
  journal      = {Theor. Comput. Sci.},
  volume       = {373},
  number       = {1-2},
  pages        = {70--91},
  year         = {2007},
  bibsource    = {dblp computer science bibliography, https://dblp.org}
}

@article{AstesianoBKKMST02,
  author       = {Egidio Astesiano and Michel Bidoit and H{\'e}l{\`e}ne Kirchner and Bernd Krieg-Br{\"u}ckner and Peter D. Mosses and Donald Sannella and Andrzej Tarlecki},
  title        = {{CASL: The Common Algebraic Specification Language}},
  journal      = {Theoretical Computer Science},
  volume       = {286},
  number       = {2},
  pages        = {153--196},
  year         = {2002}
}

@book{Diaconescu2025,
  author       = {R{\u{a}}zvan Diaconescu},
  title        = {Institution-Independent Model Theory},
  series       = {Studies in Universal Logic},
  edition      = {2},
  publisher    = {Birkh{\"{a}}user},
  address      = {Berlin, Germany},
  year         = {2025}
}

@article{dia-caf,
  author       = {R{\u{a}}zvan Diaconescu and Kokichi Futatsugi},
  title        = {{Logical foundations of CafeOBJ}},
  journal      = {Theoretical Computer Science},
  volume       = {285},
  number       = {2},
  pages        = {289--318},
  year         = {2002}
}

@article{dia-ult,
  title        = {Institution-independent Ultraproducts},
  author       = {R{\u{a}}zvan Diaconescu},
  journal      = {Fundamenta Informaticae},
  volume       = {55},
  number       = {3-4},
  pages        = {321--348},
  year         = {2003}
}

@article{LopezRuedaES23,
  author       = {Ra{\'{u}}l L{\'{o}}pez{-}Rueda and Santiago Escobar and Julia Sapi{\~{n}}a},
  title        = {An efficient canonical narrowing implementation with irreducibility and {SMT} constraints for generic symbolic protocol analysis},
  journal      = {J. Log. Algebraic Methods Program.},
  volume       = {135},
  pages        = {100895},
  year         = {2023}
}

@article{gog-ins,
  title        = {Institutions: Abstract Model Theory for Specification and Programming},
  author       = {Joseph A. Goguen and Rod M. Burstall},
  journal      = {{J. ACM}},
  number       = {1},
  volume       = {39},
  pages        = {95--146},
  year         = {1992}
}

@article{journals/sigplan/GoguenM82,
  author       = {Joseph Goguen and Jos{\'{e}} Meseguer},
  title        = {Completeness of many-sorted equational logic},
  journal      = {{ACM} {SIGPLAN} Notices},
  volume       = {17},
  number       = {1},
  pages        = {9--17},
  year         = {1982}
}

@inproceedings{MeseguerT05,
  author       = {Jos{\'{e}} Meseguer and Prasanna Thati},
  editor       = {Narciso Mart{\'{\i}}{-}Oliet},
  title        = {Symbolic Reachability Analysis Using Narrowing and its Application to Verification of Cryptographic Protocols},
  booktitle    = {Proceedings of the Fifth International Workshop on Rewriting Logic and Its Applications, {WRLA} 2004, Barcelona, Spain, March 27-28, 2004},
  series       = {Electronic Notes in Theoretical Computer Science},
  volume       = {117},
  pages        = {153--182},
  publisher    = {Elsevier},
  address      = {Amsterdam, The Netherlands},
  year         = {2004}
}

@inproceedings{EscobarMT05,
  author       = {Santiago Escobar and Jos{\'{e}} Meseguer and Prasanna Thati},
  editor       = {J{\"{u}}rgen Giesl},
  title        = {Natural Narrowing for General Term Rewriting Systems},
  booktitle    = {Term Rewriting and Applications, 16th International Conference, {RTA} 2005, Nara, Japan, April 19-21, 2005, Proceedings},
  series       = {Lecture Notes in Computer Science},
  volume       = {3467},
  pages        = {279--293},
  publisher    = {Springer},
  address      = {Berlin, Germany},
  year         = {2005}
}

@article{ThatiM06,
  author       = {Prasanna Thati and Jos{\'{e}} Meseguer},
  title        = {Complete symbolic reachability analysis using back-and-forth narrowing},
  journal      = {Theor. Comput. Sci.},
  volume       = {366},
  number       = {1-2},
  pages        = {163--179},
  year         = {2006}
}

@article{EscobarSM12,
  author       = {Santiago Escobar and Ralf Sasse and Jos{\'{e}} Meseguer},
  title        = {Folding variant narrowing and optimal variant termination},
  journal      = {J. Log. Algebraic Methods Program.},
  volume       = {81},
  number       = {7-8},
  pages        = {898--928},
  year         = {2012}
}

@article{gai-cbl,
  author       = {Daniel G{\u{a}}in{\u{a}} and Kokichi Futatsugi and Kazuhiro Ogata},
  title        = {Constructor-based Logics},
  journal      = {Journal of Universal Computer Science},
  volume       = {18},
  number       = {16},
  pages        = {2204--2233},
  year         = {2012}
}

@book{conf/maude/2007,
  editor       = {Manuel Clavel and Francisco Dur{\'{a}}n and Steven Eker and Patrick Lincoln and Narciso Mart{\'{\i}}{-}Oliet and Jos{\'{e}} Meseguer and Carolyn L. Talcott},
  title        = {All About Maude -- A High-Performance Logical Framework, How to Specify, Program and Verify Systems in Rewriting Logic},
  series       = {LNCS},
  volume       = {4350},
  publisher    = {Springer},
  address      = {Berlin, Germany},
  year         = {2007}
}

@article{EkerMMRV23,
  author       = {Steven Eker and Narciso Mart\'{\i}{-}Oliet and Jos\'{e} Meseguer and Rub\'{e}n Rubio and Alberto Verdejo},
  title        = {{The Maude strategy language}},
  journal      = {J. Log. Algebraic Methods Program.},
  volume       = {134},
  pages        = {100887},
  year         = {2023}
}

@inproceedings{Meseguer90,
  author       = {Jos{\'{e}} Meseguer},
  editor       = {St{\'{e}}phane Kaplan and Mitsuhiro Okada},
  title        = {Conditional Rewriting Logic: Deduction, Models and Concurrency},
  booktitle    = {Conditional and Typed Rewriting Systems, 2nd International {CTRS} Workshop, Montreal, Canada, June 11-14, 1990, Proceedings},
  series       = {LNCS},
  volume       = {516},
  pages        = {64--91},
  publisher    = {Springer},
  address      = {Berlin, Germany},
  year         = {1990}
}

@article{Meseguer20,
  author       = {Jos{\'{e}} Meseguer},
  title        = {Generalized rewrite theories, coherence completion, and symbolic methods},
  journal      = {J. Log. Algebraic Methods Program.},
  volume       = {110},
  year         = {2020},
  pages        = {100483}
}

@book{DBLP:books/sp/Libkin04,
  author       = {Leonid Libkin},
  title        = {Elements of Finite Model Theory},
  series       = {Texts in Theoretical Computer Science. An {EATCS} Series},
  publisher    = {Springer},
  address      = {Berlin, Germany},
  year         = {2004},
  doi          = {10.1007/978-3-662-07003-1}
}

@inproceedings{Petria07,
  author       = {Marius Petria},
  editor       = {Till Mossakowski and Ugo Montanari and Magne Haveraaen},
  title        = {An Institutional Version of {G}{\"{o}}del's Completeness Theorem},
  booktitle    = {Algebra and Coalgebra in Computer Science, Second International Conference, {CALCO} 2007, Bergen, Norway, August 20-24, 2007, Proceedings},
  series       = {LNCS},
  volume       = {4624},
  pages        = {409--424},
  publisher    = {Springer},
  address      = {Berlin, Germany},
  year         = {2007}
}

@book{Milner80,
  author       = {Robin Milner},
  title        = {A Calculus of Communicating Systems},
  series       = {LNCS},
  volume       = {92},
  publisher    = {Springer},
  address      = {Berlin, Germany},
  year         = {1980}
}

@book{Milner89,
  author       = {Robin Milner},
  title        = {Communication and concurrency},
  series       = {{PHI} Series in computer science},
  publisher    = {Prentice Hall},
  address      = {New Jersey, United States},
  year         = {1989}
}

@inbook{Goguen2000,
  author       = {Joseph Goguen and Timothy Winkler and Jos{\'{e}} Meseguer and Kokichi Futatsugi and Jean-Pierre Jouannaud},
  title        = {Introducing {OBJ}},
  booktitle    = {Software Engineering with {OBJ}: Algebraic Specification in Action},
  publisher    = {Springer},
  address      = {Berlin, Germany},
  pages        = {3--167},
  year         = {2000}
}

@book{SannellaT2012,
  author       = {Donald Sannella and Andrzej Tarlecki},
  title        = {{Foundations of Algebraic Specification and Formal Software Development}},
  series       = {{Monographs in Theoretical Computer Science. An EATCS Series}},
  publisher    = {Springer},
  address      = {Berlin, Germany},
  year         = {2012}
}

@article{CodescuG08,
  author       = {Mihai Codescu and Daniel G{\u{a}}in{\u{a}}},
  title        = {Birkhoff Completeness in Institutions},
  journal      = {Logica Universalis},
  volume       = {2},
  number       = {2},
  pages        = {277--309},
  year         = {2008}
}

@inproceedings{CodescuMRM10,
  author       = {Mihai Codescu and Till Mossakowski and Adri{\'{a}}n Riesco and Christian Maeder},
  editor       = {Michael Johnson and Dusko Pavlovic},
  title        = {{Integrating Maude into Hets}},
  booktitle    = {Algebraic Methodology and Software Technology - 13th International Conference, {AMAST} 2010, Lac-Beauport, QC, Canada, June 23-25, 2010. Revised Selected Papers},
  series       = {LNCS},
  volume       = {6486},
  pages        = {60--75},
  publisher    = {Springer},
  address      = {Berlin, Germany},
  year         = {2010}
}

@inproceedings{Diaconescu12,
  author       = {R{\u{a}}zvan Diaconescu},
  title        = {Three decades of institution theory},
  editor       = {Jean-Yves B{\'{e}}ziau},
  booktitle    = {Universal Logic: An Anthology},
  pages        = {309--322},
  publisher    = {Springer},
  address      = {Berlin, Germany},
  year         = {2012}
}

@inproceedings{go-icalp24,
  author       = {Go Hashimoto and Daniel G{\u{a}}in{\u{a}} and Ionu{\c{t}} {\c{T}}u{\c{t}}u},
  editor       = {Karl Bringmann and Martin Grohe and Gabriele Puppis and Ola Svensson},
  title        = {Forcing, Transition Algebras, and Calculi},
  booktitle    = {51st International Colloquium on Automata, Languages, and Programming, {ICALP} 2024, July 8-12, 2024, Tallinn, Estonia},
  series       = {LIPIcs},
  volume       = {297},
  pages        = {143:1--143:17},
  publisher    = {Schloss Dagstuhl - Leibniz-Zentrum f{\"{u}}r Informatik},
  address      = {Wadern, Germany},
  year         = {2024}
}

@inproceedings{DBLP:conf/mfcs/GoG25,
  author       = {Go Hashimoto and
                  Daniel G\u{a}in\u{a}},
  editor       = {Pawel Gawrychowski and
                  Filip Mazowiecki and
                  Michal Skrzypczak},
  title        = {Model-Theoretic Forcing in Transition Algebra},
  booktitle    = {50th International Symposium on Mathematical Foundations of Computer
                  Science, {MFCS} 2025, Warsaw, Poland, August 25-29, 2025},
  series       = {LIPIcs},
  volume       = {345},
  pages        = {55:1--55:18},
  publisher    = {Schloss Dagstuhl - Leibniz-Zentrum f{\"{u}}r Informatik},
  address      = {Wadern, Germany},
  year         = {2025},
  url          = {https://doi.org/10.4230/LIPIcs.MFCS.2025.55},
  doi          = {10.4230/LIPICS.MFCS.2025.55}
}

@inproceedings{GoguenM87,
  author       = {Joseph A. Goguen and Jos{\'{e}} Meseguer},
  editor       = {Hartmut Ehrig and Robert A. Kowalski and Giorgio Levi and Ugo Montanari},
  title        = {Models and Equality for Logical Programming},
  booktitle    = {TAPSOFT'87: Proceedings of the International Joint Conference on Theory and Practice of Software Development, Pisa, Italy, March 23-27, 1987, Volume 2: Advanced Seminar on Foundations of Innovative Software Development {II} and Colloquium
                  on Functional and Logic Programming and Specifications {(CFLP)}},
  series       = {Lecture Notes in Computer Science},
  volume       = {250},
  pages        = {1--22},
  publisher    = {Springer},
  address      = {Berlin, Germany},
  year         = {1987}
}

@article{CazanescuD09,
  author       = {Virgil Emil C{\u{a}}z{\u{a}}nescu and Denisa Diaconescu},
  title        = {Completeness of paramodulation without lifting lemma},
  journal      = {Romanian journal of information science and technology},
  volume       = {12},
  number       = {1},
  pages        = {25--34},
  year         = {2009}
}

@book{BaaderN98,
  author       = {Franz Baader and Tobias Nipkow},
  title        = {Term rewriting and all that},
  publisher    = {Cambridge University Press},
  address      = {Cambridge, United Kingdom},
  year         = {1998}
}

@book{Terese03,
  author       = {``Terese'' {(Marc Bezem, Jan Willem Klop, and Roel de Vrijer, eds.)}},
  title        = {Term rewriting systems},
  series       = {Cambridge tracts in theoretical computer science},
  volume       = {55},
  publisher    = {Cambridge University Press},
  address      = {Cambridge, United Kingdom},
  year         = {2003}
}

@incollection{Dershowitz90,
  author       = {Nachum Dershowitz and Jean-Pierre Jouannaud},
  editor       = {Jan van Leeuwen},
  title        = {Rewriting systems},
  booktitle    = {Handbook of Theoretical Computer Science},
  volume       = {{B}: {F}ormal Methods and Se- mantics},
  pages        = {243--320},
  publisher    = {North-Holland},
  address      = {Amsterdam, The Netherlands},
  year         = {1990}
}

@incollection{Klop92,
  author       = {Jan Willem Klop},
  editor       = {Samson Abramsky and Dov Gabbay and Tom Maibaum},
  title        = {Term rewriting systems: from Church-Rosser to Knuth-Bendix and beyond},
  booktitle    = {Handbook of Logic in Computer Science},
  pages        = {1--117},
  publisher    = {Oxford University Press},
  address      = {Oxford, United Kingdom},
  year         = {1992}
}

@incollection{Plaisted93,
  author       = {David Plaisted},
  editor       = {Dov Gabbay and J{\"{o}}rg Siekmann},
  title        = {Equational reasoning and term rewriting systems},
  booktitle    = {Handbook of Logic in AI and Logic Programming},
  publisher    = {Oxford University Press},
  address      = {Oxford, United Kingdom},
  pages        = {274--364},
  year         = {1992}
}

@book{Goguen06,
  author       = {Joseph A. Goguen},
  title        = {Theorem Proving and Algebra},
  year         = {2006},
  publisher    = {arXiv},
  address      = {9500 Gilman Drive, La Jolla CA 92093-0114 USA}
}

@article{MiddeldorpH94,
  author       = {Aart Middeldorp and Erik Hamoen},
  title        = {Completeness Results for Basic Narrowing},
  journal      = {Appl. Algebra Eng. Commun. Comput.},
  volume       = {5},
  pages        = {213--253},
  year         = {1994}
}

@incollection{Goguen00,
  author       = {Joseph A. Goguen and Timothy Winkler and Jos{\'{e}} Meseguer and Kokichi Futatsugi and Jean-Pierre Jouannaud},
  editor       = {Joseph A. Goguen and Grant Malcolm},
  title        = {Introducing {OBJ}},
  booktitle    = {Software Engineering with OBJ: Algebraic Speciﬁcation in Action},
  pages        = {3--167},
  publisher    = {Kluwer Academic},
  address      = {Dordrecht, The Netherlands},
  year         = {2000}
}

@inproceedings{DBLP:conf/concur/Noll99,
  author       = {Thomas Noll},
  editor       = {Jos C. M. Baeten and
                  Sjouke Mauw},
  title        = {On Coherence Properties in Team Rewriting Models of Concurrency},
  booktitle    = {{CONCUR} '99: Concurrency Theory, 10th International Conference, Eindhoven,
                  The Netherlands, August 24-27, 1999, Proceedings},
  series       = {Lecture Notes in Computer Science},
  volume       = {1664},
  pages        = {478--493},
  publisher    = {Springer},
  address      = {Berlin, Germany},
  year         = {1999},
  doi          = {10.1007/3-540-48320-9\_33}
}

@inproceedings{Tutu22,
  author       = {Ionu{\c{t}} {\c{T}}u{\c{t}}u},
  editor       = {Alexandre Madeira and Manuel A. Martins},
  title        = {SpeX: {A} Rewriting-Based Formal Specification Environment},
  booktitle    = {Recent Trends in Algebraic Development Techniques - 26th {IFIP} {WG} 1.3 International Workshop, {WADT} 2022, Aveiro, Portugal, June 28-30, 2022, Revised Selected Papers},
  series       = {Lecture Notes in Computer Science},
  volume       = {13710},
  pages        = {163--178},
  publisher    = {Springer},
  address      = {Berlin, Germany},
  year         = {2022}
}

@inproceedings{Meseguer:General-logics-1989,
  author       = {Jos{\'{e}} Meseguer},
  title	       = {General logics},
  booktitle    = {Logic Colloquium '87},
  crossref     = {Ebbinghaus-FernandezPrida-Garrido-Lascar-RodriquezArtalejo:LC-1987},
  pages	       = {275--329},
  year	       = {1989}
}

@proceedings{Ebbinghaus-FernandezPrida-Garrido-Lascar-RodriquezArtalejo:LC-1987,
  editor       = {Heinz-Dieter Ebbinghaus and Jos{\'{e}} Fern{\'{a}}ndez-Prida and Manuel Garrido and Daniel Lascar and Mario Rodriquez-Artalejo},
  title	       = {Logic Colloquium '87},
  series       = {Studies in Logic and the Foundations of Mathematics Series},
  volume       = {129},
  publisher    = {Elsevier},
  address      = {Amsterdam, The Netherlands},
  year	       = {1989},
}

@book{Harel00,
  author       = {David Harel and Dexter Kozen and Jerzy Tiuryn},
  title        = {Dynamic Logic},
  publisher    = {The MIT Press},
  address      = {Cambridge, MA, United States},
  year         = {2000}
}

@article{VIRY2002487,
  title        = {Equational rules for rewriting logic},
  author       = {Patrick Viry},
  journal      = {Theoretical Computer Science},
  volume       = {285},
  number       = {2},
  pages        = {487-517},
  year         = {2002}
}

@article{DeanG08,
  author       = {Jeffrey Dean and Sanjay Ghemawat},
  title        = {MapReduce: simplified data processing on large clusters},
  journal      = {Commun. {ACM}},
  volume       = {51},
  number       = {1},
  pages        = {107--113},
  year         = {2008}
}
\end{document}